\documentclass[10pt,reqno]{amsart}
\usepackage{amsmath}
\usepackage{amssymb}
\usepackage{amsfonts}
\usepackage{graphicx}
\usepackage{enumerate}
\usepackage[all]{xy}
\usepackage[authoryear,square]{natbib}
\usepackage{hyperref}
\hypersetup{
  colorlinks = true,
  linkcolor=red,   % color of internal links
  citecolor=blue,   % color of links to bibliography
  urlcolor=blue,    % color of external links
  pagebackref=true,
  implicit=false,
  bookmarks=true,
  bookmarksopen=true,
  pdfdisplaydoctitle=true
}
\allowdisplaybreaks

\newtheorem{theorem}{Theorem}[section]

\newtheorem{lemma}[theorem]{Lemma}
\DeclareMathOperator*{\Rem}{Rem}
\DeclareMathOperator*{\Tay}{Tay}

\def\qed{\raise1pt\hbox{\vrule height5pt width5pt depth0pt}}

\def\brm#1\erm{\vskip.8em{\bf\0Remark.}\ #1\vskip.8em}
\def\brrm#1\errm{\vskip.8em{\bf\0Remarks.}\ \bd#1{\hfill\bf erm}\ed}
\def\bpr#1{{\it\0Proof\ #1.}\ }
\def\bpr{{\it\0Proof.}\ }
\def\epr{\quad\qed\vskip.8em}
\def\bac#1\eac{\vskip.8em{\bf\0Acknowledgements}\ #1\vskip.8em}

\numberwithin{equation}{section}

\def\a{\alpha}        \def\b{\beta}         \def\d{\delta}     
\def\e{\varepsilon}   \def\z{\zeta}         \def\h{\eta}     
        \def\l{\lambda}       \def\m{\mu}           
\def\n{\nu}           \def\p{\pi}           \def\r{\rho}
\def\s{\sigma}        \def\t{\tau}                
          \def\ps{\psi}              
    \def\g{\gamma}        \def\G{\Gamma}        
\def\D{\Delta}        \def\L{\Lambda}       
                           
\def\th{\vartheta}    \def\O{\Omega}        
\def\x{\xi}           \def\f{\varphi}       \def\o{\omega}        
       \def\F{\Phi}

\def\PP{{\mathcal P}}\def\EE{{\mathcal E}}\def\MM{{\mathcal M}}\def\VV{{\CMcal V}}
\def\CC{{\mathcal C}}\def\FF{{\mathcal F}}
\def\NN{{\mathcal N}}\def\BB{{\mathcal B}}
\def\RR{{\mathcal R}}\def\LL{{\mathcal L}}
\def\SS{{\mathcal S}}

\def\xx{{\bf x}}

\def\hK{{\widehat  K}}

\def\RRR{\mathbb{R}}
\def\NNN{\mathbb{N}}

\def\CCC{\mathbb{C}}
\def\EEE{\mathbb{E}}

\def\ZZZ{\mathbb{Z}}
  
\def\VV{\mathcal{V}} 

\let\dpr=\partial                  \let\bs=\backslash           
\let\io=\infty                     \let\==\equiv

\def\lft{\left}                \def\rgt{\right}
\def\la{{\langle}}             \def\ra{{\rangle}}

       \def\tilde#1{{\widetilde #1}}

\def\sms{{\SS\hskip-.6em /}}

\def\wh#1{\widehat{#1}}
\def\hat#1{\wh{#1}}
\def\bar#1{\overline {#1}}

\def\lb#1{\label{#1}}
\def\mp#1{\marginpar{\tiny\bf#1}}
\def\mp#1{}
\def\pref#1{(\ref{#1})}

\def\*{{\hfill\break\null\hfill\break}}
\let\0=\noindent

\def\be{\begin{equation}}    \def\ee{\end{equation}}
\def\bea{\begin{eqnarray}}   \def\eea{\end{eqnarray}}
\def\bean{\begin{eqnarray*}} \def\eean{\end{eqnarray*}}
\def\bfr{\begin{flushright}} \def\efr{\end{flushright}}
\def\bc{\begin{center}}      \def\ec{\end{center}}
\def\bal#1\eal{\begin{align}#1\end{align}}
\def\ba#1{\begin{array}{#1}} \def\ea{\end{array}}
\def\bd{\begin{enumerate}[a.\;\leftmargin=1cm]}  \def\ed{\end{enumerate}}
\def\bv{\begin{verbatim}}    \def\ev{\end{verbatim}}

\def\insertplot#1#2#3#4#5#6{%
\begin{figure}[h!]
\begin{center}
\includegraphics[bb=0 0 #1 #2, width=#1pt, height=#2pt, 
clip=true, viewport=0 0 #1 #2]{#4.ps}
\caption{#5}
\end{center}
\end{figure}
}
\begin{document}
\title[Critical exponents of the Coulomb gas]
{Critical exponents of the two dimensional Coulomb gas at the 
 Berezinskii-Kosterlitz-Thouless transition}
\author{Pierluigi Falco}
\address{Department of Mathematics\\
California State University,  
18111 Nordhoff Street\\ 
Northridge, CA 91330}
\email{pierluigi.falco@csun.edu}
\urladdr{http://www.csun.edu/~pfalco/index.html}
\thanks{This work was supported by NSF Grant
PHY-1306571}

%%\dedicatory{Dedicated to Professor
%%X on the occasion of his eightieth birthday}

\maketitle
\begin{abstract}
The 
two dimensional Coulomb gas is the prototypical  model of statistical
mechanics  displaying a special kind of phase transition, 
named after Berezinskii,  Kosterlitz and Thouless. Physicists and
mathematicians proposed several predictions about this system. 
Two of them, valid  along the  
phase transition curve and for  small 
activity, are:  a) the long-distance decay of the ``fractional charge''
correlation is  power law, with
a multiplicative logarithmic correction; b) in such a decay, 
the exponent of the power
law, as well as the exponent of the logarithmic correction, have  a 
certain precise dependence upon  the charge value. 
In this paper we provide a proof of these two
 long standing conjectures. 
\end{abstract}
\setcounter{tocdepth}{2}
\tableofcontents
%\pagina
\section{Introduction}
The Coulomb gas is the infinite system of point particles which carry  positive or  
negative unit electric charges, interact via the electrostatic potential and are
subject to thermal disorder. In this paper we consider the {\it neutral} 
case, in which the total charge of the particle system is zero.  
(This is the case of  major importance in physics; a
non-neutral Coulomb gas could also be defined,  
see Section II.B.2 of \citep{Min87}, and  has a different phenomenology.)
The mathematical difficulty of the model, as well as the reason of physical interest,
stem from the fact that the electrostatic potential in dimension two is
very long range: for large $|x|$ it is 
\be\lb{0}
V(x)=-\frac{1}{2\p} \ln |x| + c_E + o(1),
\ee
where the constant $c_E$ depends on the microscopic regularization. 

The study of the two-dimensional Coulomb gas began in theoretical physics 
with  the suggestion of \cite{Ber71} and of \cite{KoTh73} that this model, 
as well as the related classical XY model,    
undergo a new kind of phase transition, named after them.
Shortly after,
the Berezinskii-Kosterlitz-Thouless (BKT) transition
became one of the fundamental  paradigms of the theory of critical phenomena:
on the one hand, BKT transitions were predicted for 
several other two-dimensional toy models, 
including  solid-on-solid models, 
vertex models, interacting dimers  and other lattice systems 
that can be described in terms of a 
``height function'' (see \citep{JKKN77,Kad78,dN83,Nie84,AJMPMT05}); 
on the other hand,  the BKT transition  turned out to explain the outcomes of
several experiments  on  real-world systems, such as trapped atomic gases, liquid helium films 
and arrays of Josephson junctions (see
\citep{NeKo77,RGSN81,Min87,HKCBD06,HZGC11}
and references therein). 

A precise description of the phase diagram of the Coulomb gas 
was elaborated by \cite{Kos74,JKKN77,GiSc89}. 
The properties of the gas are determined by two parameters: the 
activity $z$ (large $z$ corresponds to  high density of particles) and the inverse temperature 
$\b$ (large $\b$ corresponds to small thermal disorder).
With a non-rigorous renormalization group (RG) argument, 
Kosterlitz found the picture given in Fig.\ref{f0}, which has the following interpretation.
\insertplot{420}{100}
{}%
{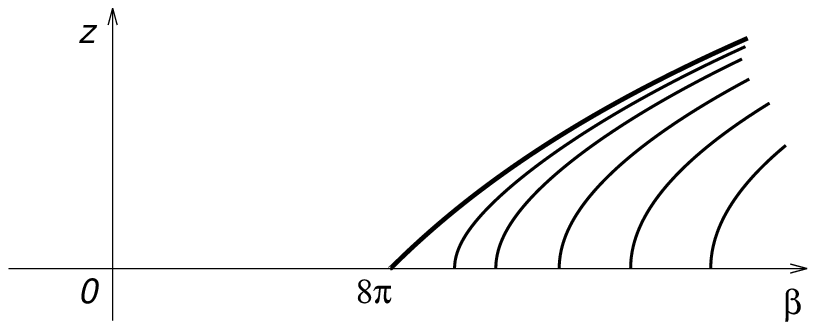}{Diagram of phases: the thicker curve is the BKT transition line.\lb{f0}}{0}  
The thicker line, $\b=\b_{\rm BKT}(z)$,  called 
BKT transition line, divides the $\b$--$z$ plane into two regions,  the 
{\it dipole phase} on the right and the {\it plasma phase} on the left,  
which are characterized by a different behavior of the correlations. 
Let us call {\it charge--$\h$ correlation}, 
$\r_\h(x-y)$, the system response to a probe  of 
charge $\h\in (0,1]$ in position $x$ and a probe of charge $-\h$ in position $y$;
and let us call {\it charge--$\h$ density}, $\r_{1,\h}$, the system response to 
a probe of charge $\h$ at a point $x$.
(A more precise definitions of the former quantity will be given below. 
The latter quantity is by definition non-zero only if $\h=1$). 
The truncated charge correlation is $\r^T_\h(x-y)=\r_\h(x-y)-\r_{1,\h}\r_{1,-\h}$.
It is expected that: 
\bd
\item For $\b>\b_{\rm BKT}(z)$,
the truncated  charge correlation display a power law decay
\be\lb{1}
\r^T_\h(x-y)\sim \frac{C}{|x-y|^{2\kappa}}\;, 
\ee
where $C\=C(z,\b)$ is a prefactor and $\kappa$ is
the {\it correlation critical exponent}. Each thinner line is  
Fig.\ref{f0} is the locus of $(\b,z)$ corresponding to a constant value
of the critical exponent  
\be\lb{2cr}
\kappa=\frac{\b_{\rm eff}}{4\p}\h^2
\ee
with  a  $\b_{\rm eff}\=\b_{\rm eff}(z,\b)>8\p$.
\item
Along the BKT line  $\b=\b_{BKT}(z)$
the truncated charge correlations 
 decay as a  power law, but with a  multiplicative logarithmic correction
\be\lb{1m}
\r^T_\h(x-y)\sim 
\begin{cases}
\frac{C}{|x-y|^{2\kappa}}(\ln|x-y|)^{\frac12}\qquad &\text{for $\h=\frac12$},
\\
\frac{C}{|x-y|^{2\kappa}\; }(\ln|x-y|)^{-\kappa}\qquad&\text{otherwise}.
\end{cases}\;
\ee
The critical exponent $\kappa$ is  constant and given by 
\pref{2cr} for $\b_{\rm eff}=8\p$.
\item  
For  $\b<\b_{BKT}(z)$, truncated charge correlations 
decay exponentially (but only if specific boundary conditions are imposed).
\ed
%For sake of comparison with the constructions that will appear below in this paper, 
%we mention that 
The 
curves in Fig.\ref{f0} were obtained by Kosterlitz as orbits of the ODE
\bal\lb{ds}
&\dot{s}(\ell)=-8\p^2 e^{8\p c_E} z(\ell)^2
\notag\\
&\dot{z}(\ell)=-2 s(\ell) z(\ell)
\eal
where  $\ell$ 
is a length parameter and $s(\ell)=1-\frac{8\p}{\b(\ell)}$.  
$\b(\ell)$ and $z(\ell)$ are effective parameters, obtained 
by averaging  fluctuations over $\ell$-size subparts of the systems: 
hence $\b(0)=\b$ and $z(0)=z$ are the true parameters
of the model; while $\b(\io)$ and $z(\io)$ are the parameters that determine 
the long-distance asymptotic behavior of the correlations.    
The orbits of \pref{ds} are hyperbolas in the $s,z$ variables; 
a sketch of them is in Fig.\ref{fig1}.
\insertplot{600}{150}
{}%
{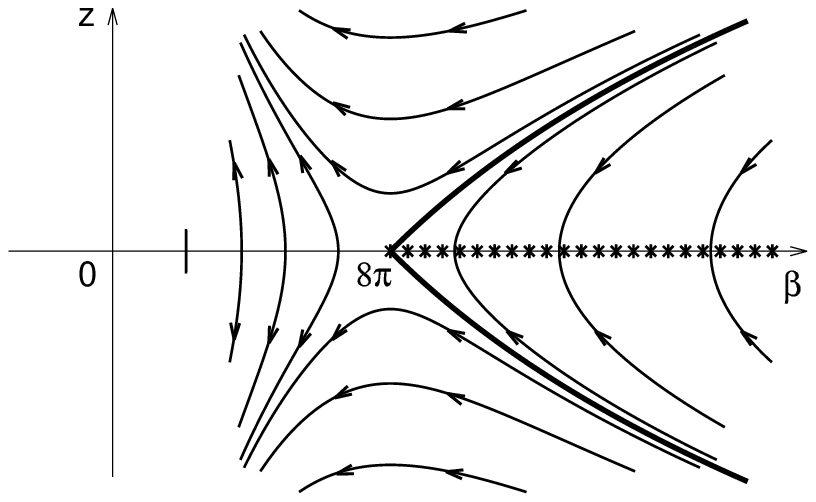}{Diagram of phases. The BKT line is the separatrix of the dynamical system; 
the asterisks denote the 
semi-line of fixed points. \lb{fig1}}{0}  
Only the initial data $(s(0), z(0))$ on the right of a separatrix asymptotically 
evolve to one of the fixed points of the horizontal axis.  
The separatrix is then identified as the BKT line. 
The speed of convergence towards the fixed point turns out to be exponential, 
except when the initial data are along the separatrix: 
in this case the convergence is much slower
\be\lb{negotiate}
s(\ell)=2\p e^{4\p c_E} z(\ell)= \frac{s(0)}{1+2s(0) \ell}
\ee
and this  explains  
the appearance of a logarithmic correction 
in the truncated charge correlations along the BKT line.  

This description of the  phases diagram %for the two dimensional Coulomb gas  
was a breakthrough discovery in physics for the theoretical and the experimental 
implications  mentioned at the beginning of this Introduction; 
however, it  has eluded a  mathematical validation 
for a long time. Indeed   
physicists' results relied on an RG computation at second order in $z$ only; 
higher 
orders are difficult to be taken into account for it is not known whether the 
perturbation theory is ultimately convergent even for small $z$ (see \citep{GN85}).
Besides, in the plasma region,  the second order approximation of the 
 RG flow is divergent and so scarcely reliable. 

Remarkably, the exponential decay of the charge correlations 
in the plasma phase was proven 
to hold by  
\cite{Yan87}, although only  
in a region of the $\b$--$z$ plane that is 
far from the BKT line and only for $\h=1$. 
His approach was not based on an RG argument, but rather 
on an expansion about mean field theory
which was used  by \cite{BryF80} to prove  the Debye screening in 
dimension three. 
That said, from now on we will focus on the dipole phase and the BKT line.  

The fundamental step towards the mathematical understanding 
of  the dipole phase was made by 
\cite{FrSp81b}:  first, by Jensen's inequality, 
they obtained a power law lower bound for $\r_\h(x-y)$; second, 
they developed  a sophisticated {\it multi-scale} decomposition of $\r_\h(x-y)$ 
that  provides an upper bound that is 
also power law. Their result, among the most celebrated ones in rigorous statistical mechanics, 
 had however three substantial limitations: 1) 
the multi-scale method
applied only to {\it fractional charge} correlation, i.e. for  $\h\in (0,1)$, 
and 
2) only in a  region
of the dipole phase that is far from the BKT line;  3) the upper and lower 
bounds, being power-laws with different exponents, 
cannot rule out the presence of multiplicative logarithmic corrections.  
Their  multi-scale method was later improved in a series of papers 
\citep{MKP90,Ma90,Br91,MK91} so to make it applicable 
in a region of dipole phase that touches the BKT line at $z=0$;
but the other limitations remained. 
Noteworthily, Fr\"ohlich-Spencer's calculations
suggested an important refinement of the conjectures: for $\b\ge \b_{\rm BKT}(z)$,
the correct formula for the critical exponent $\kappa$ cannot 
be \pref{2cr}, but one should rather expect that  
\be\lb{2crc}
\kappa=
\begin{cases}
\frac{\b_{\rm eff}}{4\p}\h^2 & {\rm if\ } \h\in (0,\frac12]\\
\frac{\b_{\rm eff}}{4\p}(1-\h)^2 & {\rm if\ } \h\in [\frac12,1)\\
4& {\rm if\ } \h=1.
\end{cases}
\ee
To our understanding, the second and third of \pref{2crc} were overlooked by physicists, 
who mostly had in mind applications with $\h\in (0,\frac12]$.

Several authors advocated the use of a rigorous RG approach 
to have a more direct access to the conjectures.  
This direction was followed  by  \cite{DH00}, who  
used the general RG approach of \cite{BY90} and some new bounds for 
 the charged clusters of particles  to obtain a convergent series 
representation of the {\it free energy} of the Coulomb gas.
This was an important work   
because it provides a method to obtain, in the RG scheme, some of the  
``power counting estimates'' which are implicit in Kosterlitz's analysis. 
However, it  is based on some technical ideas 
that appear to be applicable neither  to  the study of charge correlations 
anywhere
in the dipole phase, 
nor to the evaluation of the free energy at the  
BKT transition line.
These technical problems have 
prevented further mathematical progress in the study the two dimensional Coulomb gas  
for the last ten years.   

The aim of this and of a previous paper, \citep{Fa12}, is to show that the
Brydges-Yau's technique is truly an effective method to deal with the BKT line of
the Coulomb gas. 
In \citep{Fa12}, building on   a technical suggestion  due to D. Brydges 
and on the general scheme of \citep{Br07} (see also \citep{Dim09,BrSl10}),
we already showed that 
some difficulties of \citep{DH00} can be avoided; and that 
a convergent series representation 
for the free energy 
{\it along } the BKT line, for $z$ small enough, can be provided. 
In this paper we take up the mathematically more sophisticated and physically more 
interesting objective of 
studying the long-distance decay of  fractional 
charge correlations \pref{1m}, 
again along  the BKT curve  and for $z$ small enough. 
 
Sharp upper bounds for correlations had
already been obtained  in the general Brydges-Yau's scheme 
in the case of a  different model, the Dipole gas, 
\citep{DH92,BrKe94};
however, those approaches  do not appear to be directly applicable 
to correlations displaying an 
{\it anomalous} decay, such as the power law 
with logarithmic factors that is expected along the BKT line.
Besides, our interest here is the critical exponents, 
therefore we  rather need  
 exact  long-distance asymptotic formulas. 
For these reasons   we introduce in this 
paper a new method to deal with correlations,    
which is inspired, partially, 
on the study of \cite{BGPS94} of fermion systems with anomalous critical exponents. 

For clarity's sake in this paper we only consider the most interesting aspect
of the dipole phase: the correlation of two   
fractional charges for $(\b,z)$ along the BKT line. 
However,  our approach is also applicable to the case of integer charges 
and everywhere in the dipole phase, at least if $z$ is small enough. Furthermore, 
we believe that results on the $n$-points correlations and their scaling limits  
can also be obtained  building on  a method which was 
introduced in  \citep{Fa06,BFM07} to deal with fermion systems $n$-points correlations.

\section{Definition and Results}
The electrostatic interaction is usually defined as the inverse Laplacian; 
in dimension two, however,  the subtraction of a divergent term
is needed to make sense of it.  
For $L$ an odd integer and $R$ another integer, 
consider the finite square lattice 
$$\L=\Big\{(x_0,x_1)\in \ZZZ^2: 
\max\{|x_0|, |x_1|\}< \frac{L^{R}}{2}\Big\}
$$ 
endowed with periodic boundary condition. 
Define the Yukawa interaction on $\L$ with inverse Debye screening length 
$m>0$ as
\be\lb{yuk}
W_\L(x;m):={1\over |\L|}\sum_{k\in \L^*}{e^{ikx}\over m^2-\hat\D(k)},
\ee
where: $\L^*=\{\frac{2\p}{L^{R}}(n_0,n_1): (n_0,n_1)\in \L\}$
is the reciprocal lattice of $\L$;  $|\L|=L^{2R}$ is the volume of $\L$; 
$\hat\D(k)=-2\sum_{j=0,1}(1-\cos k_j)$ is the Fourier transform of  
the discrete Laplacian on $\L$. The {\it two dimensional electrostatic potential} 
is
\be\lb{elp}
W_\L(x|0):=\lim_{m\to 0} \lft[W_\L(x;m)-W_\L(0;m)\rgt]=
{1\over |\L|}\sum_{k\in \L^*\bs \{0\}}{e^{ikx}-1\over -\hat\D(k)}.
\ee
%(Without the subtraction of $W_\L(0;m)$ the limit $m\to 0$ would be divergent.)
It is a classical result, \citep{Sto50},  that  the 
large $|x|$ asymptotic formula for  the infinite volume limit of
$W_\L(x|0)$ is \pref{0}, for 
the $o(1)$ term that is actually $O(\frac{1}{|x|^2})$ and for
$c_E=-\frac{2\g_E +\ln 8}{4\p}$, where $\g_E$ is the Euler's constant.

We can now define the probabilistic model. 
Consider a system of point particles labeled with numbers
$j=1, 2, 3,\ldots,n$;
a configuration $\o$ is the assignment   
to each particle $j$  of a charge
$\s_j=\pm1$ and of  a position $x_j\in \L$.
Let  $\O^0_n$ be the set of the {\it neutral} configurations of $n$ particles, 
i.e. the configurations of $n$ particles
such that $\s_1+\cdots+\s_n=0$.  
The total energy of
$\o\in\O^0_n$ is
\be\lb{ene}
H_\L(\o):=\sum_{i<j=1}^n\s_i\s_j W_\L(x_i-x_j|0).
\ee
We  consider  $\O^0_0$ as made of one configuration, the 
``no particle'' one, with  zero total energy.
For activity $z\ge 0$ and inverse temperature $\b>0$, 
the Grand Canonical partition function of the two dimensional Coulomb gas
is
\be\lb{gf}
Z_\L(\b,z):=\sum_{n\ge 0}
{z^n\over n!}
\sum_{\o\in \O^0_n}
e^{-\b H_\L(\o)}.
\ee  
In the previous paper, \citep{Fa12}, we studied the free energy, 
\be\lb{freeen}
p(\b,z):=-\lim_{\L\to \io} \frac1{\b |\L|} \ln Z_\L(\b,z).
\ee
In this paper we focus on the  fractional charge correlation, which is  defined as a ratio 
of partition functions.
Consider two probes: $p_1$, which is a particle of charge $\h\in (0,1)$ at 
the lattice site
$x$;  and $p_2$, which is a particle of charge $-\h$ at the  lattice site  
$y$. Let $\o\wedge \{p_1,p_2\}$ be the configuration $\o$ augmented of the 
two probes.  Set
\be\lb{gfprobe}
Z^{p_1,p_2}_\L(\b,z):=\sum_{n\ge 0}
{z^n\over n!}
\sum_{\o\in \O^0_n}
e^{-\b H_\L(\o\wedge \{p_1,p_2\})}
\ee
(namely the probes contribute to the energy but not to the entropy of the system). 
The precise definition of $\r_\h(x-y)$ in the Introduction is then
\be\lb{infvollim}
\r_\h(x-y):=\lim_{\L\to \io} \frac{Z^{p_1,p_2}_\L(\b,z)}{Z_\L(\b,z)}.
\ee
The invariance of  \pref{infvollim}
under translations of  the probes is a consequence of the definition. 
The existence of the infinite volume limits will be proved in  the theorem below. 
When $z=0$, the BKT point is at $\b=8\p$; at these values of the parameters 
and for every  $\h$, a simple computation gives 
\be\lb{0dec}
\r_\h(x)=
\frac{e^{8\p \h^2 c_E}}{|x|^{4\h^2} }(1+o(1)),
\ee
where $o(1)$ is vanishing in the limit of $|x|\to \io$. 
When $z\neq 0$ the situation is more complicated. 
\begin{theorem}\lb{t1}
Fixed $\h\in(0,1)$,  there exist an $L_0\=L_0(\h)>1$,  
a $z_0\=z_0(\h)>0$ and an inverse temperature  
$\b_{BKT}(z)\ge 8\p$ such that if $L\ge L_0$, $0<z\le z_0$ and 
$\b=\b_{BKT}(z)$,  the limit \pref{infvollim}
exists and: 
\bd
\item If  $\h\neq \frac12$, then
\be\lb{zdec}
\r_\h(x)=\r^{(a)}_{\h}(x) + \r^{(b)}_{\h}(x),
\ee
where, for  $x$-independent $f_a,f_b,f$, 
\bal\lb{zdec1}
&\r^{(a)}_{\h}(x)=\frac{e^{8\p \h^2 c_E} +f_a}{|x|^{4\h^2} \lft(1+f\ln|x|\rgt)^{2\h^2} }\lft(1+o(1)\rgt),
\notag\\
&\r^{(b)}_{\h}(x)=\frac{f_b}{|x|^{4(1-\h)^2} \lft(1+f\ln|x|\rgt)^{2(1-\h)^2} }\lft(1+o(1)\rgt). 
\eal
\item 
If $\h=\frac12$, then, for $x$-independent $f_a,f$, 
\be\lb{zdec2}
\r_{\frac12}(x)=\frac12\frac{e^{2\p c_E}+f_a}{|x| }\lft(1+f\ln|x|\rgt)^{\frac12} \lft(1+o(1)\rgt).
\ee
\ed
In the above formulas, $o(1)$ are vanishing terms for $|x|\to\io$; 
$f=c z$ for $c>0$, $f_b=c(\h)^2 z^2(1+\tilde f_b)$ for $c(\h)>0$;   
$f_a$, $\tilde f_b$ are vanishing in the limit $z\to 0$. Besides $z_0(\h)$ is such that,
for every $[a,b]\subset(0,1)$, one has
$\inf\{z_0(\h):\h\in [a,b]\}>0$.
\end{theorem}
This is the main result of the paper. 
\brrm
\item 
In the limit $|x|\to \io$, \pref{zdec} and \pref{zdec2} coincide with \pref{1m} 
for exponent \pref{2crc} and $\b_{\rm eff}=8\p$. 
For this reason,  we identify 
the curve $\b=\b_{BKT}(z)$ with the Berezinskii-Kosterlitz-Thouless transition line. 
Whether $\b<\b_{BKT}(z)$ implies an exponential decay of truncated correlations
is an open problem; the only available rigorous result,  \citep{Yan87},
is for $\b\ll \b_{BKT}(z)$. 
\item  A heuristic interpretation of the result, neglecting 
for a moment the logarithmic corrections,  is the following. 
A probe charge $\h$ at an inverse temperature $\b=\b_{BKT}(z)$ 
placed inside the interacting system 
is equivalent  to a ``virtual'' point charge $\h+m$ at inverse temperature $8\p$ 
placed inside a free  system.  
Here $m$ represents a local fluctuation of unit charges and can 
be any positive or negative integer value.  
By choosing the two smallest values 
of the virtual charge critical exponent,  
$2(\h+m)^2$, one obtains the leading parts of $\r^{(a)}$  and $\r^{(b)}$ in \pref{zdec}.
\item A justification of the logarithmic factor in \pref{zdec} is more subtle 
and will emerge from the multi-scale approach used in the proof. 
The different formula for the case $\h=\frac12$ is related to the fact that, 
continuing with  the argument in the previous point, 
only at this value of $\h$ there are two different 
values of $m$ that minimize the virtual charge correlation exponent.
\item
If $\h\in(\frac12, 1)$, the critical  exponent of the 
free case, $2\h^2$,  differs from the one of  the interacting case, $2(1-\h)^2$; 
despite that, 
there is no  discontinuity in the behavior of the correlation at $z=0$. 
Indeed,
note that $\r_\h^{(a)}$ has a prefactor $O(1)$, whereas 
$\r_\h^{(b)}$ has a prefactor $O(z)$. 
Therefore, 
the smaller $z$, the larger the threshold distance passed which $\r_\h^{(b)}$ dominates 
over $\r_\h^{(a)}$; for $z\to 0$ such threshold distance 
is infinite, and the free case critical exponent  is recovered.
\item
Since the logarithmic corrections have $O(z)$ prefactors,
by the same argument of the previous point,  
in the limit $z\to 0$ the purely power law decay of the free case is 
recovered.
\item The prefactor $\frac12$ in \pref{zdec} is absent in formula for the correlation 
in the case $z=0$. Again, this is not a sign of discontinuity: as it can be traced 
in the proof of the Theorem,   among 
the $o(1)$ terms in \pref{zdec} there is one that in the limit $z\to0$ does not vanish, 
ceases to be subleading 
and, with its contribution,  restores the  prefactor $1$ in the leading term.   
\errm
In the next section we provide the detailed renormalization group construction  
that directly implies Theorem \ref{t1}. The reader with some 
familiarity with physicists' renormalization group jargon will 
recognize in the right hand side of \pref{lk} the {\it beta function} of the model; 
and in the right hand side of \pref{sk} the {\it gamma function}. 
The major technical novelty of \citep{Fa12}, with respect to \citep{DH00}, 
was the derivation,  in the setting of the Brydges-Yau's expansion, 
of the dynamical system \pref{lk} and of new bounds to control it. That 
allowed us to obtain a convergent series representation 
of the free energy at the BKT transition.   
The most important contribution of this paper 
is the introduction, again in the framework of the Brydges-Yau's technique, 
of {\it renormalization constants} for the observables  --namely for 
the {\it fractional charges}--  which are described by the dynamical system \pref{lk}. 
That allows us to obtain \pref{zdec} and \pref{zdec2}, partly by
bounds and partly by an 
explicit computation of the leading term of the solution of
\pref{lk}. 

In the forthcoming analysis, we will work with five  parameters: given 
a charge $\h\in (0,1)$ and $0<\t\le \t_0$,  we will need  $L\ge L_0(\h,\t)$,  
$A\ge A_0(\h,\t, L)$ 
and $0<z\le z_0(\h,\t, L, A)$ in order for the results  to be valid.  
We will also have other two parameters, $\a$ and $h\=h(\a)$, which 
however are eventually fixed by the condition $\a^2=8\p$.  Finally, 
in our notation,  $C$, $C_0$, $C_1$ or $c_0$
might represent different prefactors  when they appear in  different bounds.

\section{Strategy of the Proof}\lb{strategy}
\subsection{Multiscale approach}
Since $W_\L(x-y; m)$ has strictly positive Fourier transform, 
a Gaussian field  $\{\f_x:x\in \L\}$  is defined by assigning 
zero mean and covariance 
\be\lb{wruble}
\EEE_{m,\b}\lft[\f_x \f_y\rgt]=\b W_\L(x-y; m).
\ee
By means of the {\it sine-Gordon transformation},
such a finite-dimensional measure provides a functional integral representation 
for the partition function
\be\lb{whitman0}
Z_\L(\b,z)=\lim_{m\to 0}\EEE_{m,\b}\lft[e^{2z\sum_{x\in\L}\cos \f_x}\rgt],
\ee
as well as for the %charge-$\h$ density and  
correlation 
\be\lb{whitman}
%\r_{1,\s\h}=\lim_{\L\to\io}\la e^{i\h\s\f_{x}}\ra_{\L},\qquad
\r_\h(x-y)=\lim_{\L\to\io}\la e^{i\h(\f_{x}-\f_{y})}\ra_{\L},
\ee
where
$$
\la\cdot\ra_\L:=
\frac{\lim_{m\to 0}\EEE_{m,\b}\lft[e^{2z\sum_{x\in\L}\cos \f_x}\;\cdot\;\rgt]}
{\lim_{m\to 0}\EEE_{m,\b}\lft[e^{2z\sum_{x\in\L}\cos \f_x}\rgt]}.
$$
The proof of \pref{whitman0} and \pref{whitman} is in  Appendix \ref{appA}.
In the RG approach it is  natural to study \pref{whitman0} and 
 \pref{whitman} through 
the  generating functional of the correlations of $e^{i\h \f_{x}}$: 
define 
\be\lb{skf}
\O(J,\L):=\lim_{m\to 0}\EEE_{m,\b}\Big[
e^{2z\sum_{x\in\L} \cos\f_x+ \sum_{x\in\L}\lft(J_{x,+}e^{i\h\f_x}+
J_{x,-} e^{-i\h\f_x}\rgt)}\Big]
\ee
where $\{J_{x,\s}: x\in \L, \s=\pm1\}$ are real variables;  then
\bal 
&p(\b,z)=-\lim_{R\to \io} \frac{1}{\b |\L|}\ln \O(J,\L)\Big|_{J\=0},
\lb{pres0}
\\
\lb{correlazione0}
%&\r_{1,\s\h}
%=
%\lim_{R\to \io}\frac{1}{Q(J,\L)}
%\frac{\dpr\O(J,\L)}{\dpr J_{x,\s}}\Big|_{J\=0},
%\notag\\
&\r_\h(x-y)
=
\lim_{R\to \io}\frac{1}{Q(J,\L)}
\frac{\dpr^2\O(J,\L)}{\dpr J_{x,+}\dpr J_{y,-}}\Big|_{J\=0}.
\eal
The point of departure of the RG analysis is a multi-scale representation 
of $\O(J,\L)$.
We need some further notations. The two independent unit vector of the 
lattice are  $e_0=(1,0)$ and $e_1=(0,1)$. Consider the set of unit vectors 
$\hat u=\{\pm e_0, \pm e_1\}$:
for any $\m\in \hat u$ define 
the discrete partial derivative as $\dpr^\m \f_x:=\f_{x+\m}- \f_{x}$ if $\m=e_0, e_1$, 
or as   $\dpr^\m \f_x:=\f_{x}- \f_{x+\m}$ if $\m=-e_0, -e_1$.
Correspondingly define the vector component $x^\m:=x\cdot \m$ if $\m=e_0, e_1$, 
and  $x^\m:=-x\cdot \m$ if $\m=-e_0, -e_1$. 
In our notation, 
every sum  $\sum_{\m\in \hat u}$ will also imply a factor $\frac12$ that we do
not write
explicitly.  This means,  for example, that the Fourier transform of 
$\sum_{\m\in \hat u} \dpr^{-\m} \dpr^\m$ coincides with $\hat \D(k)$ defined after \pref{yuk}; 
and that  the discrete form 
of the first order Taylor expansion of a lattice function 
$f_y-f_x$ 
is $\sum_{\m\in \hat u} (\dpr^\m f_x)(y^\m-x^\m)$%
\footnote
{in the sense that, for any lattice path $p_{x,x}$ that joins $x=(x_0,x_1)$ 
with $y=(y_0, y_1)$ and 
has length $|y_0-x_0| +|y_1-x_1|$,  
$$
\Big|f_y-f_x-\sum_{\m\in \hat u} (\dpr^\m f_x)(y^\m-x^\m)\Big|
\le 4\max_{j=0,1} |y_j-x_j|^2 
\max_{z\in p_{x,y}}\max_{\m_1,\m_2\in \hat u}|\dpr^{\m_1}\dpr^{\m_2} f_z|
$$
}.   
In Appendix \ref{appA} we prove the multiscale functional 
integral representation 
\be\lb{pot0}
\O(J,\L)= e^{E|\L|}\lim_{m\to 0} \EEE_R \cdots \EEE_0
\lft[e^{\VV(J,\z^{(0)}+\z^{(1)}+ \cdots+ \z^{(R)} )}\rgt],
\ee
where,  fixed any $s\in (0,\frac12)$ and for $\a^2:=\b(1-s)$: 
\bd
\item   
$E=\frac12\ln (1-s)$ and the interaction $\VV(J,\f)$ is
\bal\lb{pot}
\VV(J,\f):={s\over 2}\sum_{x\in \L\atop \m\in \hat u}(\dpr^\m\f_x)^2+
z\sum_{x\in\L\atop\s=\pm1}e^{i\s\a\f_x}+\sum_{x\in \L\atop\s=\pm1}
J_{x,\s} e^{i\h\a\s\f_x}.
\eal
\item $\z^{(0)}, \ldots,\z^{(R)}$ are two-by-two independent 
Gaussian fields,  each of which has zero mean and covariance
\be\lb{expectations}
\EEE_j[ \z^{(j)}_x\z^{(j)}_y]=
\begin{cases}
\G_j(x-y)\qquad &\text{for } j=0,1,\ldots, R-1\\
\G'_R(x-y) \qquad & \text{for } j=R.
\end{cases}
\ee
Each $\G_{j}$ is  independent of $m$ and $\L$ and,   
for positive $C_q$ and $c$,
\bal
\lb{p1}&\G_j(x)=0
&{\rm for\ }|x|\ge L^{j+1}/2,
\\
\lb{p2}&|\dpr^{\m_1}\cdots\dpr^{\m_q} \G_j(x)|\le C_q L^{-jq}
&\text{for any $\m_j\in \hat u$ and any $q\ge1$},
\\
\lb{p3}&\G_j(0)={1\over 2\p}\ln L + c_j(L)
&
{\rm for\ } |c_j(L)|\le c L^{-\frac{j}{4}}, 
\eal
The covariance $\G'_R$, instead,  depends upon $m$ and $\L$. 
One has  
\be\lb{p4a}
\lim_{m\to 0} \G'_R(0)=+\io, 
\ee
while, if $\G'_R(x|0):=\G'_R(x)-\G'(0)$, 
\be\lb{p4b}
\lim_{R\to \io}\lim_{m\to 0} \G'_R(x|0)=0. 
\ee
The limit \pref{p4b} implies 
\be\lb{p4}
\G_{\io,0}(x|0):=\sum_{j=0}^\io \lft[\G_{j}(x)-\G_{j}(0)\rgt]=
-\frac{1}{2\p} \log |x| +c_E+o(1).
\ee
\ed
The meaning of \pref{p2} and \pref{p1}
is  that  $\G_j$ carries a typical momentum $O(L^{-j})$ and 
has a compact support of side length $O(L^{j+1})$.
The precise construction  of  
$\G'_{R}$ and of $\G_0, \ldots, \G_{R-1}$ was given in \citep{Fa12} 
building on \citep{BrGuMi04}; a review is in   
Appendix \ref{appA}. 

Note that the expectations in \pref{expectations} 
are independent of $\b$, while the interaction in \pref{pot} is dependent on the 
new parameters $\a$ and $s$. The relationship among $\a$, $s$ and $\b$ and their role
in the forthcoming analysis is the following.   
The parameter $s\=s(z)$ is introduced so that 
the curve in Fig.\ref{f0} that corresponds to a system 
with effective inverse temperature 
$\a^2$ has graph $\b=\b_{\a}(z)$, where
\be
\lb{p4}
\b_{\a}(z)=\frac{\a^2}{1-s(z)}.
\ee
Although in many sub-results we will leave an explicit 
dependence on $\a$, for Theorem \ref{t1} we will eventually set
$\a^2=8\p$, which means that in the statement of that Theorem 
$\b_{ BKT}(z)\=\b_{\sqrt{8\p}}(z)$.  

The RG approach consists in computing the integrals in \pref{pot0} 
progressively from the random variable with highest momentum  to the
one with  lowest. First, set
\bal\lb{itr}
&\O_{1}(J,\f,\L):=e^{E|\L|}\EEE_0\lft[e^{\VV(J,\f+\z^{(0)})}\rgt];
\eal
then, inductively for $j=2,\ldots, R$, set
\bal\lb{itrj}
&\O_{j}(J,\f,\L)
:=\EEE_{j-1}\lft[\O_{j-1}(\L;J,\f+\z^{(j-1)})\rgt];
\eal
at last, one finds
\bal\lb{0itr}
&\O(J,\L)=\lim_{m\to 0}\EEE_{R}\lft[\O_{R}(\L;J,\z^{(R)})\rgt].
\eal
In this way the evaluation of the partition function
is transformed into the evaluation of a sequence of 
{\it effective generating functionals}
$\O_1,\ldots, \O_{R}, \O$. 
\subsection{Polymer gas representation}
Following \citep{BY90,Br07,BrSl10}, each $\O_j$
can be efficiently represented as a {\it polymer gas}. 
Before describing this formulation,  we have to 
introduce a multiscale decomposition of the lattice 
and, correspondingly, special types of  lattice domains. 
\\
{\it\0a) Blocks.} Set $|x|:=\max\{|x_0|,|x_1|\}$.
Recall that each side of the square lattice $\L$ 
is made of $L^{R}$ sites, where $L$ is odd; for $j=0,1,\ldots,R$,  
pave the periodic lattice $\L$ with $L^{2(R-j)}$ disjoint 
squares of $L^{2j}$ sites, in such a way that there is a central square,
$$
\big\{x\in \L:|x|\le L^{j}/2 \big\}
$$
and all the other squares are translations of this one by 
vectors in $L^{j}\ZZZ$. An example is in  Fig. \ref{fig2}.
\insertplot{420}{220}
{}%
{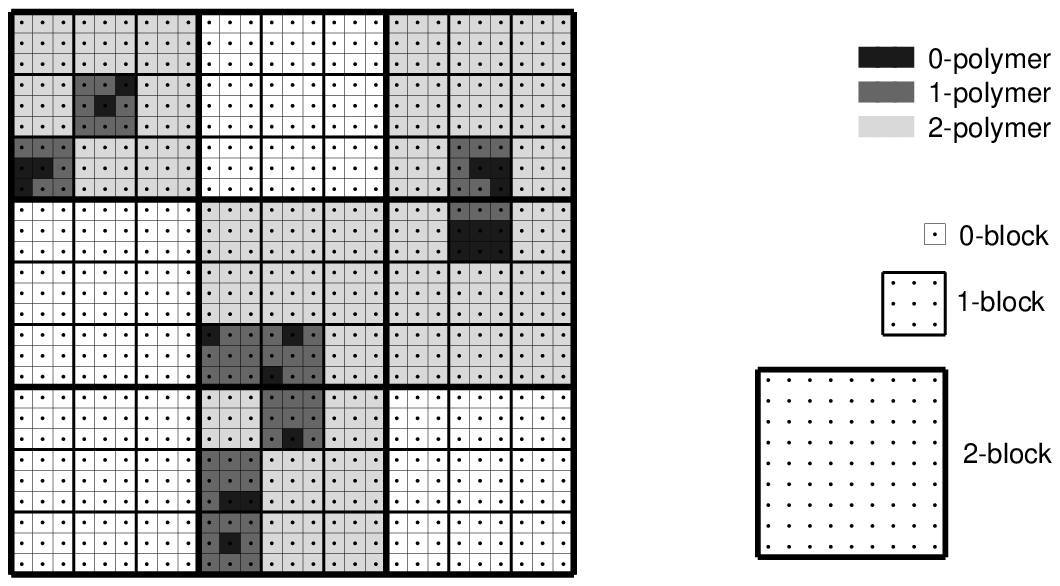}{Lattice paving with blocks of different sizes
  in the  case $L=3$ and $R=3$. \lb{fig2} }{0}
We call such squares $j$--blocks, and
we denote the set of all $j$--blocks by $\BB_j\=\BB_j(\L)$. $0$--blocks
are made of  single points: $\BB_0=\L$. 
%The other extreme case is 
%$\BB_R=\{\L\}$.  
\\
{\it\0b) Polymers.} 
A union of two-by-two different $j$--blocks is called $j$--polymer, 
and the set of all $j$--polymers 
in $\L$ is denoted $\PP_j\=\PP_j(\L)$. Suppose $X$ is a  $j$--polymer:  
$\dpr X$ is the set of sites in $X$ with a nearest neighbor outside $X$; 
$\dpr_{ext} X$ is the set of sites outside  $X$ with a nearest neighbor inside 
$X$;
$\BB_j(X)$ is the set of the $j-$blocks in $X$;
$|X|_j$ is the cardinality of $\BB_j(X)$;
the {\it closure} $\bar X$ is the smallest polymer in $\PP_{j+1}(\L)$ 
that contains $X$.  
\\
{\it\0c) Connectivity.} 
A polymer made of  two different blocks, 
$B,B'\in \BB_j$, 
is connected
if there exist $x\in B$ and $x'\in B'$ s.t. $|x-x'|=1$; the
definition extends to connected polymers of more blocks in the usual
way.  For example, in Fig. \ref{fig2} there is one connected 2-polymer, which is the 
closure of three connected $1$--polymers, which in turn are the closure of  
ten connected $0$--polymers.  
$\PP_j^c\=\PP^c_j(\L)$ is the  set of the connected $j$--polymers; 
the collection of the maximal connected parts of a $j$--polymer $X$
(each of which is a $j-$polymer by construction) is called 
$\CC_j(X)$.
\\
{\it\0d) Small polymers.} 
The polymer $X$ is {\it small} if it is connected and $|X|_j\le 4$.
The set of the small $j$--polymers  
will be called $\SS_j\=\SS_j(\L)$;    
the set of the connected  $j$--polymers  that are not small
will be called $\sms_j\=\sms_j(\L)$; the  number of 
small $j$--polymers that contain a given $j$--block is 
independent of $j$ and will be called  $S$. The 
{\it small set neighborhood} of a $j$--polymer $X$ is 
the set $X^*:=\cup\{Y\in \SS_j:Y\cap X\neq \emptyset\}$.
\\
{\it\0d) Empty set.}  
The empty set is considered as an element of $\PP_j$, 
but not of $\PP^c_j$.

We will assume that $L\ge 16$ so that, 
if $X\in \PP_j^c$, then the set $X^*\bs X$ is a ``small 
margin''  around $X$, in the following sense: if $X, Y\in \PP_j^c$ and $\bar X$, $\bar Y$
are separated by at least one $j+1$ block, 
then 
\be\lb{separation}
\min\{|x-y|: x\in X^*, y\in Y^*\}\ge L^{j+1}- 8L^{j}\ge \frac12 L^{j+1}
\ee
which, by \pref{p1},  is larger than the range of $\G_j$. 

Now we pass to the polymer gas representation of the generating functional. 
Set $\F=(J,\f)$. For each scale $j=1, \ldots, R$, assume that five real parameters, 
$E_j$ and  $t_j:=(s_j, z_j, Z_j, \bar Z_j)$ are given; and 
assume that $\O_j(\F,\L)$ has the form
\bal\lb{pfr}
\O_j(\F,\L)=e^{|\L|E_j}
\sum_{X\in \PP_j} e^{U_j(\F,\L\bs X)} 
\prod_{Y\in \CC_j(X)} K_j(\F,Y),
\eal
where
the definitions of the interaction $U_j$ and of the polymer activity $K_j$ follow. 
Given a $j$--block  $B$, the interaction is 
\be\lb{uu}
U_j(\F,B)=V_j(\F,B)+W_j(\F,B).
\ee
The first term, $V_j$, is similar to the initial interaction \pref{pot} 
and is the sum of $V_{0,j}$ and $V_{1,j}$, for 
\bal \lb{vv}
&V_{0,j}(\f,B)={s_j\over 2}\sum_{x\in B\atop \m\in \hat u}
(\dpr^\m\f_x)^2+z_jL^{-2j}\sum_{x\in B\atop \s=\pm1}e^{i\a\s\f_x},
\notag\\
&V_{1,j}(\F,B)=
Z_j L^{-2j}\sum_{x\in B\atop \s=\pm1}J_{x,\s}e^{i\h\a\s\f_x}+
\bar Z_j L^{-2j}\sum_{x\in B\atop \s=\pm1}J_{x,\s} e^{i\bar\h\a\s\f_x}.
\eal
Here, $\bar \h:=\h-1$; therefore, as $\h\in (0,1)$, also  $-\bar\h\in
(0,1)$.   The factors $L^{-2j}$ make $V_j$ explicitly
dependent on the scale $j$; besides, $V_j$   
depends upon the fields 
 $\{\f_x: x\in B\cup \dpr_{ext} B\}$ and $\{J_{x,\s}: x\in B, \s=\pm1\}$
and upon the parameters
$t_j$.  Note that $z_j$ and $s_j$ play the role of the 
the effective parameters discussed in the Introduction;  whereas  
$Z_j$ and $\bar Z_j$ are the ``fractional charge renormalization constants''. 
 
The second term in \pref{uu}, $W_j(\F,B)$, is generated by the 
multi-scale integration: $W_0(\F,B)=0$; while, for $j\ge 1$,  
inductively assume that  $W_j(\F,B)$
depends upon the scale $j$, upon the fields $\{\f_x, J_{x,\s}:x\in B^*, \s=\pm1\}$, 
and upon the parameters $t_j$.  
We give now a partially explicit formula for $W_j$; the $w$'s functions that
appear in \pref{wj0}, \pref{wj1} and \pref{wj2} will be defined in Section \ref{sc20}.
$W_j$ is the sum of  three terms:
$W_{0,j}(\f,B)$, $W_{1,j}(\F,B)$ and $W_{2,j}(\F,B)$, 
where the enumeration corresponds to the powers of $J$ as we now explain. 
 $W_{0,j}$ contains terms that are independent of $J$ and quadratic in
$s_j$, $z_j$:
\bal\lb{wj0}
&W_{0,j}(\f,B)
=
-s^2_j
\sum_{y\in \ZZZ^2\atop \m, \n\in \hat u} w_{0,a,j}^{\m\n}(y)
\sum_{x\in B}(\dpr^\m\f_x)
\Big[(\dpr^\n\f_{x+y})- (\dpr^\n\f_x)\Big]
\notag\\
&+z^2_j\sum_{y\in \ZZZ^2}w_{0,b,j}(y)
\sum_{x\in B\atop\s=\pm} 
\Bigg[ e^{i\s\a(\f_x-\f_{x+y})}
-1
+|y|^2\frac{\a^2}4\sum_{\m\hat u}(\dpr^{\m}\f_x)^2
\Bigg]
\notag\\
&+z^2_j\sum_{y\in \ZZZ^2} 
w_{0,c,j}(y)
\sum_{x\in B\atop\s=\pm}
e^{i\s\a(\f_x+\f_{x+y})}
\notag\\
&+z_j s_j
\sum_{y\in \ZZZ^2\atop \m\in \hat u} 
w^\m_{0,d,j}(y)
\sum_{x\in B\atop\s=\pm}
i\s\lft[
e^{i\s\a\f_x}(\dpr^\m\f_{x+y})-
e^{i\s\a\f_{x+y}}(\dpr^{-\m}\f_x)\rgt]
\notag\\
&-z_j s_j
\sum_{y\in \ZZZ^2}
w_{0,e,j}(y)\sum_{x\in B\atop\s=\pm}
\lft(e^{i\s\a\f_{x+y}}-e^{i\s\a\f_x}\rgt).
\eal
$W_{1,j}$ contains terms linear in  $J$, and linear  in $s_j$ or $z_j$:
\bal\lb{wj1}
&W_{1,j}(\F, B)=
z_j Z_j L^{-2j}\sum_{y\in \ZZZ^2} w_{1,b,j}(y)\sum_{x\in B\atop \s=\pm}
J_{x,\s} e^{i\a\s(\h\f_x+\f_{x+y})}
\notag\\
&+z_j \bar Z_j L^{-2j}\sum_{y\in \ZZZ^2} \bar w_{1,b,j}(y)\sum_{x\in B\atop \s=\pm}
J_{x,\s} e^{i\a\s(\bar \h\f_x-\f_{x+y})} 
\notag\\
&+z_j Z_j L^{-2j}\sum_{y\in \ZZZ^2} w_{1,c,j}(y)\sum_{x\in B\atop \s=\pm}
J_{x,\s} e^{i\a\s\bar \h\f_x}
\lft[e^{-i\a\s(\f_{x+y}-\f_x)}-1+i\a\s y^\m\sum_{\m\in \hat u}(\dpr^\m \f_{x})\rgt]
\notag\\
&
+z_j \bar Z_j L^{-2j}\sum_{y\in \ZZZ^2} \bar w_{1,c,j}(y)\sum_{x\in B\atop \s=\pm}
J_{x,\s}e^{i\a\s\h\f_x}\lft[ e^{i\a\s(\f_{x+y}-\f_x)} -
1-i\a\s y^\m\sum_{\m\in \hat u}(\dpr^\m \f_{x})\rgt]
\notag\\
&
+
s_j Z_j L^{-2j} \sum_{y\in \ZZZ^2\atop\n\in \hat u}w^\n_{1,d,j} (y)
\sum_{x\in B\atop  \s=\pm}J_{x,\s} e^{i\h\a\s\f_x} 
\s \lft[(\dpr^\n\f_{x+y})- (\dpr^\n\f_{x})\rgt]
\notag\\
&
+
s_j \bar Z_j L^{-2j} \sum_{y\in \ZZZ^2\atop\n\in \hat u}\bar w^\n _{1,d,j} (y)
\sum_{x\in B\atop  \s=\pm}J_{x,\s} e^{i\bar\h\a\s\f_x} 
\s \lft[(\dpr^\n\f_{x+y})- (\dpr^\n\f_{x})\rgt].
\eal
Finally, $W_{2,j}$ contains the terms quadratic in $J$, and
independent of $s$ or $z$: 
\bal\lb{wj2}
&W_{2,j}(\F, B)=
\sum_{y\in\ZZZ^2\atop\e=\pm}w^\e_{2,a,j}(y)
\sum_{x\in B\atop \s=\pm}J_{x,\s}J_{x+y,\s\e} 
e^{i\h\a\s(\f_x+\e\f_{x+y})}
\notag\\
&+\sum_{y\in\ZZZ^2\atop\e=\pm}\bar w^\e_{2,a,j}(y)
\sum_{x\in B\atop \s=\pm}J_{x,\s}J_{x+y,\s\e} 
e^{i\bar\h\a\s(\f_x+\e \f_{x+y})}
\notag\\
&+\sum_{y\in\ZZZ^2\atop\e=\pm}w^\e_{2,b,j}(y)
\sum_{x\in B\atop \s=\pm}
J_{x,\s}J_{x+y,\s\e} 
\lft[e^{i\a\s(\h\f_x+\e\bar\h \f_{x+y})}+
e^{i\a\s(\bar\h\f_x+\e\h \f_{x+y})}\rgt]
\notag\\
&+\sum_{y\in\ZZZ^2\atop \e=\pm}w^\e_{2,c,j}(y)
\sum_{x\in B\atop \s=\pm}
J_{x,\s}J_{x+y,\e\s} e^{i\a\s (1+\e)(\h-\frac12)\f_x}.
\eal
We extend these definitions from $j$--blocks to $j$--polymers additively: 
for $X\in \PP_j$:
\be\lb{nt}
U_j(\F,X):=
\sum_{B\in \BB_j(X)}U_j(\F,B);
\ee
$V_j(\F,X)$ and $W_j(\F,X)$ are defined in the same way.

Returning to the explanation of \pref{pfr},
the polymer activity, 
$K_j(\F,X)$, is also generate by the multi-scale integration: 
$K_0(\F,X)=0$; while, for $j\ge 1$, 
$K_j(\F,X)$ depends upon  $\{\f_x, J_{x,\s}:x\in X^*, \s=\pm1\}$ and 
is the sum  of four terms, 
\be\lb{pinning0}
K_j(\F,X)=K_{0,j}(\f,X)+K_{1,j}(\F,X)+K_{2,j}(\F,X)
+K_{\ge3,j}(\F,X)
\ee
where, again, the enumeration refers to  the
powers of $J$. The last term 
is proportional to the third power or an higher power of $J$:  
it will not play any role in the analysis of this paper, 
since eventually we are only interested in up to
two derivatives in $J$ at $J=0$. The second and third terms  
can be further decomposed: 
\bal\lb{pinning}
K_{1,j}(\F,X)=
&L^{-2j}\sum_{x\in X\atop \s=\pm1} 
J_{x,\s} \lft[Z_j K_{1,j}(\f,X, x,\s)
+\bar Z_j K^\dagger_{1,j}(\f,X, x,\s)\rgt],
\notag\\
K_{2,j}(\F,X)=
&\sum_{x_1\in X, x_2\in X^*\atop \s_1,\s_2=\pm1}
J_{\s_1, x_1}J_{\s_2, x_2}K_{2,j}(\f,X,x_1,\s_1,x_2,\s_2).
\eal
%
%(That means  that $K_{1,j}$ is assumed to depend on $J_{\s,\xx}$ for $\xx\in X$, 
%not in the bigger $X^*$).  
Note that   
$K_{1,j}(\f,X, x,\s)$ and $K^\dagger_{1,j}(\f,X, x,\s)$ are ``pinned'' at $x$
in the sense that they are defined by \pref{pinning} 
only for $x\in X$; we extend 
their definitions by setting   
$K_{1,j}(\f,X, x,\s)=K^\dagger_{1,j}(\f,X, x,\s)=0$ whenever $x\not\in X$. 
In the same way, $K_{2,j}(\f,X,x_1,\s_1,x_2,\s_2)$ is pinned at 
$x_1$ and $x_2$ and we set 
$K_{2,j}(\f,X,x_1,\s_1,x_2,\s_2)=0$ if $x_1\not\in X$ or $x_2\not\in X^*$. 
Besides note that at least one power of $J$ is assumed to be 
restricted to the set $X$ (indeed, the same sort of dependence in $J$ is 
assumed in \pref{wj1} and \pref{wj2}).  

This completes the explanation of the inductive assumption \pref{pfr}.
As we read from \pref{pot0} and \pref{pot}, 
 \pref{pfr} holds  at $j=0$, 
for 
$$ E_0\=E=\frac12\ln(1-s),
\qquad
(s_0,z_0, Z_0, \bar Z_0)
=(s,z,1,0),\qquad
W_0\=K_0\=0.
$$
We shall see that it also holds 
by induction for any $j=1,2, \ldots, R$,  with:
\bd
\item 
Effective couplings 
$(s_j, z_j)$ and effective polymer activity $K_{0,j}$  given by
\bal\lb{lk}
&s_{j+1}=s_j-a_j z_j^2+\FF_j
\notag\\
&z_{j+1}=L^2 e^{-\frac{\a^2}2 \G_j(0)}\lft[z_j-b_j s_jz_j+\MM_j\rgt]
\notag\\
&K_{0,j+1}=\LL_{0,j} + \RR_{0,j},
\eal
for coefficients $a_j$,$b_j$, and  functionals
$\FF_j\=\FF_j(K_{0,j})$, $\MM_j\=\MM_j(K_{0,j})$, 
$\RR_{0,j}\=\RR_{0,j}(z_j, s_j, K_{0,j})$ and $\LL_j\=\LL_j(K_{0,j})$. 
The functionals  $\FF_j$, $\MM_j$ and $\RR_{0,j}$ will play the role
of ``remainder parts''  with respect to the 
other terms in the equation.  The  functional $\LL_{0,j}$ 
will be a contraction with respect to suitable norms. 
\item Effective free energy $E_j$ given by 
\be\lb{elk}
E_{j+1}=E_j + L^{-2j} \lft[\EE_{1,j}+
 s_j \EE_{2,j}+ s^2_j \EE_{3,j}+ z^2_j \EE_{4,j}\rgt],
\ee
for coefficients $\EE_{2,j}$, $\EE_{3,j}$, $\EE_{3,j}$ and for a functional 
$\EE_{1,j}\=\EE_{1,j}(K_{0,j})$.
\item Fractional charge renormalization constants  $Z_j$ and $\bar Z_j$, 
\bal\lb{sk}
&Z_{j+1}= L^2 e^{-\h^2\frac{\a^2}2\G_j(0)} 
\lft[(1 -s_j m_{1,1,j}+\MM_{1,1,j})
Z_j+\lft(z_j m_{1,2,j} +\MM_{1,2,j}\rgt)\bar Z_j\rgt],
\notag\\
&\bar Z_{j+1}= L^2 e^{-\bar\h \frac{\a^2}2\G_j(0)} 
\lft[(1- s_j m_{2,2,j}+\MM_{2,2,j}) \bar Z_j
+\lft(z_j  m_{2,1,j} +\MM_{2,1,j}\rgt)Z_j\rgt],
\notag\\
&K_{1,j+1}=\LL_{1,j} + \RR_{1,j},
\eal
for coefficients $\{m_{p,q,j}:p,q=1,2\}$ and functionals  
$\{\MM_{p,q,j}\=\MM_{p,q,j}(K_{1,j}):p,q=1,2\}$, $\LL_{1,j}\=\LL_{1,j}(K_{1,j})$
and $\RR_{1,j}\=\RR_{1,j}(s_j,z_j,K_{0,j}, K_{1,j})$.  The  functional $\LL_{1,j}$ 
will be a contraction with respect to suitable norms. 
\ed
For every  $j=0,1,\ldots, R$,  
all the coefficients and functionals appearing in \pref{lk}, \pref{elk} and 
\pref{sk} are independent of $\L$: this will simplify  the discussion of the 
calculation  of the limit $\L\to \io$. 
Note that at $\a^2=8\p$, because of \pref{p3}, 
$L^2 e^{-\frac{8\p}2 \G_j(0)}\sim 1$ and  
the map \pref{lk} 
is our rigorous counterpart of Kadanoff's ODE for the effective coupling constants, 
\pref{ds}. Note also that \pref{elk} and \pref{sk} 
depend on the flow \pref{lk}, but do not affect it; therefore
the study of \pref{lk} done in \citep{Fa12} remains valid for 
the  developments of this paper.

The last step of the RG is 
\bal\lb{0pfr}
\O(J,\L)
&=e^{E_R|\L|}
\lim_{m\to 0}\EEE_{R}\lft[e^{U_{R}(J,\z^{(R)},\L)}+K_R(J,\z^{(R)}, \L)\rgt].
\eal
Suppressing the dependence in the set $\L$
of interactions and polymer activities, and setting $\d E_R:=E_{R+1}-E_R$, 
$\x_x=\x_x^{(R)}$, $\F_x=(J_x, \x_x^{(R)})$,    
we have:
\bd
\item For the free energy, 
\bal\lb{finalen}
&\frac1{|\L|}\ln \O(0,\L):=E_{R+1}
\notag\\
&=E_R +L^{-2R}\lim_{m\to 0}\ln\EEE_R\lft[1+\lft(e^{V_{0,R}(\z) + W_{0,R}(\z)}-1\rgt) + K_{0,R}(\z)\rgt].
\eal
%
%\item for the fractional charge density
%\bal\lb{finalden}
%&\frac{\dpr \O}{\dpr J_{x,\s}}(0,\L)
%=e^{-\d E_R|\L|}
%\notag\\
%&\qquad\times 
%\lim_{m\to 0}\EEE_R\lft[e^{V_{0,R}(\z) + W_{0,R}(\z)} 
%\lft(\frac{\dpr V_{1,R}(\F)}{\dpr J_{x,+}}
%+\frac{\dpr W_{1,R}(\F)}{\dpr J_{x,+}}\rgt)+ 
%\frac{\dpr K_{1,R}(\F) }{\dpr J_{x,+}}\rgt]_{J=0};
%\eal
\item For the fractional charge correlation 
\bal\lb{finalcorr}
&\frac{\dpr^2 \O}{\dpr J_{x,+}\dpr J_{0,-}}(0,\L)
=e^{-\d E_R|\L|}
\notag\\
&\qquad\times
\lim_{m\to 0}\EEE_R\lft[e^{V_{0,R}(\z) + W_{0,R}(\z)} \lft(\frac{\dpr V_{1,R}(\F)}{\dpr J_{x,+}}
+\frac{\dpr W_{1,R}(\F)}{\dpr J_{x,+}}\rgt)\lft(\frac{\dpr V_{1,R}(\F)}{\dpr J_{0,-}}
+\frac{\dpr W_{1,R}(\F)}{\dpr J_{0,-}}\rgt)\rgt]_{J=0}
\notag\\
&
+e^{-\d E_R|\L|}
\lim_{m\to 0}\EEE_R\lft[e^{V_{0,R}(\z) + W_{0,R}(\z)} 
\frac{\dpr^2 W_{2,R}(\F)}{\dpr J_{x,+}\dpr J_{0,-}}
+\frac{\dpr^2 K_{2,R}(\F)}{\dpr J_{x,+}\dpr J_{0,-}}\rgt]_{J=0}.
\eal
\ed
\subsection{Bounds on the RG map}
To control the limit $R\to \io$ of \pref{finalen} and 
\pref{finalcorr}, we need bounds for all the intermediate steps of the RG map. 
In the previous paper, \citep{Fa12}, we dealt with \pref{lk}, \pref{elk} and 
the formula for the free energy \pref{pres0}.   
We showed that there exists a unique choice of the initial value 
$s$ as function of $z$
such that the limit for $j\to \io$ of $s_j$ $z_j$ and $K_j$ is vanishing. 
More precisely,
we found the following results. 
\begin{lemma}[\citep{Fa12}]
\lb{scott}
Consider the coefficients $a_j$ and $b_j$ in \pref{lk}.
For $\a^2=8\p$, there exists a $j$-independent 
$C\=C(L)$ and a number $\tilde c_E$ such that
\bal\lb{scott1}
&|a_j-a|\le C L^{-\frac14 j},
&&|b_j-b|\le C L^{-\frac14 j},
\eal
where $a=8\p^2 e^{-8\p \tilde c_E} \ln L$ and $b=2\ln L$.
\end{lemma}
The constant $\tilde c_E$ in this Lemma is not the same as $c_E$ in \pref{0}
--although it has a similar origin; note, however,  that 
$\tilde c_E$ will not explicitly appear 
in the final results \pref{zdec} and \pref{zdec2}.
For stating  the next results, set, for any $j\ge 1$,  
\be\lb{negotiate0}
q_j:=\frac {q_1}{1+|q_1| (j-1)}, \qquad q_1:= \sqrt{ab} z_1.
\ee
Hence $q_1=z_1 4\p e^{4\p \tilde c_E} \ln L $ and $q_j$ 
is almost a discrete version of $2 s(\ell)$ in \pref{negotiate}. 
Given two parameters, $h>0$ and $A>1$, in  Section \ref{s5}
we will introduce the   
norm  $\|\cdot\|_{h,T_j}\=\|\cdot\|_{h,T_j}(A)$,  
that will measure the size of polymer activities.  
\begin{theorem}[\citep{Fa12}]\lb{t.sm}
Given  a $\t>0$ small enough, for   $L$ and  $A$ large enough, 
there exists an $\e\=\e(A,L,\t)$ such that the following statement holds. 
If  $0<z\le \e$, there exists a unique $s\=s(z)$ such that 
the solution of  \pref{lk} with initial
data $(z_0,s_0)=(z, s(z))$ satisfies 
\bal\lb{slk}
&\lft|s_j-\frac{|q_j|}{b}\rgt|\le \frac{\t}{b}\frac{|q_1|}{[1+|q_1|(j-1)]^{\frac32}},
\notag\\
&\lft|z_j-\frac{q_j}{\sqrt {ab}}\rgt|\le 
\frac{\t}{\sqrt {ab}}\frac{|q_1|}{[1+|q_1|(j-1)]^{\frac32}},
\notag\\
&\|K_{0,j}\|_{h,T_j}\le \frac{\t^2 |q_1|^2}{[1+|q_1|(j-1)]^3},
\eal
for  all $j=1, \ldots, R$. 
Besides, the choice of the parameters $L$,  $A$, $\e$ and 
the function $s(z)$ are 
independent of $|\L|$.
\end{theorem} 
As anticipated, the  $s(z)$ found in this Theorem 
determines the graph of the  BKT transition line,  $\b=\b_{\rm BKT}(z)$, 
 via \pref{p4}. 
This result was instrumental to control \pref{elk} and 
to prove  the convergence of \pref{pres0}. 
\begin{theorem}[\citep{Fa12}]\lb{t3.5}
There exists $C\=C(\a,L)$ such that, given  any  $j=0,1,\ldots, R$,  
if  $|s_j|, |z_j|,\|K_{0,j}\|_{h,T_j}\le c_0|q_j|$, then
\be\lb{r2}
|E_{j+1}-E_{j}|\le C L^{-2j}|q_j|.
\ee
Besides, $E_0, \ldots, E_{R}$ (but not $E_{R+1}$) 
are independent of $|\L|$. 
\end{theorem}
The consequence of this result is a convergent series representation of the 
free energy
$$
p(\b,z)=-\frac1{2\b} \log(1-s(z))-\frac1\b\sum_{j\ge 0} (E_{j+1}-E_j),
$$
which was the main result of \citep{Fa12}.
In this paper we study \pref{sk} and \pref{finalcorr}. For this task, 
we need to introduce  a norm for activities with one pinning point, 
$\|\cdot\|_{1,h,T}$, and a norm for activities with two pinning points, 
$\|\cdot\|_{2,h,T}$, see discussion after \pref{pinning}; such norms will 
be defined in Section \ref{s5}.
In the following result, we control  the activities 
$K_{1,j}$ and $K^\dagger_{1,j}$. 
\begin{theorem}\lb{leibler}
There exists a $C\=C(\a)>0$ such that, under 
the same hypothesis of Theorem \ref{t.sm},
\be\lb{leibler0}
\|K_{1,j}\|_{1,h,T_j}\le C |q_j|^2,\qquad 
\|K^\dagger_{1,j}\|_{1,h,T_j}\le C |q_j|^2.
\ee
\end{theorem} 
The proof is in Section \ref{7.1}. 
Next, we study the coefficients in the flow \pref{sk}.  
\begin{lemma}\lb{t3.1bb}
There exists  a   $j$-independent 
$C\=C(\a,L)$ such that, for any  $p=1,2$, 
\be\lb{scott2ab}
|\MM_{p,1,j}|\le C A^{-1} \|K_{1,j}\|_{1,h,T_j},\qquad
|\MM_{p,2,j}|\le C A^{-1} \|K^\dagger_{1,j}\|_{1,h,T_j}.
\ee
\end{lemma} 
\begin{lemma}\lb{t3.1b}
Consider  $a$ and $b$  given in Lemma \ref{scott}.  
There exists  a   $j$-independent 
$C\=C(\a,L)$ such that: if $\a^2\ge 8\p$ and $p,q=1,2$, 
\be\lb{scott2a}
|m_{q,p,j}|\le C;
\ee
besides,  if $\a^2=8\p$,  
\bal
|m_{1,1,j}-\h^2  b|\le C L^{-\frac j4},
\qquad
|m_{2,2,j}-\bar \h^2 b|\le C L^{-\frac j4};
\lb{scott2}
\eal
finally,  if $\a^2=8\p$ and $\h=-\bar \h= \frac12$,
then $\MM_{1,1,j}=\MM_{2,2,j}$, $\MM_{1,2,j}=\MM_{1,2,j}$ and 
\bal
|m_{2,1,j}-\frac{\sqrt{ab}}2|\le C L^{-\frac j4},
\qquad
|m_{1,2,j}-\frac{\sqrt{ab}}2|\le C L^{-\frac j4}.
\lb{scott3}\eal 
\end{lemma}
This Lemma does not provide  the exact asymptotic values of $m_{2,1,j}$ and $m_{1,2,j}$ 
if $\h\neq \frac12$; however, they will not be  necessary for studying 
\pref{sk}. 
To formulate the next result, set  $Z^+_j:=Z_j+\bar Z_j$,   $Z^-_j:=Z_j-\bar Z_j$
and
$$
g_j:=-\p \sum_{k=1}^j[\G_k(0)-\frac 1{2\p}\log L], 
$$
which is a bounded sequence because of \pref{p3}.
\begin{theorem}\lb{leiblerbis}
In the same hypothesis of Theorem \ref{t.sm}, for $j=1,\ldots,R$:
\bd
\item If $\h=-\bar \h=\frac12$, there exist two coefficients  $\{c_\s: \s=\pm\}$ that 
are vanishing for $z\to 0$ and are 
such that
\bal\lb{leibler1}
&Z^+_{j+1}=Z_{1}^+ L^{\frac32 j}(1+|q_1| j)^{\frac14}e^{g_j+c_++r_{1,j}},
\notag\\
&Z^-_{j+1}=Z^-_1 L^{\frac32 j}(1+|q_1| j)^{-\frac34}e^{g_j+c_{-}+r_{2,j}};
\eal
in the above formulas,  
for a constant $C$ and  for $m=1,2$,  
$$
|r_{m,j}|\le C \frac{\t}{\sqrt{1+|q_1|j}}.
$$
\item If $0\le \h<\frac12$,  there exist two coefficients, 
$c_1$, $c_2$, which are vanishing in the limit $z\to 0$ and 
are such that
\bal\lb{leibler2}
&Z_{j+1}=L^{2j(1-\h^2)}(1+|q_1| j)^{-\h^2} e^{4\h^2 g_j+c_{1}}
\lft[e^{r_{1,j}}Z_1+c_{2}e^{s_{1,j}}\bar Z_1 \rgt],
\notag\\
&\bar Z_{j+1}=L^{2j(1-\h^2)}(1+|q_1| j)^{-\h^2}
e^{4\h^2 g_j}\lft[r_{2,j}Z_1+s_{2,j}\bar Z_1 \rgt],
%\notag\\
%&\phantom{****}+ 
%L^{2j(1-\bar \h^2)}(1+|q_1| j)^{-\bar \h^2}e^{4\bar \h^2 g_j+c_{3}+r_{4,j}}\bar Z_1
\eal 
where, for a $C_0\=C_0(\h)$ and any $m=1,2$, 
$$
|r_{m,j}|\le C_0 \frac{\t}{\sqrt{1+|q_1|j}},
\qquad
|s_{m,j}|\le C_0 \frac{1}{\sqrt{1+|q_1|j}}+C_0 L^{-2(\bar \h^2-\h^2)j}.
$$
A formula for $c_2$ is, for a $c(\h)>0$, 
$$
c_2=z e^{4\p (\bar\h^2-\h^2)\G_{0}(0)}
\lft[c(\h)
- m_{1,2,0}\rgt]+ O(z^{\frac32}).
$$
\item If $\frac12 <\h <1$, \pref{leibler2}
holds after interchanging $Z_j$ with $\bar Z_j$ and $\h$ with $-\bar \h$ 
(hence the formula for $c_2$ becomes 
$z e^{4\p (\h^2-\bar \h^2)\G_{0}(0)}[c(-\bar \h)-m_{2,1,0}] + O(z^{\frac32})$).
%: 
%\bal\lb{leibler3}
%&\bar Z_{j+1}=L^{2j(1-\bar \h^2)}(1+|q_1| j)^{-\bar \h^2} e^{-4\h^2 g_j+c_{1}}
%\lft[e^{r_{1,j}}\bar Z_1+c_{2}e^{s_{1,j}} Z_1 \rgt],
%\notag\\
%& Z_{j+1}=L^{2j(1-\bar \h^2)}(1+|q_1| j)^{-\bar \h^2}
%e^{-4\bar \h^2 g_j}\lft[r_{2,j}\bar Z_1+r_{3,j} Z_1 \rgt]
%\notag\\
%&\phantom{****}+ 
%L^{2j(1-\h^2)}(1+|q_1| j)^{-\h^2}e^{-4\h^2 g_j+c_{3}+r_{4,j}}Z_1.
%\eal 
\ed
Finally,  for every $\h\in (0,1)$, 
\bal\lb{12:57am}
&Z_1=L^{2}e^{-\h^2 \frac{\a^2}{2}\G_0(0)}(1+O(z)), 
\notag\\
&\bar Z_1=L^{2}e^{-\bar \h^2 \frac{\a^2}{2}\G_0(0)} m_{2,1,0}z. 
\eal 
\end{theorem}
\begin{theorem}\lb{pherson} 
Under the same hypothesis of Theorem \ref{t.sm} and 
if $A\ge e^2$, there exists a $C>0$ such that,
for any $j=1,2,\ldots, R$ 
(suppressing the dependence in the variables $\f,X,x_1,\s_1,x_2,\s_2$),
\bal\lb{maldacena}
K_{2,j}=&\sum_{k=0}^j 2^{-(j-k)} e^{-L^{-k}|x_1-x_2|}
 L^{-4k}
\notag\\
&\times\lft[Z_k^2 K^{(a,k)}_{2,j}+\bar Z_k^2 K^{(\bar a,k)}_{2,j}+
Z_k\bar Z_k K^{(b,k)}_{2,j}\rgt],
\eal
where,  for any $\d=a,\bar a, b$, 
\be\lb{maldacena1}
\|K^{(\d,k)}_{2,j}\|_{2,h,T_j}
\le C|q_k|.
\ee
\end{theorem}
As a consequence of the above Theorems we can finally turn to the 
 calculation of the fractional charge  correlation. 
Consider the $w$'s function in \pref{wj2}.
\begin{theorem} \lb{finale}
The limits 
\bal\lb{canaletto0}
&w^-_{2,a}(x):=\lim_{R\to \io}w^-_{2,a, R}(x)\qquad 
w^-_{2,\bar a}(x):=\lim_{R\to \io}\bar w^-_{2,a, R}(x)
\notag\\
&w^-_{2,b}(x):=\lim_{R\to \io}w^-_{2,b, R}(x)\qquad 
w^-_{2,c}(x):=\lim_{R\to \io} w^-_{2,c, R}(x)
\eal
exist and,  
under the same hypothesis of Theorem \ref{t.sm},
%\be
%\lim_{R\to \io}\frac{\dpr \O}{\dpr J_{x,\s}}(0,\L)=0,
%\ee 
\be\lb{canaletto}
\lim_{R\to \io}\frac{\dpr^2 \O}{\dpr J_{x,+}\dpr J_{0,-}}(0,\L) 
=2w_{2,a}^-(x)+ 2 w_{2,\bar a}^-(x) +2w_{2,c}^-(x).
\ee
(While $w^-_{2,b}(x)$ does not contribute to the correlation.) 
\end{theorem}
The last ingredient for the proof of the main Theorem 
is then an exact evaluation of the long $|x|$ asymptotic formulas 
for the functions in \pref{canaletto0}.
\begin{theorem}\lb{finalevero}
For coefficients 
$f, f_a, f_{\bar a}, \tilde f_{b}$ that are vanishing for $z\to 0$, 
and for a  constant $C$:
\bd
\item If $\h=-\bar\h=\frac12$, then, for $\d=a,\bar a$, 
\bal
&w^-_{2,\d}(x)=
\frac{e^{2\p c_E} + f_{\d}}{8|x|} \lft(1+f \ln |x|\rgt)^{\frac12}(1+o(1)),
\\
&|w^-_{2,c}(x)|\le \frac{C}{|x|}\lft(1+f\ln |x|\rgt)^{-\frac12}.
\eal
\item 
If $\h\neq \frac12$, then,  for the same  $c(\h)$ of Theorem \ref{leiblerbis},  
\bal\lb{finale1}
&w^-_{2,a}(x)+w^-_{2,\bar a}(x)
=\frac{e^{8\p \h^2 c_E}+f_a}{2|x|^{4\h^2}}\lft(1+f\ln |x|\rgt)^{-2\h^2}(1+o(1))
\notag\\
&\qquad\qquad\qquad+\frac{c(\h)^2 z^2(1+\tilde f_b)}{2|x|^{4\bar \h^2}}
\lft(1+f\ln |x|\rgt)^{-2\bar \h^2}(1+o(1)),
\\
&|w^-_{2,c}(x)|\le \frac{C}{|x|^{4\h^2}}\lft(1+f\ln |x|\rgt)^{-2\h^2-1}
+\frac{C}{|x|^{4\bar \h^2}}\lft(1+f\ln |x|\rgt)^{-2\bar \h^2-1}.
\eal
\ed
(While, for every $\h\in (0,1)$, $w^-_{2,b}(x)=0$.)
Besides, $f=4\p e^{4\p \tilde c_E}L^2 e^{-4\p\G_0(0)} z$.
\end{theorem}
Our main result, Theorem \ref{t1}, is then a direct consequence of Theorem
\ref{finale}
and  Theorem \ref{finalevero}.

\section{Dimensional bounds}\lb{s5}
Here we set up scale dependent  norms that we will use to control the size of the 
polymer activities. We will also show how norms encode the {\it dimensional analysis}
used in physics to adapt renormalization group ideas  to this model.  
\subsection{Norms and regulators: definitions}\lb{nr}
We mainly follow \citep{Br07}. Let $j\in \NNN$. 
For $n=0,1,2$ and for $\dpr^\m$ the discrete derivative introduced
before \pref{pot0}, define
\be\lb{ln}
\|\nabla^n_j\f\|_{L^\io (X^*)}:=\max_{\m_1,\ldots,\m_n\atop \m_j\in \hat u}\max_{x\in X^*}
L^{nj}\big|\dpr^{\m_1}\cdots \dpr^{\m_n} \f_x\big|.
\ee
For $X$ a connected $j$--polymer, let  $\CC^2_j(X)$ be 
the linear space of the 
functions $\f:X^*\to \CCC$ with norm 
$$
\|\f\|_{\CC^2_j(X)}:=\max_{n=0,1,2}\|\nabla_j^n\f\|_{L^\io(X^*)}.
$$ 
Observe that $\nabla_j$ is $L^j \dpr$, which makes the norm 
explicitly scale dependent; besides, 
we are using the notation $\CC^2_j(X)$ even though  
the domain involved in the definition of the norm 
is the set $X^*$.
Let $\NN_j(X)$ be the space of the smooth complex activities of the
polymer $X^*$,
i.e. the set of  $C^\io$ functions $F(\f,X):\CC^2_j(X)\rightarrow \CCC$.
The $n$-order derivative of  $F$ along the directions 
$f_1,\ldots,f_n\in \CC^2_j(X)$ 
is 
\be\lb{fdiff}
D^n  F(\f,X)\cdot (f_1,\ldots,f_n)
:=\sum_{x_1,\ldots,x_n\in X^*}(f_1)_{x_1}\cdots (f_n)_{x_n}
{\dpr^n F\over \dpr \f_{x_1}\cdots\dpr \f_{x_n}}(\f,X).
\ee
Again, despite the notation $\NN_j(X)$,  
the relevant set here is the  bigger set $X^*$.
The size of the differential of order $n$ is given by
\be\lb{tjn}
\|D^n F(\f,X)\|_{T^n_j(\f,X)}:=\sup_{\|f_i\|_{\CC^2_j(X)}=1}
\big|D^n F(\f,X)\cdot (f_1,\ldots,f_n)\big|.
\ee
Then, given any  $h>1$, define the norm
\bal\lb{sn}
&\|F(\f,X)\|_{h,T_j(\f,X)}
:=\sum_{n\ge 0}{h^n\over n!}\|D^n F(\f,X)\|_{T^n_j(\f,X)}. 
\eal
In order to control the norm of the activities as function of the field $\f$,
for any scale $j$ and any $X\in \PP^c_j$
introduce the {\it field regulators}, $G_j(\f,X)\ge 1$, 
that  depends upon  derivatives of $\f$ only. An explicit choice will be provided below.
Then, define 
\bal\lb{rn}
\|F(X)\|_{h,T_j(X)}
&:=\sup_{\f\in \CC^2_j(X)}{\|F(\f,X)\|_{h,T_j(\f,X)}\over G_j(\f,X)}.
\eal
Finally, we have to weight the polymer activity w.r.t. the size of the
set. Given a parameter $A>1$, define 
\bal\lb{an}
\|F\|_{h,T_j}\=\|F\|_{h,T_j}(A):=\sup_{X\in \PP_j^c} A^{|X|_j}\|F(X) \|_{h,T_j(X)}.
\eal
Inspired by the discussion after \pref{pinning}, given a charge $\s=\pm$ and 
a lattice point $x$,   
we will call an activity of the form $F(\f,X, x, \s)$ pinned
at the lattice point 
$x$ if   $F(\f,X, x, \s)=0$ whenever $x\not\in X$. 
For such  activities, we define
$$
\|F\|_{1,h,T_j}:=\sup_{x\in \ZZZ^2,\s=\pm}\|F(\cdot,\cdot, x, \s)\|_{h,T_j}(A).  
$$
Likewise, given two charges $\s, \s'=\pm1$ and two lattice points
$x$ and $x'$,   an activity  $F(\f,X, x, \s, x', \s')$ is pinned at $x$ and $x'$
if $F(\f,X, x, \s, x', \s')=0$ whenever $x\not\in X$ 
or $x'\not\in X^*$. For such activities we set  
$$
\|F\|_{2,h,T_j}:=\sup_{x,x'\in \ZZZ^2\atop \s,\s'=\pm}
\|F(\cdot,\cdot, x, \s, x', \s')\|_{h,T_j} (A^{\frac12}). 
$$
In the last definition, note that  the weight in the size of the 
polymer  has been reduced to $A^{\frac12}$.

This concludes the set up  of the norms, except for the choice
of some  parameters and functions that were involved in the definition.
The parameter $h\=h(\a)$ is chosen to be 
$h:=\max\{1, 2 \mathfrak{h}_j(\a):j\ge 0\}$, 
where 
\be\lb{hut}
\mathfrak{h}_j(\a):=\max\{\|h_j\|_{\CC^2_j(X)}: 0\in X\in \SS_j\} 
\ee
and $h_j(x)$ is the function $\a[\G_j(x)-\G_j(0)]$. The 
usefulness of this choice will become clear in Appendix \ref{ptdh}. 
It is not difficult 
to see that, by \pref{p2}, $\mathfrak{h}_j$ is bounded in $j$ 
and so the definition of the 
constant $h$ makes sense. 
The parameter $A$ will be chosen large enough in various points below. 
Next, we have to choose $G_j$. Here we follow \citep{Fa12}.
Given two positive constants $c_1$, $c_3$, 
and a positive function of $L$, $\kappa_L$, 
if $X\in \PP_j^c$, the function $G_j$ is such that
\be\lb{6.69}
\ln G_j(\f,X)=c_1\kappa_L
\|\nabla_j\f\|^2_{L^2_j(X)}+c_3\kappa_L\|\nabla_j\f\|^2_{L^2_j(\dpr X)}
+c_1\kappa_L W_j(\nabla^2_j\f,X)^2,
\ee
where we have used $L^2$-type norms
\bal\lb{ln2}
&\|\nabla^n_j\f\|^2_{L^2_j(X)}:=
L^{-2j}\sum_{x\in X}\sum_{\m_1,\ldots,\m_n}
L^{2nj}\big|\dpr^{\m_1}\cdots \dpr^{\m_n} \f_x\big|^2,
\notag\\
&\|\nabla^n_j\f\|^2_{L^2_j(\dpr X)}:=
L^{-j}\sum_{x\in \dpr X}\sum_{\m_1,\ldots,\m_n}
L^{2nj}\big|\dpr^{\m_1}\cdots \dpr^{\m_n} \f_x\big|^2,
\eal
\be\lb{6.109}
W_j(\f, X)^2:=\sum_{B\in \BB_j(X)} \|\f\|^2_{L^{\io}(B^*)}.
\ee
To control the field dependence of $U_j$ we shall
occasionally use an auxiliary field regulator, called 
{\it strong field regulator}, $G^{\rm str}_j$: for
$B\in \BB_j$ and $X\in \PP_j$, 
\be\lb{6.66}
\ln G^{\rm str}_j(\f,B):=\kappa_L
\max_{n=1,2}\|\nabla^n_j\f\|^2_{L^\io(B^*)},
\quad
G_j^{\rm str}(\f,X):=\prod_{B\in \BB_{j}(X)}G_j^{\rm str}(\f,B).
\ee
\subsection{Norms and regulators: properties}
First, it is important to observe that 
$\NN_j(X)$ with the norm $\|\cdot\|_{h,T_j(X)}$ is a
Banach space.  We now list some useful features of the field regulators. 
As apparent from the definition, if $X\in P_{j+1}$, 
\be\lb{nw}
G^{\rm str}_j(\f',X)\le G^{\rm str}_{j+1}(\f',X). 
\ee
Consider a  polymer  $X\in \PP_j$. 
From the definitions, we have  $L^2_j(X)=\sum_{Y\in \CC_j(X)} L^2(Y)$. 
Besides, since two $Y$'s in $\CC_j(X)$ 
have disjoint boundaries, we also have  $L^2_j(\dpr X)=\sum_{Y\in \CC_j(X)} L^2(\dpr Y)$.
Therefore
\be\lb{6.51}
\prod_{Y\in \CC_j(X)}G_j(\f,Y)=G_j(\f,X).
\ee
For the following results to hold, 
$c_3$ and $c_1$ must be large enough, but independently of the
scale $j$ and the size $L$.  
Unless otherwise stated, $j=0,1,\ldots, R-1$.
\begin{lemma}\lb{l6.100b} 
For any polymer   $X\in \PP_j$,
\bal\lb{6.100b}
G^{\rm str}_j(\f,X)\le G_j(\f,X).
\eal
For any  polymer $X\in \PP_j$ and any block $B\in \BB_j$, but $B$  not inside $X$,
\bal\lb{6.52}
G^{\rm str}_j(\f,B)G_j(\f,X)\le G_j(\f,B\cup X).
\eal
\end{lemma} 
This Lemma corresponds to  formula (6.52) 
of \citep{Br07}: the proof can be found in that paper after Lemma 6.21. 
The role of the field regulators in the forthcoming analysis is to have 
a standard function to integrate with respect to 
the Gaussian measures.
\begin{lemma}\lb{l6.53} 
Let $\kappa_L=c(\log L)^{-1}$ with $c>0$  and small enough.
\bd
\item For $j=0, 1, \ldots, R-1$ and any connected polymer $X\in \PP_j^c$, 
\be\lb{6.54}
\EEE_j \lft[G_j(\f,X)\rgt]\le  2^{|X|_j} G_{j+1}(\f',\bar X);
\ee
if instead $j=R$, 
\be\lb{ex6.53bis}
\EEE_R \lft[G_R(\f,\L)\rgt]\le 2.
\ee 
\item For $j=0, 1, \ldots, R-1$, $m=1,2,3$ and any small polymer  $X\in \SS_j$, 
there exists a $C_m>1$ such that  
\bal\lb{6.58}
\lft(1+\max_{n=1,2}\|\nabla^n_{j+1}\f'\|_{L^\io(X^*)}\rgt)^{m}
\EEE_j
\lft[G_{j}(\f,X)\rgt]\le
\frac{C_m}{\kappa_L^{m/2}} 2^{|X|_j} G_{j+1}(\f',\bar X);
\eal
besides the last formula holds even if $G_{j}(\f,X)$ on the left hand side member 
is replaced by 
$\sup_{t\in[0,1]} G_{j}(t\f'+\z,X)$.
\ed
\end{lemma} 
The proof is in   Section D 
of \citep{Fa12}. 
From the definitions set up so far, we can  derive some
simple bounds that will be needed in the next section. 
For any $\f\in \CC^2_{j+1}(X)$, we have 
$\|\f\|_{\CC^2_j(X)}\le \|\f\|_{\CC^2_{j+1}(X)}$, so that, 
for any $F\in \NN_j(X)$
\be\lb{6.8}
\|F(\f,X)\|_{h,T_{j+1}(\f,X)}
\le\|F(\f,X)\|_{h,T_{j}(\f,X)} .
\ee
If $Y\subset X$, for any $\f\in \CC^2_j(X)$ we have
$\|\f\|_{\CC^2_j(Y)}\le \|\f\|_{\CC^2_j(X)}$, so that 
$\CC^2_j(X)\subset \CC^2_j(Y)$ and
\be\lb{6.36}
\|F(\f,X)\|_{h,T_{j}(\f,X)}
\le\|F(\f,X)\|_{h,T_{j}(\f,Y)}.
\ee
For any two  polymers $Y_1, Y_2$ not
necessarily disjoint and such
that $Y_1\cup Y_2\subset X$, and any two polymer activities, 
$F_{1}\in\NN_j(Y_1)$ and $F_{2}\in\NN_j(Y_2)$, we have:
a generalized {\it triangular inequality}
\be\lb{6.5}
\|F_{1}(\f,Y_1)+F_{2}(\f,Y_2)\|_{h,T_{j}(\f,X)}
\le\|F_{1}(\f,Y_1)\|_{h,T_{j}(\f, Y_1)} 
+ \|F_{2}(\f, Y_2)\|_{h,T_{j}(\f, Y_2)} ,
\ee
(which is stronger than the usual triangular inequality because 
different norms appear in the two members); 
and the {\it factorization property}
\be\lb{6.37}
\|F_{1}(\f,Y_1)F_{2}(\f,Y_2)\|_{h,T_{j}(\f,X)}
\le\|F_{1}(\f,Y_1)\|_{h,T_{j}(\f, Y_1)} 
\|F_{2}(\f, Y_2)\|_{h,T_{j}(\f, Y_2)}.
\ee
Details of the  proofs of these  inequalities are in \citep{Br07}.

Finally, given $\|F(X)\|_{h,T_{j}(X)}$,  in order to have an estimate of the size of
$\|F\|_{h,T_{j+1}}$ one needs to sum over the position of 
the polymer.  Let us consider separately the case of configurations
on small sets and on large sets. 
For $\l\in (0,1)$ and $\r=s,l$, set
\bal\lb{gld}
&k_\r(A,\l):=\sup_{V\in P^c_{j+1}}A^{|V|_{j+1}}
\sum_{Y\in O_\r}^{\bar Y=V}(\l A)^{-|Y|_j},
\eal
where  $O_s=\SS_j$ and $O_l= \not\!\SS_j$. Besides, consider also the case of a pinning point 
in the sum and set
\bal\lb{gld2}
k^*_s(A,\l):=\sup_{V\in P^c_{j+1}}A^{|V|_{j+1}}\sup_{x\in V}
\sum_{Y\in \SS_j\atop Y\ni x}^{\bar Y=V}(\l A)^{-|Y|_j}.
\eal
Note that $k_s(A,\l)$, $k_l(A,\l)$ and $k^*_s(A,\l)$ are
$j$-independent, and so the notation is consistent. 
\begin{lemma}\lb{l6.90}
There exist $c >0$ and $\th>0$  such that, for $A$ large enough
\be\lb{6.90}
k_s(A,\l)\le c L^2,\qquad
k^*_s(A,\l)\le c,\qquad 
k_l(A,\l)\le A^{-\th}.
\ee
\end{lemma}
For the proof  see Lemma 6.19  and Lemma 6.18 in \citep{Br07}.
In brief, when the sum is over small sets and there is no pinning point, 
the bound is proportional
to a volume factor $L^2$; when the sum is over large sets, 
the bound is finite in $L$ and vanishing for large $A$. 

\subsection{Dimensional analysis}
We now return to the actual polymer activities of our RG treatment of the Coulomb Gas. 
To reproduce the physicists' analysis we first need to decompose the
polymer activity into terms which represents clusters of particles 
with given  total charge.  
To do so, note that
$K_{0,j}$ contains terms that, as functions of the fields,  
either are periodic of
period $2\p/\a$ or are  derivative terms; therefore
$K_{0,j}$ is invariant under $\f_\xx\to \f_\xx+ \frac{2\p}{\a}t$ for any 
constant, integer field $t$. As explained Appendix \ref{abCG}, 
such invariance provides via a Fourier analysis 
the following decomposition  into
{\it charged components} for $K_{0,j}$ as well as for $K_{1,j}$, $K^\dagger_{1,j}$
and $K^{(\d,k)}_{2,j}$.
\begin{lemma}\lb{ldec3}
For $j=0, 1, \ldots, R$ and for any $X\in \PP^c_j$,  
\bal\lb{dec3}
&K_{0,j}(\f,X)=\sum_{q\in \ZZZ} \hK_{0,j}(q,\f,X),
\notag\\
&K_{1,j}(\f,X,x,\s)=
\sum_{q\in \ZZZ} \hK_{1,j}(q,\f,X,x,\s),
\notag\\
&K^\dagger_{1,j}(\f,X,x,\s)=
\sum_{q\in \ZZZ} \hK^\dagger_{1,j}(q,\f,X,x,\s),
\eal
and, for $\d=a,\bar a, b$
\be\lb{dec3.0}
K^{(\d,k)}_{2,j}(\f,X,x,\s,x',\s')=
\sum_{q\in \ZZZ} \hK^{(\d,k)}_{2,j}(q,\f,X,x,\s,x',\s').
\ee
The above series are absolutely convergent and, 
if $\th$ is a constant field, 
\bal\lb{dec4}
&\hK_{0,j}(q,\f,X)
= e^{i q\a\th}\hK_{0,j}(q,\f-\th,X),
\notag\\
&\hK_{1,j}(q,\f,X,x,\s)= 
e^{i (q+\h\s)\a\th}\hK_{1,j}(q,\f-\th,X,x,\s),
\notag\\
&\hK^\dagger_{1,j}(q,\f,X,x,\s)= 
e^{i (q+\bar \h \s)\a\th}\hK^\dagger_{1,j}(q,\f-\th,X,x,\s),
\notag\\
&\hK^{(a,k)}_{2,j}(q,\f,X,x,\s,x',\s')= 
e^{i (q+\h \s+\h \s')\a\th}\hK^{(a,k)}_{2,j}(q,\f-\th,X,x,\s,x',\s'),
\notag\\
&\hK^{(\bar a,k)}_{2,j}(q,\f,X,x,\s,x',\s')= 
e^{i (q+\bar \h \s+\bar \h\s')\a\th}\hK^{(a,k)}_{2,j}(q,\f-\th,X,x,\s,x',\s'),
\notag\\
&\hK^{(b,k)}_{2,j}(q,\f,X,x,\s,x',\s')= 
e^{i (q+\h \s+\bar \h\s')\a\th}\hK^{(a,k)}_{2,j}(q,\f-\th,X,x,\s,x',\s').
\eal
Besides,
\be\lb{cin0}
\|\hK_{0,j}\|_{h,T_j}\le \|K_{0,j}\|_{h,T_j},
\ee
\be\lb{cin1}
\|\hK_{1,j}\|_{1,h,T_j}\le \|K_{1,j}\|_{1,h,T_j},\qquad 
\|\hK^\dagger_{1,j}\|_{1,h,T_j}\le \|K^\dagger_{1,j}\|_{1,h,T_j},
\ee
and for $\d=a,\bar a, b$
\be\lb{cin2}
\|\hK^{(\d,k)}_{2,j}\|_{2,h,T_j}\le \|K^{(\d,k)}_{2,j}\|_{2,h,T_j}.
\ee
\end{lemma}
The meaning of 
\pref{dec4} is:  $\hK_{0,j}(q,\f,X)$ represents  clusters of
particles with a total charge  $q$; 
$\hK_{1,j}(q,\f,X,x,\s)$ and $\hK^\dagger_{1,j}(q,\f,X,x,\s)$ 
represent clusters of particle with total charge 
 $q+\h \s$ and  $q+\bar \h \s$ respectively; similarly for 
$\hK^{(\d,k)}_{2,j}(q,\f,X,x,\s,x',\s')$.

Now we can discuss the typical bound we need 
in the rest of the paper. 
By \pref{6.8}  and \pref{6.54},  
for any connected polymer $X\in \PP_j^c$ 
\bal\lb{6.4}
\|\EEE_j\lft[K_{0,j}(\f,X)\rgt]\|_{h,T_{j+1}(\f',X)}
\le \|K_{0,j}\|_{h,T_j}\lft({A\over 2}\rgt)^{-|X|_j}
G_{j+1}(\f',\bar X); 
\eal
and, by \pref{cin0},  for the charged component $\hK_{0,j}(q,\f,X)$,  
\bal\lb{6.4bis}
\|\EEE_j\lft[\hK_{0,j}(q,\f,X)\rgt]\|_{h,T_{j+1}(\f',X)}
\le \|K_{0,j}\|_{h,T_j}\lft({A\over 2}\rgt)^{-|X|_j}
G_{j+1}(\f',\bar X);
\eal
similar bounds can be derived  for $\hK_{1,j}(q,\f,X,x,\s)$
and $\hK^\dagger_{1,j}(q,\f,X,x,\s)$;  and also for $\hK^{(\d,k)}_{2,j}(q,\f,X,x,\s,x',\s')$.
Then we could  use \pref{6.90} to sum over the polymer $X$. However, following this procedure, 
the sum over the small polymers $X$ will generate a bound proportional to the 
{\it volume factor} $L^2$, which would exponentially  
increase the size of the bound for $\|K_{0,j}\|_{h,T_j}$ at each step. 
To avoid that, 
we need to improve \pref{6.4} and \pref{6.4bis} whenever $X$ is a small set 
to beat such an $L^2$.
Observe that we passed from scale $j+1$ to scale $j$ by
the bound \pref{6.8} which is of  general validity.
Under special circumstances, this step can
be done in a more efficient way.
To formulate the next results in a simplified notation, in general 
we will say that $F(\f,X)$ is  a {\it charge $p$ activity}  if, 
for any constant complex field $\th$,  one has
$$
F(\f,X)=e^{i\a p \th} F(\f-\th,X).
$$   
\begin{theorem}\lb{dh}
Consider a charge $p$ activity  $F(\f,X)$, with $X\in \SS_j$. 
There exists a $C\=C(\a)$ such that, 
\be\lb{eqdh}
\|\EEE_j\lft[F(\f,X)\rgt]\|_{h,T_{j+1}(\f',X)}\le\r(p,\a) \|F\|_{h,T_j} \lft({A\over 2}\rgt)^{-|X|_j} 
G_{j+1}(\f',\bar X)
\ee 
for a  ``dimensional factor''
$$
\r(p,\a)= C^{1+|p|} 
L^{-d(p)\frac{\a^2}{4\p}},
$$
where $d(p)=p^2$ if $|p|\le 1$ and $d(p)=2|p|-1$ otherwise. 
\end{theorem}
\pref{eqdh} differs from \pref{6.4} by the prefactor
$\r(p,\a)$.
The proof, mostly borrowed from  \citep{DH00},  is in Appendix \ref{ptdh}.
As an application consider 
the charged components of $K_{0,j}$ and of $\hat K_{1,j}$ with total charge
$p:|p|> 1$.  Setting $F(\f):=\hat K_{0,j}(q, \f,X)$, 
the hypothesis of the theorem is satisfied for $p=q$; therefore  
\bal
&\|\EEE_j\lft[\hat K_{0,j}(q, \f,X)\rgt]\|_{h,T_{j+1}(\f',X)}
\notag\\
&\le C^{1+|q|} 
L^{-(2|q|-1)\frac{\a^2}{4\p}}\|K_{0,j}\|_{h,T_j} \lft({A\over 2}\rgt)^{-|X|_j} 
G_{j+1}(\f',\bar X).
\eal
Considering that $\a^2\ge 8\p$, if  $|q|\neq 0,1$ and $L$ is large
enough, the prefactor  $C^{1+|q|}  L^{-(2|q|-1)\frac{\a^2}{4\p}}\le (C^2L^{-3})^{|q|}$ beats
the volume factor $L^2$ that will be generated by \pref{6.90} once we sum 
the above bound over $X\in \SS_j$.
The same conclusion holds for
$\hat K_{1,j}(q,\f,X,x,\s)$. Indeed, the theorem
applies with $p=q+\h \s$ and we have 
\bal\lb{eqdh1}
&\|\EEE_j\lft[\hat
  K_{1,j}(q,\f,X,x,\s)\rgt]\|_{h,T_{j+1}(\f',X)}
\notag\\
&\le C^{2|q+\h \s|} 
L^{-d(q+\h \s)\frac{\a^2}{4\p}}\|K_{1,j}\|_{1,h,T_j} \lft({A\over 2}\rgt)^{-|X|_j} 
G_{j+1}(\f',\bar X).
\eal
Therefore, if $|q+\h\s|>1$, the prefactor is
$C^{2|q+\h \s|} L^{-(2|q+\h \s|-1)\frac{\a^2}{4\p}}\le
(C^2L^{-2})^{|q+\h\s|}$ and beats the volume factor $L^2$. For completeness,
we also state that 
\bal\lb{eqdh1bar}
&\|\EEE_j\lft[\hat K^\dagger_{1,j}(q,\f,X,x,\s)\rgt]\|_{h,T_{j+1}(\f',X)}
\notag\\
&\le C^{2|q+\bar \h \s|} 
L^{-d(q+\bar \h \s)\frac{\a^2}{4\p}}\|K^\dagger_{1,j}\|_{h,T_j} \lft({A\over 2}\rgt)^{-|X|_j} 
G_{j+1}(\f',\bar X).
\eal
Finally, for $\d=a,\bar a, b$ and $0\le k\le j$, 
\bal\lb{lili}
&\|\EEE_j\lft[\hat K^{(\d,k)}_{2,j}(q,\f,X,x,\s,x',\s')\rgt]\|_{h,T_{j+1}(\f',X)}
\notag\\
&\le C^{1+|p|} 
L^{-d(p)\frac{\a^2}{4\p}}\|K^{(\d,k)}_{2,j}\|_{2,h,T_j} \lft({A\over 2}\rgt)^{-|X|_j} 
G_{j+1}(\f',\bar X),
\eal
where 
$$
p=
\begin{cases}
q+\h(\s+\s')\qquad & \text{ if $\d=a$}\\
q+\bar \h(\s+\s')\qquad & \text{ if $\d=\bar a$}\\
q+(\h\s+\bar \h \s')\qquad & \text{ if $\d=b$}\\
\end{cases}
.
$$
For other terms for which the above power counting improvement is not
sufficient we need to extract some finite order of the Taylor
expansion, which we now define. 
Let $F(\x,X)$ be a smooth
function of the field $\{\x_x:x\in X^*\}$; the $n$-order
Taylor expansion of $F(\x,X)$ at $\x=0$ is
\bal\lb{defTay}
&(\Tay_{n,\x} F)(\x,X)
:=\sum_{m=0}^n {1\over m!}
\sum_{x_1\ldots, x_m\in X^*}\x_{x_1}\cdots\x_{x_m}
{\dpr^m F\over \dpr \x_{x_1}\cdots \dpr \x_{x_m}}(0,X);
\eal
the $n$-order remainder is
\be\lb{defRem}
(\Rem_{n,\x} F)(\x,X):= F(\x,X)-(\Tay_{n,\x} F)(\x,X).
\ee
 The next theorem provides the power counting improvement
in such cases. 
\begin{theorem}\lb{by}
Consider  a charge $p$ activity $F(\f,X)$ with support $X\in \SS_j$ and fix any point 
$x_0\in X$. For any $m\in \NNN$, there exist $C\=C(\a)$ and $C_m$ 
such that,  
if $(\d\f)_x:=\f_x-\f_{x_0}$
\be
\|\Rem_{m,\d\f'} \EEE_j\lft[F(\f,X)\rgt]\|_{h,T_{j+1}(\f',X)}\le \r_m(p,\a)
\|F\|_{h,T_j} \lft({A\over 2}\rgt)^{-|X|_j} 
G_{j+1}(\f',\bar X)
\ee
for a ``dimensional factor'' 
$$
\r_m(p,\a):=
 C^{1+|p|}  C_m
L^{-d(p)\frac{\a^2}{4\p}}(\sqrt{\kappa_L} L)^{-(m+1)}
$$
where, again, $d(p)=p^2$ if $|p|\le 1$ and $d(p)=2|p|-1$ otherwise. 
\end{theorem}
The  proof of this theorem, mostly borrowed from  \citep{Fa12}, 
is in Appendix \ref{callan}. $\kappa_L= c(\log L)^{-1}$ as stated in 
Lemma \ref{l6.53}.
There are various consequences of this Theorem that interest us. 
First, it applies to the neutral components of $K_{0,j}$. Setting
$F(\f,X):= \hat K_{0,j}(0,\f,X)$,  
\bal\lb{6.62}
&\|\Rem_{2,\d\f'} 
\EEE_j\lft[\hat K_{0,j}(0,\f,X)\rgt]\|_{h,T_{j+1}(\f',X)}
\notag\\
&\le\r_2(0,\a)\|K_{0,j}\|_{h,T_j} \lft({A\over 2}\rgt)^{-|X|_j} 
G_{j+1}(\f',\bar X).
\eal
For $L$ large enough,  
the dimensional factor $\r_2(0,\a)= C C_2 (\sqrt{\kappa_L} L)^{-3}$ beats  
the volume factor
$L^2$.
%
%\begin{theorem}\lb{fa}
%In the same hypothesis of Theorem \ref{dh}, 
%for any $m\in \NNN$, there exist $C\=C(\a)$ and 
%$C_m$ such that, 
%%
%\bal
%&\|\Rem_{m,\d\f'} \EEE_j\lft[F(\f)\rgt]\|_{h,T_{j+1}(\f',X)}
%\notag\\&
%\le 
% C^{1+|p|}  C_m(\sqrt{\kappa_L} L)^{-(m+1)}
%L^{-(2|p|-1)\frac{\a^2}{4\p}}\|F\|_{h,T_j} \lft({A\over 2}\rgt)^{-|X|_j} 
%G_{j+1}(\f',\bar X)
%\eal
%\end{theorem}
Second, this theorem applies to the components of  $\hat K_{0,j} $ with 
charges $q=\pm1$. Indeed, for $F(\f,X):=\hat K_{0,j}(q,\f, X)$ the
hypothesis holds for $p=q$ and then 
\bal
&\|\Rem_{0,\d\f'} \EEE_j\lft[\hat K_{0,j}(q,\f,X)\rgt]\|_{h,T_{j+1}(\f',X)}
\notag\\
&\le 
\r_0(q,\a)\|K_{0,j}\|_{h,T_j} \lft({A\over 2}\rgt)^{-|X|_j} 
G_{j+1}(\f',\bar X).
\eal
For $q=\pm1$ and $L$ large enough, the dimensional factor 
 $\r_0(1,\a)=C_0 C^2  (\sqrt{\kappa_L} L)^{-1}
L^{-\frac{\a^2}{4\p}}$ is smaller than the volume factor $L^2$.
The third application is  the charged components of $\hat K_{1,j}$. 
We find
\bal\lb{6.62b}
&\|\Rem_{1,\d\f'} 
\EEE_j\lft[\hat K_{1,j}(q,\f,X,x,\s)\rgt]\|_{h,T_{j+1}(\f',X)}
\notag\\
&\le \r_1(q+\h\s,\a)\|K_{1,j}\|_{1,h,T_j} \lft({A\over 2}\rgt)^{-|X|_j} 
G_{j+1}(\f',\bar X).
\eal
Finally, for $\d=a,\bar a, b$ and $0\le k\le j$, 
\bal\lb{lili2}
&\|\Rem_{0,\d\f'}\EEE_j\lft[\hat K^{(\d,k)}_{2,j}(q,\f,X,x,\s,x',\s')\rgt]\|_{h,T_{j+1}(\f',X)}
\notag\\
&\le C^{1+|p|} 
C_0 L^{-2d(p)}\lft(\sqrt{k_L} L\rgt)^{-1}\|K^{(\d,k)}_{2,j}\|_{2,h,T_j} \lft({A\over 2}\rgt)^{-|X|_j} 
G_{j+1}(\f',\bar X),
\eal
where 
$$
p=
\begin{cases}
q+\h(\s+\s')\qquad & \text{ if $\d=a$}\\
q+\bar \h(\s+\s')\qquad & \text{ if $\d=\bar a$}\\
q+(\h\s+\bar \h \s')\qquad & \text{ if $\d=\bar a$}.\\
\end{cases}
$$
%*****
We can now describe  the ``power counting'' argument that will drive our 
analysis in the rest of the paper: 
a) by Theorem \ref{dh}, terms with charge $q$  contract by a factor 
$L^{-\frac{\a^2}{4\p}d(q)}$; 
b) by Theorem \ref{by}, terms proportional to $(\dpr \f')^n$
contract by a factor $L^{-n}$; 
c) as  a consequence of Lemma \ref{l6.90}, 
all terms are increased by a
volume factor $L^2$. 
Therefore, at $\a^2=8\p$, the action of the RG to contract
the size of: 
i) the  terms of total  integer charge $p$, with  $|p|\ge 2$; 
ii) the terms of total  charge $p$, 
$|p|=1$, after that  the 0-th order Taylor expansion in $\dpr \f'$ has been
extracted;
iii) the terms of total charge $p$, with $|p|=\h$ or $\bar \h$, 
after that  the 1-th order Taylor expansion in $\dpr \f'$ has been
extracted;
iv)  neutral terms, after that  the
2-th order Taylor expansion in $\dpr \f'$ has been
extracted. 
The terms that are extracted at points ii), iii) and iv) 
are absorbed into $E_j, t_j$ (see definitions before \pref{pfr}) to generate  
$E_{j+1}, t_{j+1}$.  
These ideas will be  made precise in the next sections.

\section{Renormalization Group Map}\lb{RGM}
%Set 
%$\G_{j,n}(x):=\G_{n}(x)+\G_{n+1}(x)+\cdots+\G_{j}(x)$ if $n\le j$; and 
%$\G_{j,n}(x):=0$ if $n\ge j+1$. Besides, 
%$\G_{j,n}(0|x):= \G_{j,n}(0)-\G_{j,n}(x)$. Given the functions of the field
%$F_1(\F), \ldots,F_s(\F)$, the quantity $O(F_1^{p_1},\ldots,F_s^{p_s})$ will 
%indicate  terms that, 
%if $F_q(\f)$ is rescaled into $t F_q(\f)$,  contain at least a power
%$p_q$ of $t$ for each $q=1,2,\ldots, s$. 
In the present and in the following section we adopt an abridged
notation for the fields.
In general, 
we remove the labels $j$ because they will be clear from the context, 
and we label the sum of the fields on higher
scales with a prime, so that $\z_x:=\z^{(j)}_x$ and 
$\f'_x:=\z^{(R)}_x+\z^{(R-1)}_x+\cdots+\z^{(j+1)}_x$; besides,   
$\f_x:=\f'_x+\z_x$. We also set $\F=(J,\f)$ and $\F'=(J,\f')$. 
We indicate with  $O(F_1, \ldots F_n)$ a term that is 
proportional to  the fist power, at least, 
of each of $F_j$'s.
Besides in the context of the inductive hypothesis described in Section \ref{strategy}, 
we will also assume the following symmetry properties. 
Define the $\p/2$ rotation $R(x_0,x_1):=(-x_1,x_0)$ and the translation 
$T_y x:=x+y$; and extend these transformations
in a natural way to lattices subsets; besides,  let $(R\f)_x:=\f_{R x}$ and 
$(T_y\f)_x:=\f_{x+y}$. We inductively assume that, for $\SS=R, T_y$, 
\be\lb{sym1}
\hK_{0,j}(q,\SS\f,\SS Y)=\hK_{0,j}(q,\f,Y),   
\ee
% %
\be\lb{sym2}
\hK_{1,j}(q,\SS\f,\SS Y, \SS x, \s)=\hK_{1,j}(q,\f,Y, x, \s),
\ee
\be\lb{sym3}
\hK^\dagger_{1,j}(q,\SS\f,\SS Y, \SS x, \s)=\hK^\dagger_{1,j}(q,\f,Y, x, \s),
\ee
\be\lb{sym04}
\hK^{(\d,k)}_{2,j}(q,\SS\f,\SS Y,\SS x,\s,\SS x',\s')=
\hK^{(\d,k)}_{2,j}(q,\f,Y,x,\s,x',\s').
\ee
Besides, 
\be\lb{sym4}
\hK_{0,j}(-q,-\f,Y)=\hK_{0,j}(q,\f,Y),   
\ee
\be\lb{sym5}
\hK_{1,j}(-q,-\f,Y, x, -\s)=\hK_{1,j}(q,\f,Y, x, \s),
\ee
\be\lb{sym6}
\hK^\dagger_{1,j}(-q,-\f,Y, x, -\s)=\hK^\dagger_{1,j}(q,\f,Y, x, \s),
\ee
\be\lb{sym7}
\hK^{(\d,k)}_{2,j}(-q,-\f,Y,x,-\s,x',-\s')=
\hK^{(\d,k)}_{2,j}(q,\f,Y,x,\s,x',\s').
\ee
We now discuss the RG procedure at a generic scale $j=1, \ldots, R-1$; subsequently 
we will discuss the slightly different procedure at scales $j=0$.  

\subsection{General RG  step}\lb{s4.2} 
Assume by induction that at a given scale 
$j=1,2,\ldots,R-1$ the formula \pref{pfr}  holds. 
We want to provide a useful way to recast  
$\O_{j+1}=\EEE_j\lft[\O_j(J,\f'+\z)\rgt]$ into the 
same form of \pref{pfr}:
\be\lb{pfr2}
\O_{j+1}(\F')=e^{E_{j+1}|\L|}
\sum_{X\in \PP_{j+1}}e^{U_{j+1}(\F',\L\bs X)}
\prod_{Y\in \CC_{j+1}(X)} K_{j+1}(\F',Y).
\ee
We have the freedom to decide what to include  in 
$K_{j+1}$ and what  in $U_{j+1}$.
Our aim will be to have a formula for $K_{j+1}$ of the form $\LL_j + \RR_j$
where $\LL_j$ contains the  linear order in $K_j$ and 
the linear and quadratic orders in $s_j$ and $z_j$; besides, we want $\LL_j$
to be a contraction. 
To obtain that, as
explained in the end of the previous section, 
we need to implement the extraction 
based on the power counting argument.  
The next Lemma can be read in this way: there is a natural tentative 
choice for $K_{j+1}$, which 
at lowest orders contains  the terms 
$\EEE_j [K_j]$ and $\EEE^T_j[V_j;V_j]$; 
from such a choice,
a term $Q_j= O(K_{j})$ is extracted from $\EEE_j [K_j]$  and a term $Q^*_j=O(V^2_{j})$ 
is extracted  
from $\EEE^T_j[V_j;V_j]$; next,   $Q_j$ and $Q^*_j$ are stored into 
$U_{j+1}$ and generate the new--scale parameters,  $E_{j+1}, t_{j+1}$,  
from the old ones, $E_j, t_{j}$. 

Before stating the Lemma, we need some definitions. 
%If $Y$ is a $j+1$ polymer, 
%in this section we will denote $[Y]^*_{j}$ the small set neighborhood of $Y$ seen 
%as a $j$ polymer, i.e. $[Y]^*_{j}=\cup_{B\in\BB_j(Y)}B^* $. 
Introduce the two ``extraction activities'': 
\bd
\item  
The activity $Q_j(\F',B,X)$, which  is nonzero only for 
$X\in \SS_j$ and  $B\in \BB_j(X)$.  It is assumed to  
depend upon the fields  $\{\f'_x, J_{x,\s}:x\in X^*, \s=\pm1\}$;
however, it is also 
assumed that the dependence in at least one power of $J$ is restricted to 
the block $B$ (as opposed to the larger $X^*$).    
\item 
The activity  $Q^*_j(\F',D,Y)$, which is nonzero only for 
$|Y|_{j+1}\le 2$ and  $D\in \BB_{j+1}(Y)$. It is assumed to 
depend upon the fields $\{\f'_x, J_{x,\s}:x\in D^*, \s=\pm1\}$; 
but, again,  one power of  $J$ is in fact restricted to the set $D$
(as opposed to $D^*$). 
\ed
Then define  a new polymer activity  $J_j$, 
which contains the  extraction activities: 
\bal\lb{dfj}
&J_j(\F',D,Y):=Q^*_{j}(\F',D, Y)
+\sum_{B\in \BB_j(D)}
\sum_{X\in \SS_j\atop X\supset B}^{\bar X=Y}
Q_{j}(\F',B,X)
\cr
&\qquad-\d_{D,Y}\sum_{Y'\in \SS_{j+1}}^{Y'\supset D}
\lft[Q^*_{j}(\F',D,Y')+\sum_{B\in \BB_j(D)}
\sum_{X\in \SS_j\atop X\supset B}^{\bar X=Y'}
Q_j(\F',B,X)\rgt].
\eal
Hence $J_j(\F',D,Y)$ is zero unless $Y\in \SS_{j+1}$ and $D\in \BB_{j+1}(Y)$. 
As the conditions $X\in \SS_j$ and $X\supset B$ together imply $X^*\subset D^*$
for $D=\bar B$, 
then  $J_j(\F',D,Y)$ depends upon  
$\{\f'_x, J_{x,\s}: x\in D^*, \s=\pm1\}$;  however, one power of $J$ 
is actually restricted to $D$.
The second line 
of \pref{dfj} (with $\d_{D,Y}=1$ if $Y=D$ and $\d_{D,Y}=0$ otherwise) has been included 
so to obtain the crucial property of zero average:  
\bal\lb{5.23}
\sum_{Y\in \PP^c_{j+1}} J_j(\F',D,Y)=0.
\eal
For $Y\in \PP^c_{j+1}$,
define 
\bal\lb{5.23ancora}
\tilde K_j(\F, Y):=\sum_{X'\in \PP_{j}(Y)}^{\overline  {X'}=Y} 
e^{U_{j}(\F,Y\bs X')} 
\prod_{Y'\in \CC_{j}(X')} K_j(\F, Y'),
\eal 
which depends on $\{\f_x, J_{x,\s}: x\in Y^*, \s=\pm1\}$, 
Now we are ready for the extractions. 
For every  block $D\in \BB_{j+1}$, 
define
\bal\lb{4.19bis}
&P_j(\F',\z,D):=e^{U_j(\F,D)}-e^{U_{j+1}(\F',D)+(E_{j+1}-E_j)|D|},
\eal
which depends on $\{\z_x:x\in \cup_{B\in \BB_j(D)} B^*\}$ and 
on $\{\f'_x,J_{x,\s}:x\in D^*,\s=\pm1\}$.
For every connected polymer $Y\in \PP_{j+1}^c$, define
\bal
\lb{5.9bis}
&R_j(\F',\z,Y):=\tilde K_j(\F,Y)-\sum_{D\in \BB_{j+1}(Y)}J_j(\F',D,Y),
\eal
which depends on $\{\z_x:x\in Y^*\}$ and 
on $\{\f'_x, J_{x,\s}:x\in Y^*, \s=\pm1 \}$.
Note that in   \pref{5.23ancora} and  \pref{5.9bis} one power of $J$ is 
restricted to $Y$; likewise, in \pref{4.19bis} one power of $J$ is restricted to $D$.
\begin{lemma}
\lb{l5.3}
Given formula \pref{pfr} with certain $t_j$, $E_j$ and $K_j$; 
given any two extraction activities    as defined above
and such that 
\be\lb{cond0}
Q_{j}(\F',B,X)=O(K_j), \qquad Q^*_{j}(\F',D,Y)=O(V_j^2); 
\ee
and given parameters 
$E_{j+1},t_{j+1}$ that satisfy
\bal
\lb{cond1}
(E_{j+1}-E_{j})|D|+V_{j+1}(\F',D)-\EEE_j\lft[V_{j}(\F,D)\rgt]
&=O(K_{j}, V_{j}^2), 
\eal 
the following holds. A possible choice for 
$K_{j+1}$ in \pref{pfr2} is 

\bal\lb{j+1}
K_{j+1}(\F', Y')=&\sum_{X_0,X_1\atop  Z, (D) }^{\to Y'}
e^{-(E_{j+1}-E_j)|W|+U_{j+1}(\F',Y'\bs W)}
\notag\\
&\qquad\times   
\EEE_j\lft[P_j(\F',\z)^ZR_j(\F',\z)^{X_1}\rgt]
J_j(\F')^{X_0,(D)},
\eal
where the notation is:
\bd 
\item The sum with 
label $\to Y'$ indicates the sum  
over three $j+1$--polymers $X_0$, $X_1$,  $Z$, contained in $Y'$,
and over one $j+1$--block, $D_Y\in \BB_{j+1}(Y)$, per  each polymer 
$Y\in \CC_{j+1}(X_0)$, such
that: a) $X_0$ and $X_1$ are separated by at least 
by one $j+1$--block, namely $\CC_{j+1}(X_0\cup X_1)=
\CC_{j+1}(X_0)+ \CC_{j+1}(X_1)$; b) $Z\in \PP_{j+1}(Y'\bs (X_0\cup X_1))$;
c) each connected 
component of $X_0$ is $j+1$-small; 
d) $\cup_{Y}D^*_Y\cup Z\cup X_1=Y'$. Besides,  $W\=X_0\cup X_1\cup Z$.
\item For polymers $Z, X\in \PP_{j+1}$, we set
\be\lb{mixed}
P_j(\F',\z)^Z:=\prod_{D\in\BB_{j+1}(Z)}P_j(\F',\z,D),\qquad 
R_j(\F',\z)^X:=\prod_{Y\in\CC_{j+1}(X)}R_j(\F',\z,Y).
\ee
\item Given $X_0\in \PP_{j+1}$ and one 
$D_Y\in \BB_{j+1}(Y)$ for each $Y\in \CC_{j+1}(X_0)$, we set
\be\lb{mixed2}
J_j(\F')^{X_0,(D)}:=\prod_{Y\in \CC_{j+1}(X_0)} J_j(\F', D_Y, Y).
\ee
\ed
Such  choice of  
$K_{j+1}(\F',Y')$ can be decomposed in the sum of two parts, the leading one,
$\LL_j (\F',Y')$ and the remainder one $\RR_j(\F',Y')$ in a way  that: the latter 
is an higher order correction in the sense that 
if $V_j$, $V_{j+1}$ are scaled by  $t$ and $W_j$, $W_{j+1}$, 
$K_j$ are scaled by 
$t^2$, for small parameter $t$, then $\RR_j(\F',Y')=O(t^3)$; while the former
has an explicit formula
$$
\LL_j(\F',Y')=\LL^{(a)}_j(\F',Y')+\LL^{(b)}_j(\F',Y')+\LL^{(c)}_j(\F',Y'),
$$
where, for $\d E_j:=E_{j+1}-E_j$, 
\bal\lb{5.25}
&\LL^{(a)}_j(\F',Y')
=\sum_{X\in \PP^c_j(Y')}^{\overline X=Y'}
\lft[\EEE_j[K_j(\F,X)]-\sum_{B\in \BB_j(X)} Q_{j}(\F',B,X)\rgt],
\notag\\
&\LL^{(b)}_j(\F',Y')
=
\frac12\sum_{B_0, B_1\in \BB_{j}(Y')}^{\overline{B_0\cup B_1}=Y'}
\EEE^T_j\lft[V_j(\tilde t_j, \F,B_0);V_j(\tilde t_j,\F,B_1)\rgt]
-\sum_{D\in \BB_{j+1}(Y')}Q^*_{j}(\F',D, Y'),
\notag\\
&\LL^{(c)}_j(\F',Y')
=-\sum_{B\in \BB_{j}}^{\bar B=Y'}
\Bigg[\d E_j|B|+V_{j+1}(\F',B)
-\EEE_j\lft[V_j(\F,B)\rgt]
-\sum_{X\in \SS_j}^{X\supset B}
Q_{j}(\F',B,X)\Bigg]
\notag\\
&-\sum_{D\in \BB_{j+1}}^{D=Y'}
\Bigg[W_{j+1}(\F',D)-
\EEE_j\lft[W_j(\tilde t_{j},\F,D)\rgt]
-\sum_{Y\in \SS_{j+1}}^{Y\supset D} 
Q^*_{j}(\F',D,Y)\Bigg]
\eal
for any $\tilde t_j$ such that $\tilde t_j-t_j= (O(z^2), O(z^2), O(z), O(z))$. 

Besides the scale $j+1$ activity, $K_{j+1}$,  
can be decomposed into charged terms 
as stated in  \pref{pinning0}, \pref{pinning}, \pref{dec3} and \pref{dec4} for 
the scale $j$ activity.  
\end{lemma} 
\bpr
Starting from \pref{pfr} and  re-blocking the polymers 
on scale $j+1$, we obtain an  equivalent formulation
for $\O_j$:
\bal\lb{pfj+1}
\O_j(\F)=e^{E_j|\L|}\sum_{X\in \PP_{j+1}} \Big[\prod_{D\in \BB_{j+1}(\L\bs X)} e^{U_{j}(\F,D)}\Big]
\prod_{Y\in \CC_{j+1}(X)} \tilde K_j(\F, Y)
\eal
for $\tilde K_j$ given by \pref{5.23ancora}.  
Plugging \pref{4.19bis} and \pref{5.9bis} 
in \pref{pfj+1} and expanding,  we find
\pref{pfr2}, for $K_{j+1}$ given by \pref{j+1}.
Observe that, to derive it,  we also used the factorization of $\EEE_j$ over
sets that are in two different connected components of a $j+1$--polymer
as explained in \pref{separation}.
Besides,  in some terms we have the parameters $\tilde t_j$ instead of the more 
natural $t_j$
because the difference
can be  
left inside $\RR_j$.
Finally, by construction, $W\subset Y'$
so that that $K_{j+1}(\F',Y)$ depends 
on the fields $\{\f'_x,J_{x,\s}: x\in Y^*,\s=\pm1\}$; and, in particular, 
one power of $J$ is restricted to $Y$, as required. 

We have to prove that the linear part in $K_j$, $Q_j$, $Q^*_j$ and second
order part in $V_j$ of this choice of
$K_{j+1}$ is \pref{5.25}: expanding formula \pref{j+1}, using \pref{cond0} and 
\pref{cond1}, 
we obtain  \pref{5.25}  via 
two simple identities,
\bal\lb{*id1}
&\sum_{D\in \BB_{j+1}(Y')}J_j(\F',D,Y')=
\sum_{D\in \BB_{j+1}(Y')}Q^*_{j}(\F',D,Y')
+\sum_{X\in\SS_j}^{\bar X=Y'} 
\sum_{B\in \BB_{j}(X)}Q_{j}(\F',B,X)
\cr
&\qquad\qquad-
\sum_{D\in \BB_{j+1}}^{D=Y'}\sum_{Y\in \SS_{j+1}}^{Y\supset D}
Q^*_{j}(\F',D,Y)
-\sum_{B\in \BB_j}^{\bar B=Y'}
\sum_{X\in \SS_j}^{X\supset B}
Q_{j}(\F',B,X);
\eal
and, by \pref{5.23}, 
\bal\lb{*id2}
&\sum_{Y\in \SS_{j+1}}
\sum_{D\in \BB_{j+1}(Y)}^{D^*=Y'}\; J_j(\F',D,Y)
=\sum_{D\in \BB_{j+1}}^{D^*=Y'}
\sum_{Y\in \SS_{j+1}}^{Y\supset D}\; J_j(\F',D,Y)=0.
\eal
This completes the proof of the Lemma.\epr
The usefulness of  \pref{5.25} is that,  
as planned before, in $\LL^{(a)}_j(\F',Y')$ and $\LL^{(b)}_j(\F',Y')$
we read the extraction of 
$Q_{j}$  and $Q^*_{j}$ from $\EEE_j[K_j]$ and $\EEE^T_j[V_j;V_j]$ respectively;  
in $\LL^{(c)}_j(\F',Y')$ the same terms are re-absorbed into 
$E_{j},t_{j}$ 
so generating  $E_{j+1},t_{j+1}$.

Note that by construction $\LL_j$ depends on 
$t_j$, $K_j$, $Q_{j}$, $Q^*_{j}$, $\tilde t_j$, $\d E_j$ and  $t_{j+1}$; however, in 
Section \ref{s6},  we will determine the last five of them as function of $t_j$ and $K_j$,
so that also $\LL_j$ is ultimately only a function on $t_j$ and $K_j$. In fact, 
as stated in the next Theorem, also the dependence on $t_j$ disappears from $\LL_j$. 
Decompose 
$$
\LL_j(\F',Y)= \LL_{0,j}(\f',Y) + \LL_{1,j}(\F',Y)+ \LL_{2,j}(\F',Y)
+ \LL_{\ge 3,j}(\F',Y)
$$
where  the enumeration refers to the powers of $J$. The term that in linear
in $J$ is  
\bal
\LL_{1,j}(\F',Y)
&=L^{-2(j+1)} Z_{j+1} \sum_{x\in Y\atop\s=\pm} J_{x,\s} \LL_{1,j}(\f',Y,x,\s)
\notag\\
&+L^{-2(j+1)} \bar Z_{j+1} \sum_{x\in Y\atop\s=\pm} J_{x,\s} \LL^{\dagger}_{1,j}(\f',Y,x,\s).
\eal
The term that is quadratic in $J$ is 
\bal
\LL_{2,j}(\F',Y)&=\sum_{x_1\in Y, x_2\in Y^*\atop \s_1,\s_2=\pm1}
J_{\s_1, x_1}J_{\s_2, x_2}\LL_{2,j}(\f',Y,x_1,\s_1,x_2,\s_2)
\eal
and can be further decomposed into (suppressing the dependence in 
$\f',Y,x_1,\s_1,x_2,\s_2$) 
$$
\LL_{2,j}=\sum_{k=0}^{j} 
2^{-(j-k)}L^{-4k } e^{-L^{-k}|x_1-x_2|}
\lft[Z^2_k\LL^{(a,k)}_{2,j}+ \bar Z^2_k\LL^{(\bar a,k)}_{2,j}
+Z_k \bar Z_k\LL^{(b,k)}_{2,j} 
\rgt]. 
$$
\begin{theorem}\lb{faculty2}
For a suitable choice of  $Q_{j}$, $Q^*_{j}$, $\tilde t_j$, $\d E_j$ and  $t_{j+1}$
as functions of $t_j$, $K_j$, the leading part $\LL_j$ is independent of $t_j$ and
is linear in  $K_j$. 
Besides,  under the inductive assumption that  
$$
\lft|\frac{Z_j}{Z_{j+1}}\rgt|\le 1, 
\qquad \lft|\frac{\bar Z_j}{\bar Z_{j+1}}\rgt|\le 1,
$$
$\LL_j$ satisfies the following bounds: 
\bd
\item for the term of $\LL_j$ that is independent of $J$, 
\be\lb{allen2}
\|\LL_{0,j}\|_{h, T_{j+1}}\le  \r(L,A)\|K_{0,j}\|_{h, T_j};
\ee
where $\r(L,A)$ is arbitrarily small for $L$ and  $A$ large enough;
\item
for the terms of $\LL_j$ that are linear or quadratic in $J$, 
\bal\lb{arkani2}
&\|\LL_{1,j}\|_{1,h,T_{j+1}}\le \r(L,A,\h) \|K_{1,j}\|_{1,h,T_{j}}
\notag\\
&\|\LL^{\dagger}_{1,j}\|_{1,h,T_{j+1}}\le \r(L,A,\h) \|K^\dagger_{1,j}\|_{1,h,T_{j}}
\eal
\be\lb{goddar2}
\|\LL^{(\d, k)}_{2,j}\|_{2,h,T_{j+1}}\le \r(L,A,\h) \|K^{(\d, k)}_{2,j}\|_{2,h,T_{j}}
\ee
for a $\r(L,A,\h)$ that is arbitrarily small for any $\h\in (0,1)$ if
$L$ and $A$ are large enough. 
\ed
\end{theorem}
The first point of the result was already proven in \cite{Fa12}. 
The proof of the second point is a direct consequence of Lemma \ref{faculty}, Lemma \ref{bois}, 
Lemma \ref{bourgain} and Lemma \ref{bois2} in Section \ref{s6}. There 
we will also explain how to obtain \pref{lk}, 
\pref{elk}, \pref{sk} and the following formula for $\tilde t_j$,
\be\lb{5.25def}
\tilde t_j=(s_{j+1}, z_{j+1} L^{-2} e^{\frac{\a^2}2 \G_j(0)}, Z_{j+1} L^{-2} e^{\h^2\frac{\a^2}2 \G_j(0)}, 
\bar Z_{j+1} L^{-2} e^{\bar \h^2\frac{\a^2}2 \G_j(0)}).
\ee

Consider now the remainder part. 
Using \pref{j+1} for $K_{j+1}$ and formula \pref{5.25} for its
leading part, we obtain the following formula for $\RR_j$:
\be\lb{exprem}
\RR_j(\F', Y'):=%\RR^{(0)}_j(\F', Y')= 
\sum_{n=1}^9 \RR^{(n)}_j(\F', Y')
\ee
where, suppressing the dependence in the field (again
$\d E_{j}:=E_{j+1}-E_j$), 
\bal
\RR^{(1)}_j(Y')
&=
\sum_{D\in \BB_{j+1}}^{D=Y'}\Big[\EEE_j\lft[P_j(D)\rgt]+V_{j+1}(D)- \EEE_j[V_j(D)]-\d E_{j+1}|D|
\notag\\
&\qquad\qquad
-\frac12 \EEE^T[V_j(D);V_j(D)]+W_{j+1}(D)-\EEE_j[W_j(D)]\Big],
\notag\\
\RR^{(2)}_j(Y')&=
\frac12\sum_{D_1, D_2\in \BB_{j+1}\atop D_1\neq D_2}^{D_1\cup D_2=Y'}
\Big[\EEE_j\lft[P_j(D_1) P_j(D_2)\rgt]- 
\EEE^T_j\lft[V_j(D_1); V_j(D_2)\rgt]\Big],
\notag\\
\RR^{(3)}_j(Y')
&=
\sum_{D\in \BB_{j+1}}^{D=Y'}\Big[\EEE_j[W_j(D)]-\EEE_j[W_j(\tilde t_j, D)]\Big],
\notag\\
\RR^{(4)}_j(Y')
&=
\frac12\sum_{D_1, D_2\in \BB_{j+1}}^{D_1\cup D_2=Y'}
\Big[\EEE^T_j\lft[V_j(D_1); V_j(D_2)\rgt]- 
\EEE^T_j\lft[V_j(\tilde t_j, D_1); V_j(\tilde t_j, D_2)\rgt]\Big],
\notag\\
\RR^{(5)}_j(Y')&=
\sum_{|\CC_{j+1}(X_0\cup X_1)|\ge 1\atop |Z|_{j+1}+|\CC_{j+1}(X_0\cup X_1)|\ge 2}^{\to Y'} 
\EEE_j\lft[P_j^Z R_j^{X_1}\rgt] J_j^{X_0, (D)},
\notag\\
\RR^{(6)}_j(Y')&=
\sum_{|\CC_{j+1}(X_0\cup X_1)|\ge 1}^{\to Y'} 
\lft(e^{-\d E_j|Y'|+U_{j+1}(Y'\bs W)}-1\rgt)\EEE_j\lft[P_j^Z R_j^{X_1}\rgt] J_j^{X_0, (D)},
\notag\\
\RR^{(7)}_j(Y')&=
\sum_{Z\in \PP_{j+1}\atop 
|Z|_{j+1}\ge 3}^{Z=Y'}\EEE_j\lft[P_j^Z\rgt]+ \lft(e^{-\d E_j|Y'|}-1\rgt)
\EEE_j\lft[P_j^{Y'}\rgt],
\notag\\
\RR^{(8)}_j(Y')&=
\sum_{X\in \PP_j\atop |\CC_j(X)|\ge 2}^{\bar X=Y'}\EEE_j\lft[\prod_{Y\in \CC_j(X)} K_j(Y)\rgt],
\notag\\
\RR^{(9)}_j(Y')&=\sum_{X\in \PP_j}^{\bar X=Y'} 
\EEE_j\lft[\lft(e^{U_j(Y'\bs X)}-1\rgt)\prod_{Y\in \CC_j(X)} K_j(Y)\rgt].
\eal
$\RR_j$, as well as each $\RR^{(n)}_j$, can be  decomposed in terms with increasing 
powers of $J$, 
\be\lb{rdec1}
\RR_j(\F',Y)=\RR_{0,j}(\f',Y)+\RR_{1,j}(\F',Y)+\RR_{2,j}(\F',Y)
+\RR_{\ge 3,j}(\F',Y).
\ee 
The term that is linear in $J$ is
\bal\lb{rdec2}
\RR_{1,j}(\F',Y)
&=
L^{-2(j+1)}Z_{j+1}\sum_{x\in Y\atop \s=\pm1} 
J_{x,\s}\RR_{1,j}(\f',Y, x,\s)
\notag\\
&+ L^{-2(j+1)}\bar Z_{j+1}
\sum_{x\in Y\atop \s=\pm1} J_{x,\s}\RR^{\dagger}_{1,j}(\f',Y, x,\s).
\eal
The term that  is quadratic in $J$ is 
\bal\lb{rdec3}
\RR_{2,j}(\F',Y)&=\sum_{x_1\in Y, x_2\in Y^*\atop \s_1,\s_2=\pm1}
J_{\s_1, x_1}J_{\s_2, x_2}\RR_{2,j}(\f',Y,x_1,\s_1,x_2,\s_2)
\eal
and can be further decomposed into (suppressing the dependence in 
$\f',Y,x_1,\s_1,x_2,\s_2$) 
\be\lb{rdec4}
\RR_{2,j}=\sum_{k=0}^{j} 
2^{-(j-k)}L^{-4k } e^{-L^{-k}|x_1-x_2|}
\lft[Z^2_k\RR^{(a,k)}_{2,j}+ \bar Z^2_k\RR^{(\bar a,k)}_{2,j}+Z_k \bar Z_k\RR^{(b,k)}_{2,j} 
\rgt]. 
\ee
\begin{theorem}\lb{crone00}
If $z>0$ is small enough and  $|s_j|,|z_j|\le c_0|q_j|$, 
$\|K_{0,j}\|_{h,T_j}\le c_0|q_j|^2$, there exists $C\=C(A, L, \a)$ such that, 
\bd
\item for the term of $\RR_j$ that is independent of $J$
\bal\lb{old3.22}
&\|\RR_{0,j}- \dot\RR_{0,j}\|_{h, T_{j+1}}
\notag\\
&\le C 
\lft[|q_j|^2|s_j-\dot s_j|+  |q_j|^2|z_j-\dot z_j|+ |q_j|\|K_{0,j}-\dot K_{0,j}\|_{h,T_j}\rgt]
\eal
where $\dot R_{0,j}$ is obtained from $R_{0,j}$ by  replacing $s_j,z_j,K_{0,j}$ with any 
$\dot s_j, \dot z_j, \dot K_{0,j}$ that satisfy   $|\dot s_j|,|\dot z_j|\le c_0|q_j|$
and $\|\dot K_{0,j}\|_{h,T_j}\le c_0|q_j|^2$;
\item for the terms of $\RR_j$ that are linear in $J$,
\be\lb{cosmo00}
\|\RR_{1,j}\|_{1, h, T_{j+1}}\le C \lft[|q_j|^2
+|q_j|\|K_{1,j}\|_{1,h,T_j}+|q_j|\|K^\dagger_{1,j}\|_{1,h,T_j}\rgt],
\ee
and the same bound is valid for $\|\RR^\dagger_{1,j}\|_{1, h, T_{j+1}}$;
\item 
for the terms of $R_{j}$ that are quadratic in $J$, with the extra assumption
that $\|K_{1,j}\|\le c_0 |q_j|^2$ and $\|K^\dagger_{1,j}\|\le c_0 |q_j|^2$, 
\bal\lb{geary00}
\|\RR^{(\d,k)}_{2,j}\|_{2, h, T_{j+1}} \le
\begin{cases} 
C |q_k| &\text{for $k=j$} \\
C |q_j| \|K^{(\d,k)}_{2,j}\|_{2,h,T_j}\qquad&\text{for $0\le k\le j-1$} 
\end{cases}
\eal
\ed
\end{theorem}
The first point  was already proven in \citep{Fa12}. The second and third 
points are  consequence of Lemma \ref{crone} and Lemma \ref{fassin}.

\subsection{First RG step} 
The starting point is formula \pref{itr} that
in our current notations  reads
\be\lb{itr0}
\O_1(\F)=e^{E_0|\L|} \EEE_0\lft[e^{V_0(\F,\L)}\rgt].
\ee
As already noted, the term in the square brackets of \pref{itr0} 
has the form \pref{pfr} for $j=0$ for  $W_{0}(\F,B)=0$, $K_0(\F,Y)=0$, 
and for  parameters $E_0=E$ and $t_0=(s, z, 1, 0)$. 
We want to recast $\O_1$ into the form \pref{pfr} 
for $j=1$. For doing so, 
we apply Lemma \ref{l5.3} to the scale $j=0$: since $K_{0}=0$,
in Section \ref{s6} we will see that there exists a choice of
$Q_0^*$, $\tilde t_0$, $\d E_0$ and $t_1$ such that   
\be\lb{5.25bis}
\LL_{0}(\F,Y)\=0.
\ee
However, since the choice for $Q^*_0$ will differ from the general formula for 
$Q^*_j$ in the part that does not depend on $J$ 
(we do this for merging with the treatment of 
\citep{Fa12})  the remainder part is
slightly different from \pref{exprem} at $j=0$, $W_0=0$ and $K_0=0$; indeed we have 
\be\lb{exprem0}
\RR_0(\F', Y'):=%\RR^{(0)}_j(\F', Y')= 
\sum_{n=1}^6 \RR^{(n)}_0(\F', Y')
\ee
where, suppressing the dependence in the field, 
\bal
\RR^{(1)}_0(Y')
&=
\sum_{D\in \BB_{1}}^{D=Y'}\Big[\EEE_0\lft[P_0(D)\rgt]+V_{1}(D)- \EEE_0[V_0(D)]-\d E_0|D|
\notag\\
&\qquad\qquad
-\frac12 \EEE^T_0[V_0(D);V_0(D)]+W_{1}(D)+\frac12 \EEE^T_0[V_{0,0}(D);V_{0,0}(D)]\Big],
\notag\\
\RR^{(2)}_0(Y')&=
\frac12\sum_{D_1, D_2\in \BB_{1}\atop D_1\neq D_2}^{D_1\cup D_2=Y'}
\Big[\EEE_0\lft[P_0(D_1) P_0(D_2)\rgt]- 
\EEE^T_0\lft[V_0(D_1); V_0(D_2)\rgt]
\notag\\
&\qquad\qquad+ 
\EEE^T_0\lft[V_{0,0}(D_1); V_{0,0}(D_2)\rgt]\Big],
\notag\\
\RR^{(3)}_0(Y')
&=
\frac12\sum_{D_1, D_2\in \BB_{1}}^{D_1\cup D_2=Y'}
\Big[\EEE^T_0\lft[V_0(D_1); V_0(D_2)\rgt]- 
\EEE^T_0\lft[V_0(\tilde t_0, D_1); V_0(\tilde t_0, D_2)\rgt]\Big]
\notag\\
&-
\frac12\sum_{D_1, D_2\in \BB_{1}}^{D_1\cup D_2=Y'}
\Big[\EEE^T_0\lft[V_0(D_1); V_0(D_2)\rgt]- 
\EEE^T_0\lft[V_{0,0}(\tilde t_0, D_1); V_{0,0}(\tilde t_0, D_2)\rgt]\Big],
\notag\\
\RR^{(4)}_0(Y')&=
\sum_{|\CC_{1}(X_0\cup X_1)|\ge 1\atop |Z|_{1}+|\CC_{1}(X_0\cup X_1)|\ge 2}^{\to Y'} 
\EEE_0\lft[P_0^Z R_0^{X_1}\rgt] J_0^{X_0, (D)},
\notag\\
\RR^{(5)}_0(Y')&=
\sum_{|\CC_{1}(X_0\cup X_1)|\ge 1}^{\to Y'} 
\lft(e^{-\d E_0|Y'|+U_{1}(Y'\bs W)}-1\rgt)\EEE_0\lft[P_0^Z R_0^{X_1}\rgt] J_0^{X_0, (D)},
\notag\\
\RR^{(6)}_0(Y')&=
\sum_{Z\in \PP_{1}\atop 
|Z|_{1}\ge 3}^{Z=Y'}\EEE_0\lft[P_0^Z\rgt]+ \lft(e^{-\d E_0|Y'|}-1\rgt)
\EEE_0\lft[P_0^{Y'}\rgt].
\eal
The decompositions \pref{rdec1}, \pref{rdec2},\pref{rdec3} and \pref{rdec4}
are valid also at $j=0$. 
\begin{theorem}\lb{crone001}
Under the same hypothesis of Theorem \ref{crone00}, 
\bd
\item for the term of $\RR_0$ that is independent of $J$
\bal\lb{old3.221}
&\|\RR_{0,0}-\dot \RR_{0,0}\|_{h, T_{1}}
\le C |q_0|\lft[|s_0-\dot s_0|+ |z_0-\dot z_0|\rgt];
\eal
\item for the terms of $\RR_0$ that are linear in $J$,
\be\lb{cosmo001}
\|\RR_{1,0}\|_{1, h, T_{1}}\le C |q_0|^2,
\ee
and the same bound is valid for $\|\RR^\dagger_{1,0}\|_{1, h, T_{1}}$;
\item 
for the terms of $R_{0}$ that are quadratic in $J$,
\bal\lb{geary001}
\|\RR^{(\d,0)}_{2,0}\|_{2, h, T_{1}} \le C|q_0|. 
\eal
\ed
\end{theorem}
As for Theorem \ref{crone00}, we only need to prove the second and third 
points, which  are  consequence of Lemma \ref{crone} and Lemma \ref{fassin}.
Note that \pref{cosmo001} and \pref{geary001} coincide with 
\pref{cosmo00} and \pref{geary00} at $j=0$; while \pref{old3.221} differs
from \pref{old3.22} at $j=0$ and is the same as in \citep{Fa12}.

\section{Leading part of the  RG map}\lb{s6}
\subsection{Running coupling constants}\lb{sc20}

The choice of  $Q_{j}$ requires Taylor expansion in $\nabla \f'$. 
For any point $x_0\in X$, 
if $(\d \f')_x:= \f'_x-\f'_{x_0}$ (which is 
a sum of $\nabla \f'$'s), using \pref{dec4}, we have 
\bal
&\hK_{0,j}(q,\f,X)= e^{i\a q \f'_{x_0}} \hK_{0,j}(q,\d\f'+\z,X),
\notag\\
&\hK_{1,j}(q,\f,X, x, \s)= e^{i\a (q+\h \s) \f'_{x_0}}
\hK_{1,j}(q,\d\f'+\z,X,x, \s),
\notag\\
&\hK_{1,j}^\dagger(q,\f,X, x,\s)= 
e^{i\a (q+\bar \h \s) \f'_{x_0}}\hK^\dagger_{1,j}(q,\d\f'+\z,X, x,\s),
\notag\\
&\hK^{(a,k)}_{2,j}(q,\f,X,x,\s,x',\s')= 
e^{i (q+\h \s+\h \s')\a\f'_{x_0}}\hK^{(a,k)}_{2,j}(q,\d\f'+\z,X,x,\s,x',\s'),
\notag\\
&\hK^{(\bar a,k)}_{2,j}(q,\f,X,x,\s,x',\s')= 
e^{i (q+\bar \h \s+\bar \h\s')\a\f'_{x_0}}\hK^{(a,k)}_{2,j}(q,\d\f'+\z,X,x,\s,x',\s'),
\notag\\
&\hK^{(b,k)}_{2,j}(q,\f,X,x,\s,x',\s')= 
e^{i (q+\h \s+\bar \h\s')\a\f'_{x_0}}\hK^{(a,k)}_{2,j}(q,\d\f'+\z,X,x,\s,x',\s').
\eal
We now  choose $Q_{j}$. Set 
$Q_{j}(\F',B,X)=0$ if $X\not\in S_j$ or $B\not\in \BB_j(X)$; otherwise 
$Q_{j}(\F',B,X)$ is the sum of  the following four terms.
\bd
\item A term proportional to $K_{0,j}$:
\bal\lb{qbr}
Q_{0,j}(\f',B,X)
=&\frac{1}{|X|}\sum_{x_0\in B} \Tay_{2,\d\f'} 
\EEE_j\lft[\hK_{0,j}(0,\d\f'+\z,X)\rgt]
\notag\\
&+
\frac{1}{|X|}\sum_{x_0\in B\atop\s=\pm1}e^{i\s\a\f'_{x_0}}\Tay_{0,\d\f'}
\EEE_j\lft[\hK_{0,j}(\s,\d\f'+\z,X)\rgt].
\eal
\item Two terms proportional to $K_{1,j}$:
\bal
Q_{1,j}(\F',B,X)=
&
Z_j L^{-2j}\sum_{x\in B\atop \s=\pm1}
J_{x,\s} e^{i\h\a\s \f'_x} \Tay_{1,\d\f'}
\EEE_j\lft[\hK_{1,j}(0,\d\f'+\z,X, x,\s)\rgt]
\notag\\
+&
Z_j L^{-2j}\sum_{x\in B\atop \s=\pm1}
J_{x,\s}e^{i\bar\h\a\s \f'_x} \Tay_{1,\d\f'}
\EEE_j\lft[\hK_{1,j}(-\s,\d\f'+\z,X, x,\s)\rgt],
\notag\\\notag\\
Q^\dagger_{1,j}(\F',B,X)=&
\bar Z_j L^{-2j}\sum_{x\in B\atop \s=\pm1}
 J_{x,\s} e^{i\bar \h\a\s \f'_x}  \Tay_{1,\d\f'}
\EEE_j\lft[\hK^\dagger_{1,j}(0,\d\f'+\z,X, x,\s)\rgt]
\notag\\
+&
\bar Z_j L^{-2j}
\sum_{x\in B\atop \s=\pm1}
 J_{x,\s} e^{i \h\a\s \f'_x}\Tay_{1,\d\f'}
\EEE_j\lft[\hK^\dagger_{1,j}(\s,\d\f'+\z,X, x,\s)\rgt],
\lb{qbr1}
\eal
where the special point in $\d \f'$ is $x$. 
Even though the pinning at $x$ prevents the generation of the volume 
factor $L^2$ when we sum these terms over $X$, note that here the extraction 
is guided by the standard power counting. 
As it will be clear in Section \ref{s6.4}, 
the reason for doing so is that we want to preserve the prefactor 
$L^{-2j}$ at each scale, which costs an $L^2$ factor at each step.
\item  
A term  proportional to $K_{2,j}$:   
\bal
&Q_{2,j}(\F',B,X)
=\sum_{k=0}^j2^{-(j-k)} L^{-4k} \sum_{x_1\in B \atop x_2\in X^*}
e^{-L^{-k}|x_1-x_2|}
\notag\\
&\times \sum_{\s,\s'=\pm1}J_{x_1,\s}J_{x_2,\s'}e^{i\a(\s+\s')(\h-\frac12)\f'_{x_1}}
\notag\\
&\times \lft\{Z_k^2 \Tay_{0,\d\f'}
\EEE_j\lft[ \hat K_{2,j}^{(a,k)}(-\frac{\s+\s'}2, \d\f'+\z, X, x_1, \s, x_2, \s')\rgt]
\rgt.
\notag\\
&\qquad+\lft.
\bar Z^2_k\Tay_{0,\d\f'}\EEE_j\lft[ 
\hat K_{2,j}^{(\bar a,k)}(\frac{\s+\s'}2, \d\f'+\z, X, x_1, \s, x_2, \s')
\rgt]
\rgt.
\notag\\
&\qquad+\lft.
Z_k\bar Z_k \Tay_{0,\d\f'}\EEE_j\lft[ 
\hat K_{2,j}^{(b,k)}(-\frac{\s-\s'}2, \d\f'+\z, X, x_1, \s, x_2,\s')
\rgt]\rgt\}.\lb{qbr2b}
\eal
As opposed to what we did for \pref{qbr1}, in %\pref{qbr2}, 
\pref{qbr2b} the extraction follows a power counting 
that does {\it not} take account of the volume factor $L^2$.
Note that the above term is then irrelevant unless $\h=\frac12$ or $\s=-\s'$;
however we extract the same  $Q_{2,j}$ for every $\s,s',\h$ not 
to have an $L$ that be divergent for $\h\to\frac12$. 
\ed
Finally, we have the following result: 
$\LL^{(a)}_j(\F',Y')$ is of the form 
$$
\LL^{(a)}_j(\F',Y')= \LL^{(a)}_{0,j}(\f',Y') + \LL^{(a)}_{1,j}(\f',Y')
+ \LL^{(a)}_{2,j}(\f',Y')
$$
where  
\bal
\LL^{(a)}_{1,j}(\f',Y')
&=L^{-2(j+1)} Z_{j+1} \sum_{x\in Y'\atop \s=\pm1} J_{x,\s} \LL^{(a)}_{1,j}(\f',Y',x,\s)
\notag\\
&+L^{-2(j+1)} \bar Z_{j+1} \sum_{x\in Y'\atop \s=\pm1} J_{x,\s} \LL^{(a)\dagger}_{1,j}(\f',Y',x,\s)
\notag
\eal
and  
\bal
\LL^{(a)}_{2,j}(\f',Y')
&=\sum_{x_1\in Y', x_2\in {Y'}^*\atop \s_1,\s_2=\pm1} J_{x_1,\s_1} J_{x_2,\s_2}  
\LL^{(a)}_{2,j}(\f',Y',x_1,\s_1,x_2,\s_2)
\eal
for (neglecting the variables that are $\f',Y',x_1,\s_1,x_2,\s_2$ in each term)
\bal
\LL^{(a)}_{2,j}
&=\sum_{k=0}^j 2^{-(j-k)}L^{-4k}e^{-L^{-k}|x_1-x_2|}
\lft[Z^2_k \LL^{(a,k)}_{2,j}
+\bar Z^2_k \LL^{(\bar a,k)}_{2,j}+Z_k \bar Z_k\LL^{(b,k)}_{2,j}\rgt].
\eal
\begin{lemma}\lb{faculty}
Assume by induction that 
$$
\lft|\frac{Z_j}{Z_{j+1}}\rgt|\le 1\qquad \lft|\frac{\bar Z_j}{\bar Z_{j+1}}\rgt|\le 1;
$$
then, for large enough $L$, there exist $\r(L,A)$ and $\r(L,A,\h)$ such that 
\be\lb{allen}
\|\LL^{(a)}_{0,j}\|_{h, T_{j+1}}\le  \r(L,A)\|K_{0,j}\|_{h, T_j};
\ee
\bal\lb{arkani}
&\|\LL^{(a)}_{1,j}\|_{1,h,T_{j+1}}\le \r(L,A,\h) \|K_{1,j}\|_{1,h,T_{j}},
\notag\\
&\|\LL^{(a)\dagger}_{1,j}\|_{1,h,T_{j+1}}\le \r(L,A,\h) \|K^\dagger_{1,j}\|_{1,h,T_{j}},
\eal
\be\lb{goddar}
\|\LL^{(\d,k)}_{2,j}\|_{2,h,T_{j+1}}\le \r(L,A,\h) \|K^{(\d,k)}_{2,j}\|_{2,h,T_{j}}.
%\qquad \text{for $\d=a,\bar a, b$}.
\ee
Besides, fixed any $\h\in (0,1)$, the prefactors 
$\r(L,A)$ and $\r(L,A,\h)$  are arbitrarily small for $L$ and  $A$ large enough. 
\end{lemma}
The proof is in Section \ref{s6.4}. 
The next step is to choose $Q^*_{j}$: set
$Q^*_{j}(\F',D,Y'):=0$ if $|Y'|_{j+1}\ge 3$ or $D\not\in \BB_{j+1}(Y')$; otherwise,
if $j\ge 1$, 
\bal\lb{qqq}
Q^*_{j}(\F',D,Y')
&:=
{1\over 2}\sum_{B_0\in \BB_j(D)\atop B_1\in \BB_{j}(Y')}^{\overline{B_0\cup B_1}=Y'}
\EEE^T_j\lft[V_{0,j}(\tilde t_j,\f,B_0);V_{0,j}(\f,B_1)\rgt]
\notag\\
&+\sum_{B_0\in \BB_j(D)\atop B_1\in \BB_{j}(Y')}^{\overline{B_0\cup B_1}=Y'}
\EEE^T_j\lft[V_{1,j}(\tilde t_j,\F,B_0);V_{0,j}(\tilde t_j,\f,B_1)\rgt] 
\notag\\
&+{1\over 2}\sum_{B_0\in \BB_j(D)\atop B_1\in \BB_{j}(Y')}^{\overline{B_0\cup B_1}=Y'}\EEE^T_j\lft[V_{1,j}(\tilde t_j,\F,B_0);V_{1,j}(\tilde t_j,\F,B_1)\rgt].
\eal
If instead $j=0$, we do not include in \pref{qqq} the fist line, i.e. the one 
proportional to $V_{0,0}^2$. This was also the choice in \citep{Fa12}; 
and this explains why right hand side of \pref{old3.221} is quadratic (as opposed to cubic) 
in $s$ and $z$.   
With this definition of $Q^*_j$, the proof of the following Lemma is a  
computational verification.  
\begin{lemma} \lb{bois} 
$$
\LL^{(b)}_j(\F',Y')=0
$$
\end{lemma}
Finally, we have to deal with $\LL^{(c)}_j(\F',Y')$; namely we have to show
how $E_j, t_j$ and $Q_j, Q^*_j$  generate $E_{j+1}, t_{j+1}$ so that 
\pref{cond1} holds and $\LL^{(c)}_j(\F',Y')$ is a contraction. 
We  split this term in two pieces. 
First define  intermediate effective parameters: 
\bd
\item Intermediate effective couplings  
$s^*_{j}$ and $z^*_{j}$
\bal\lb{lk2}
&s^*_j:=s_j +\FF_j,
\notag\\
&z^*_j:=L^{2}e^{-\frac{\a^2}{2}\G_j(0)}\lft[z_j+\MM_j\rgt]
\eal
where $\FF_j$ and $\MM_j$ are functionals of the fields  
\bal
&\FF_j\=\FF_j(K_j)=\sum_{X\in \SS_j}^{X\ni0}
\frac{L^{-2j}}{|X|_{j}|X|}\sum_{x_0\in X\atop x_1,x_2\in X^*}
\EEE_j\lft[\frac{\dpr^2\hK_{0,j}}
{\dpr \f_{x_1}\dpr \f_{x_2}}(0,\z,X)\rgt]
\sum_{\m\in\hat u}(x_1-x_0)^\m(x_2-x_0)^\m,
\notag\\
&\MM_j\=\MM_j(K_j)=\frac{e^{\frac{\a^2}{2}\G_j(0)}}{2}\sum_{\s=\pm1}
\sum_{X\in \SS_j}^{X\ni0}
\frac{1}{|X|_{j}}
\EEE_j\lft[\hK_{0,j}(\s,\z,X)\rgt].
\eal
\item Intermediate effective free energy
\be\lb{lk2b0}
E^*_j=E_j+ L^{-2j}\lft[\EE_{1,j} + s_j \EE_{2,j}\rgt]
\ee
for
\bal 
&\EE_{1,j}\=\EE_{1,j}(K_j)=\sum_{X\in \SS_j}^{X\ni0}
\frac{1}{|X|_{j}}\EEE_j\lft[\hK_{0,j}(0,\z, X)\rgt],
\notag\\
&\EE_{2,j}=-{L^{2j}\over 2}\sum_{\m\in \hat u} (\dpr^{-\m}\dpr^\m\G_j)(0).
\eal
\item  Intermediate renormalization constants 
\bal\lb{lk2b}
Z^*_j&=L^2e^{-\h^2\frac{\a^2}{2}\G_j(0)}
\lft[\lft(1+\MM_{1,1,j}\rgt)Z_j
+\MM_{1,2,j}(K_j)\bar Z_j\rgt],
\notag\\
\bar Z^*_j&=L^2e^{-\bar \h^2\frac{\a^2}{2}\G_j(0)}
\lft[\MM_{2,1,j} Z_j
+\lft(1+\MM_{2,2,j}\rgt)\bar Z_j\rgt],
\eal
where  the functionals $\MM_{p,q,j}$ are
\bal\lb{dfm1}
&\MM_{1,1,j}\=\MM_{1,1,j}(K_j)=
\frac{e^{\h^2\frac{\a^2}{2}\G_j(0)}}{2}\sum_{\s=\pm1}\sum_{X\in \SS_j}^{X\ni0}
\EEE_j\lft[\hK_{1,j}(0,\z,X, 0,\s)\rgt],
\notag\\
&\MM_{1,2,j}\=\MM_{1,2,j}(K_j)=
\frac{e^{\h^2\frac{\a^2}{2}\G_j(0)}}{2}\sum_{\s=\pm1}\sum_{X\in \SS_j}^{X\ni0}
\EEE_j\lft[\hK^\dagger_{1,j}(\s,\z,X, 0,\s)\rgt],
\notag\\
&\MM_{2,1,j}\=\MM_{2,1,j}(K_j)=
\frac{e^{\bar \h^2\frac{\a^2}{2}\G_j(0)}}{2}\sum_{\s=\pm1}\sum_{X\in \SS_j}^{X\ni0}
\EEE_j\lft[\hK_{1,j}(-\s,\z,X, 0,\s)\rgt],
\notag\\
&\MM_{2,2,j}\=\MM_{2,2,j}(K_j)=
\frac{e^{\bar \h^2\frac{\a^2}{2}\G_j(0)}}{2}\sum_{\s=\pm1}
\sum_{X\in \SS_j}^{X\ni0}
\EEE_j\lft[\hK^\dagger_{1,j}(0,\z,X, 0,\s)\rgt].
\eal
Note that, in the definition of $\MM_{m,n, j}$ we 
only retained the Tay$_0$ part of \pref{qbr1}; this is  
because of cancellations due to \pref{sym2} and \pref{sym3}. 
For example, \pref{sym2} for $\SS=R^2$ gives for any $m=0,1$
\be\lb{conseq1}
\sum_{X\in \SS_j}^{X\ni0} \sum_{y\in X^*}
\EEE_j\lft[\frac{\dpr \hK_{1,j}}{\dpr \z_y}(0,\z,X, 0,\s)\rgt]y^\m=0.
\ee
Besides, we used also the symmetry under charge conjugation \pref{sym5} and \pref{sym6}, 
which implies, for example, 
\be\lb{conseq2}
\EEE_j\lft[\hK_{1,j}(0,\z,X, 0,1)\rgt]=\frac12\sum_{\s=\pm1} 
\EEE_j\lft[\hK_{1,j}(0,\z,X, 0,\s)\rgt]
\ee
These points will be detailed in Section \ref{s6.4b}.
\ed
Next, split $\LL^{(c)}_j(\F',Y')$ into two terms 
$$
\LL^{(c)}_j(\F',Y')=\LL^{(c1)}_j(\F',Y')+ \LL^{(c2)}_j(\F',Y')
$$
for 
\bal\lb{irb}
\LL^{(c1)}_j(\F',Y'):=\sum_{B\in \BB_{j}}^{\bar B=Y'}
\Bigg\{
&(E^*_{j}-E_j)|B|+V_{j+1}(t^*_j,\f',B)
-\EEE_j\lft[V_{j}(\f,B)\rgt]
\notag\\
&-\sum_{X\in \SS_j}^{X\supset B}
\lft[Q_{j}(\f',B,X)-Q_{2,j}(\f',B,X)\rgt]\Bigg\}
\eal
By construction, $\LL^{(c1)}_j(\F',Y')$ is made of a part that is $J$-independent, 
which we call $\LL^{(c1)}_{0,j}(\F',Y')$, and a part that is linear in $J$, 
which we call $\LL^{(c1)}_{1,j}(\F',Y')$ and which can be further decomposed
\bal
\LL^{(c1)}_{1,j}(\F',Y')
&=L^{-2(j+1)} Z_{j+1} \sum_{x,\s} J_{x,\s} 
\LL^{(c1)}_{1,j}(\f',Y',x,\s)\notag\\
&+L^{-2(j+1)} \bar Z_{j+1} \sum_{x,\s} J_{x,\s} 
\LL^{(c1)\dagger}_{1,j}(\f',Y',x,\s)\;.
\eal
\begin{lemma}\lb{bourgain} 
For large enough $L$, there exist $\r(L,A)$ and $\r(L,A,\h)$  such that 
\be\lb{chaniotis}
\|\LL^{(c1)}_{0,j}\|_{h, T_{j+1}}\le  \r(L,A)\|K_{0,j}\|_{h, T_j},
\ee
\bal\lb{chaniotis2}
&\|\LL^{(c1)}_{1,j}\|_{1,h,T_{j+1}}\le \r(L,A,\h) \|K_{1,j}\|_{1,h,T_j},
\notag\\
&\|\LL^{(c1)\dagger}_{1,j}\|_{1,h,T_{j+1}}\le \r(L,A,\h) \|K^\dagger_{1,j}\|_{1,h,T_j}.
\eal 
Besides,  $\r(L,A)$ and $\r(L,A,\h)$ are arbitrarily small for $L$ and  $A$ large enough. 
\end{lemma}
The proof is in Section \ref{s6.4b}. 
By subtraction, the other part of $\LL^{(c)}_j(\F',Y')$ is
\bal\lb{eW}
&\LL^{(c2)}_j(\F',Y')=\sum_{D\in \BB_{j+1}}^{D=Y'}
\Bigg\{
\lft(E_{j+1}-E^*_{j}\rgt)|D|
+V_{j+1}(t_{j+1}- t^*_j,\F',D)
\notag\\
&+
W_{j+1}(t_{j+1},\F',D)-\EEE_j\lft[W_j(\tilde t_{j},\F,D)\rgt]
\notag\\
&-\sum_{Y\in \SS_{j+1}}^{Y\supset D} 
Q^*_{j}(\F',D,Y)-\sum_{B\in \BB_j(D)}\sum_{X\in\SS_j}^{X\supset B}
Q_{2,j}(\F',B,X)
\Bigg\}.
\eal
We want to  choose $E_{j+1}$, $s_{j+1}$ and 
$z_{j+1}$ so that $\LL^{(c2)}_j(\F',Y')$ vanishes.
Because of the identity
\bal\lb{srp}
\sum_{Y\in \SS_{j+1}}^{Y\supset D}
Q^*_j(\F', D,Y)
&=\frac12\EEE^T_j\lft[V_{0,j}(\tilde t_j,\f,D);V_{0,j}(\tilde t_j,\f,D^*)\rgt]
\notag\\
&+\EEE^T_j\lft[V_{1,j}(\tilde t_j,\F,D);V_{0,j}(\tilde t_j,\f,D^*)\rgt]
\notag\\
& 
+\frac12\EEE^T_j\lft[V_{1,j}(\tilde t_j,\F,D);V_{1,j}(\tilde t_j,\F,D^*)\rgt], 
\eal 
and because of computations in Section \ref{sc2}, we finally set: 
\bd
\item Effective couplings $s_{j+1}$ and $z_{j+1}$, 
\bal\lb{lk3}
&s_{j+1}=s^*_j-a_jz^2_j
\notag\\
&z_{j+1}=z^*_j-L^2 e^{-\frac{\a^2}{2}\G_j(0)} b_js_j z_j
\eal
where, setting  $\G_{j,n}(x):=\sum_{m=n}^j\G_{m}(x)$ and 
$\G_j(0|x):= \G_j(0)-\G_j(x)$, 
the coefficients in \pref{lk3} are $a_0=0$, $b_0=0$, and, for  any $j\ge1$, 
\bal\lb{dfm3} 
&a_j:={\a^2\over 2} \sum_{y\in \ZZZ}|y|^2
\lft[
w_{b,j}(y)\lft(e^{-\a^2\G_j(0|y)}-1\rgt)
+
e^{-\a^2\G_j(0)}\lft(e^{\a^2\G_j(y)}-1\rgt)L^{-4j}\rgt],
\notag\\
&b_j:={\a^2 \over 2}\sum_{y\in \ZZZ\atop\m\in\hat u}
\lft[\lft(\dpr^\m\G_{j}\rgt)^2(y) + 
2\sum_{n=0}^{j-1}\lft(\dpr^\m\G_{n}\rgt)(y)\lft(\dpr^\m\G_{j}\rgt)(y)
e^{-{\a^2\over2}\G_{j-1,n}(0)} L^{2(j-n)}\rgt].
\eal
\item Effective free energy $E_{j+1}$
\bal
\lb{lk3b}
&E_{j+1}=E^*_{j}+L^{-2j}\lft[s^2_j\EE_{3,j}+
z^2_j\EE_{4,j}\rgt] 
\eal
where 
the coefficients in \pref{lk4} are $\EE_{3,0}=\EE_{4,0}=0$ and, for any $j\ge 1$, 
\bal\lb{dfm4}
&\EE_{3,j}:={L^{2j}\over 4}\sum_{y\in \ZZZ^2}\sum_{\m,\n\in\hat u}
\Big[(\dpr^{-\m}\dpr^\n\G_{j})(y) + 2(\dpr^{-\m}\dpr^\n\G_{j-1,1})(y)\Big]
(\dpr^{-\m}\dpr^\n\G_{j})(y),
\notag\\
&
\EE_{4,j}:=2L^{2j}\sum_{y}w_{0,b,j}(y)\lft[e^{-\a^2\G_j(0|y)}-1-{\a^2\over 2}
|y|^2\sum_{\m\in \hat u}(\dpr^{-\m}\dpr^\m \G_j)(0)\rgt]
\notag\\
&\qquad\qquad
+L^{-2j}\sum_{y}e^{-\a^2\G_j(0)}
\lft(e^{\a^2\G_j(y)}-1\rgt).
\eal
\item Fractional charge renormalization constants $Z_{j+1}$ and $\bar Z_{j+1}$
\bal\lb{lk4}
&Z_{j+1}=Z^*_{j}+L^2 e^{-\h^2\frac{\a^2}{2}\G_j(0)} 
\lft( -m_{1,1,j} s_j Z_j 
+ m_{1,2,j} z_j \bar Z_j\rgt),
\notag\\
&\bar Z_{j+1}=\bar Z^*_j+L^2 e^{-\bar\h^2\frac{\a^2}{2}\G_j(0)} 
\lft(-m_{2,2,j} s_j\bar Z_j+m_{2,1,j} z_j Z_j \rgt),
\eal
where the coefficients are, for any $j\ge 0$,  
\bal\lb{dfm4}
&m_{1,1, j}=
\frac{\a^2\h^2}{4}
\sum_{y\in \ZZZ^2\atop\n\in\hat u} 
\lft[
(\dpr^\n \G_j)^2(y)+2\sum_{n=0}^{j-1} 
  (\dpr^\n \G_n)(y)\lft[(\dpr^\n \G_j)(y)-(\dpr^\n \G_j)(0)\rgt]
L^{2(j-n)} e^{-\h^2\frac{\a^2}{2} \G_{j-1,n}(0)}\rgt],
\notag\\
&m_{2,2, j}=
\frac{\a^2\bar \h^2}{4}
\sum_{y\in \ZZZ^2\atop\n\in\hat u} 
\lft[(\dpr^\n \G_j)^2(y)+2\sum_{n=0}^{j-1}  (\dpr^\n \G_n)(y)
\lft[(\dpr^\n\G_j)(y)-(\dpr^\n \G_j)(0)\rgt]
L^{2(j-n)} e^{-\bar \h^2\frac{\a^2}{2} \G_{j-1,n}(0)}
 \rgt],
\notag\\
&m_{1,2, j}=
\sum_{y\in \ZZZ^2}
\lft[\bar w_{2,c,j}(y)\lft(e^{-\a^2\bar \h\G_j(y|0)}-1\rgt)
+L^{-2j}
e^{\bar \h\a^2\G_j(0)}\lft(e^{-\bar \h\a^2\G_j(y)}-1\rgt)
\rgt],
\notag\\
&m_{2,1, j}=
\sum_{y\in \ZZZ^2}
\lft[w_{2,c,j}(y)\lft(e^{\a^2\h\G_j(y|0)}-1\rgt)
+L^{-2j}
e^{-\h\a^2\G_j(0)}\lft(e^{\h\a^2\G_j(y)}-1\rgt)
\rgt].
\eal
\item 
The functions $w$'s in \pref{wj0} are all vanishing for $j=0,1$; while, for $j\ge 2$,
\bal\lb{lwn}
&w_{0,a,j}^{\m\n}(y)={1\over 2}\sum_{n=1}^{j-1}
(\dpr^{-\m}\dpr^\n \G_{n})(y),
\notag\\
&w_{0,b,j}(y)=
\frac12\sum_{n=1}^{j-1}
e^{-\a^2\G_{j-1,n+1}(0|y)}
e^{-\a^2\G_n(0)}\lft(e^{\a^2\G_n(y)}-1\rgt)L^{-4n},
\notag\\
&w_{0,c,j}(y)={1\over 2}\sum_{n=1}^{j-1}
e^{-\a^2[\G_{j-1,n+1}(0)+\G_{j-1, n+1}(y)]}
e^{-\a^2\G_n(0)}\lft(e^{-\a^2\G_n(y)}-1\rgt)L^{-4n},
\notag\\
&w^{\m}_{0,d,j}(y)={\a\over 2}\sum_{n=1}^{j-1}
e^{-{\a^2\over 2}\G_{j-1, n}(0)}
(\dpr^\m\G_{n})(y)
L^{-2n},
\notag\\
&w_{0,e,j}(y)={\a^2\over 4}\sum_{n=1}^{j-1}
e^{-{\a^2\over 2}\G_{j-1,n}(0)}
\sum_{\m\in \hat u} \lft[\lft(\dpr^\m\G_{j-1,n}\rgt)^2(y)-
\lft(\dpr^\m\G_{j-1,n+1}\rgt)^2(y)\rgt]
L^{-2n}.
\eal
\item 
The functions $w$'s in \pref{wj1} are 
\bal\lb{lw1n}
w_{1,b,j}(y)&=\sum_{n=0}^{j-1}L^{-2n} 
e^{-\frac{\a^2}{2}\G_{j-1,n}(0)}
e^{-\h\a^2\G_{j-1,n+1}(y)}\lft(e^{-\h\a^2\G_n(y)}-1\rgt),
\notag\\
\bar w_{1,b,j}(y)&=\sum_{n=0}^{j-1}
L^{-2n} 
e^{-\frac{\a^2}{2}\G_{j-1,n}(0)}
e^{\bar \h\a^2\G_{j-1,n+1}(y)}\lft(e^{\bar \h\a^2\G_n(y)}-1\rgt),
\notag\\
w_{1,c,j}(y)&=\sum_{n=0}^{j-1}
L^{-2n} 
e^{-\frac{\a^2}{2}\G_{j-1,n}(0)}
e^{\h\a^2\G_{j-1,n+1}(y)}\lft(e^{\h\a^2\G_n(y)}-1\rgt),
\notag\\
\bar w_{1,c,n}(y)&=\sum_{n=0}^{j-1}
L^{-2n} 
e^{-\frac{\a^2}{2}\G_{j-1,n}(0)}
e^{-\bar\h\a^2\G_{j-1,n+1}(y)}\lft(e^{-\bar \h\a^2\G_n(y)}-1\rgt),
\notag\\
w^{\n}_{1,d,j}(y)&=
i\a\h\sum_{n=0}^{j-1}
(\dpr^\n \G_n)(y),
\notag\\
\bar w^{\n}_{1,d,j}(y)&=
i\a\bar \h\sum_{n=0}^{j-1}
(\dpr^\n \G_n)(y).
\eal
\item 
the functions $w$'s in \pref{wj2} are 
\bal\lb{hofern}
w^{\e}_{2,a,j}(y)&=\frac12 \sum_{n=0}^{j-1}
Z_n^2L^{-4n}
e^{-\h^2(1+\e)\a^2\G_{j-1,n+1}(0)}
e^{-\h^2\a^2\e\G_{j-1,n+1}(y|0)}
\notag\\
&\qquad\times
e^{-\h^2\a^2\G_n(0)}\lft(e^{-\h^2\a^2\e \G_n(y)}-1\rgt),
\notag\\
\bar w^{\e}_{2,a,j}(y)&=\frac12 \sum_{n=0}^{j-1}
\bar Z_n^2L^{-4n}
e^{-\bar \h^2(1+\e)\a^2\G_{j-1,n+1}(0)}
e^{-\bar \h^2\a^2\e\G_{j-1,n+1}(y|0)}
\notag\\
&\qquad\times
e^{-\bar \h^2\a^2\G_n(0)}\lft(e^{-\bar \h^2\a^2\e \G_n(y)}-1\rgt),
\notag\\
w^{\e}_{2,b,j}(y)&=\frac12 \sum_{n=0}^{j-1}
Z_n\bar Z_n L^{-4n}
e^{-(\h+\e\bar\h)^2\frac{\a^2}{2}\G_{j-1,n+1}(0)}
e^{-\h\bar \h\a^2\e\G_{j-1,n+1}(y|0)}
\notag\\
&\qquad\times
e^{-(\h^2+\bar \h^2)\frac{\a^2}{2}\G_n(0)}\lft(e^{-\h\bar \h\a^2\e
  \G_n(y)}-1\rgt),
\notag\\
w^{\e}_{2,c,j}(y)&=
\sum_{k=0}^{j-1} L^{-4k} e^{-L^{-k} |y|}
\sum_{n=k}^{j-1} e^{-\frac{\a^2}2(1+\e)^2(\h-\frac12)^2 \G_{j-1,n+1}(0)} 2^{-(n-k)} 
\notag\\
&\times \lft\{Z_k^2 \frac12\sum_{\s=\pm1}\sum_{X\in \SS_n}
\EEE_j\lft[ \hat K_{2,n}^{(a,k)}\lft(-\s\frac{1+\e}2,\z, X,0, \s, y, \s\e\rgt)\rgt]
\rgt.
\notag\\
&\quad+\lft.
\bar Z^2_k \frac12\sum_{\s=\pm1}\sum_{X\in \SS_n}\EEE_j\lft[ 
\hat K_{2,n}^{(\bar a,k)}\lft(\s\frac{1+\e}2,\z, X,0, \s, y, \s\e\rgt)
\rgt]
\rgt.
\notag\\
&\hskip-2em+\lft.
Z_k\bar Z_k \frac12\sum_{\s=\pm1}\sum_{X\in \SS_n}
\EEE_j\lft[ 
\hat K_{2,n}^{(b,k)}\lft(-\s\frac{1-\e}2,\z, X, 0, \s, y,\s\e\rgt)
\rgt]\rgt\}.
\eal
Note that, because of the smallness condition on  $X$, $w^\e_{2,c,j}(y)=0$ 
for $|y|\ge 8L^{j-1}$.
\ed
\begin{lemma} \lb{bois2} 
$$
\LL^{(c2)}_j(\F',Y')=0.
$$
\end{lemma}
By \pref{p1}, $W_j(\f, B)$ depends on the fields $\f_x$ and $J_{x,\s}$ 
for $x$ in a
neighborhood of $B$ of diameter $L^j/2$, which is a subset of $B^*$. 
Finally, joining \pref{lk3} with \pref{lk2} we obtain
\pref{lk} and \pref{5.25def};  and condition \pref{cond1} is fulfilled.

\subsection{Proof of Lemma \ref{t3.1bb} }\lb{s6.1}
The formulas for $\{\MM_{m,n,j}:m,n=1,2\}$ are in \pref{dfm1}. The bounds 
\pref{scott2ab}  directly descend from \pref{eqdh1} and \pref{eqdh1bar}
at $\f'=0$. For example, for $\l=\frac12$ and a  $C\=C(\a,L)$,  
\bal\lb{mmj11}
&|\MM_{1,1,j}|
\le
\frac{e^{\h^2\frac{\a^2}{2}\G_j(0)}}{2}\sum_{\s=\pm1}\sum_{X\in \SS_j}^{X\ni0}
\|\EEE_j\lft[\hK_{1,j}(0,\z,X, 0,\s)\rgt]\|_{h,T_{j+1}(0,X)}
\notag\\
&\le
C\;\|K_{1,j}\|_{1,h,T_j}
\sum_{X\in \SS_j}^{X\ni0} \lft(\frac A2\rgt)^{-|X|_j} \le
CS^2\;k^*_s(A,\l) A^{-1} \|K_{1,j}\|_{1,h,T_j}.
\eal
The other $\MM_{p,q,j}$'s can be studied in a  similar way.  
\subsection{Proof of Lemma \ref{faculty}}\lb{s6.4}
In \citep{Fa12} we already proved  formula \pref{allen}, 
for $\r(L,A)= C (L^{-\th} + A^{-\th'})$, where $C>1$ and $\th,\th'>0$. 
We only need to derive \pref{arkani} and \pref{goddar}.  
Consider $\LL^{(a)}_{1,j} (\f',V,x,\s)$ and 
decompose
$$
\LL^{(a)}_{1,j} (\f',V,x,\s)=\sum_{n=1}^3\LL^{(n)}_{j} (\f',V,x,\s),
$$
where, with  Taylor expansions in $\d \f'$, 
\bal
\LL^{(1)}_j (\f',V,x,\s)
&:=\frac{Z_j}{Z_{j+1}}L^2
\sum_{Y\in\not\SS_j(V)\atop Y\ni x}^{\bar Y=V}
\EEE_j[K_{1,j}(\f,Y, x,\s)],
\lb{el1}
\\
\LL^{(2)}_j (\f',V,x,\s)
&:=\frac{Z_j}{Z_{j+1}}L^2
\sum_{Y\in \SS_j(V)\atop Y\ni x}^{\bar Y=V}
\sum_{q\in \ZZZ\atop
|q+\h \s|>1}
\EEE_j\Big[\hK_{1,j}(q,\f,Y, x,\s)\Big],
\lb{el2}
\\
\LL^{(3)}_j (\f',V,x,\s)
&:=\frac{Z_j}{Z_{j+1}}L^2\sum_{Y\in \SS_j(V)\atop Y\ni x}^{\bar Y=V}
\sum_{q=0, -\s}
\Rem_{1,\d\f'}\;\EEE_j[\hK_{1,j}(q, \f,Y, x,\s)].
\lb{el3}
\eal
Let us consider each of the terms, 
 assuming $|Z_j/Z_{j+1}|\le 1$.
\bd
\item {\it Norm of $\LL^{(1)}$.} Use \pref{6.5},  as well as 
a simple extension of \pref{6.4} to activities with a pinning point,  
to find
\bal
&\|\LL^{(1)}_{j}(\f',V,x,\s)\|_{h,T_{j+1}(\f',V)}
\le L^2 \sum_{Y\in\not\SS_j\atop Y\ni x}^{\bar Y=V} 
\|\EEE_j\lft[K_{1,j}(\f,Y, x,\s)\rgt]\|_{h,T_{j+1}(\f',Y)}
\notag\\
&\qquad\qquad\qquad\qquad
\le  G_{j+1}(\f',V)\|K_{1,j}\|_{1,h,T_j}
L^2\sum_{Y\in\not\SS_j\atop Y\ni x}^{\bar Y=V} A^{-|Y|_j}2^{|Y|_j}
\notag\\
&\qquad\qquad\qquad\qquad
\le  G_{j+1}(\f',V)A^{-|V|_{j+1}}\;\|K_{1,j}\|_{1,h,T_j}\; 
L^2 k_l(A,1/2).
\eal
By \pref{6.90}, we find  
$$
\|\LL^{(1)}_{j}\|_{1,h,T_{j+1}}\le \d_1(L,A)\|K_{1,j}\|_{1,h,T_j}
$$
with $\d_1(A,L)=L^2 A^{-\th}$: this quantity can be made as small as needed since
$A$ is chosen after $L$.
\item {\it Norm of $\LL^{(2)}$.} Use \pref{6.5} and  \pref{eqdh1}
to find, for $C\=C(\a)$ and if $\a^2\ge 8\p$, 
\bal
&\|\LL^{(2)}_j(\f',V, x,\s)\|_{h,T_{j+1}(\f',V)}
\le L^2
\sum_{Y\in \SS_j(V)\atop Y\ni x}^{\bar Y=V}
\sum_{q\in \ZZZ\atop
|q+\h \s|>1}
\|\EEE_{j}\Big[\hK_{1,j}(q,\f,Y, x,\s)\Big]\|_{h,T_{j+1}(\f',Y)}
\notag\\
&\qquad
\le G_{j+1}(\f',V)A^{-|V|_{j+1}}
\;\|K_{1,j}\|_{h,T_j} 
\; k^*_s(A,1/2)L^2\sum_{q\in \ZZZ\atop
|q+\h \s|>1}
L^{-4|q+\s\h|+2} C^{2|q+\s\h|};
\eal
by \pref{6.90} we obtain  
$$
\|\LL^{(2)}_{j}\|_{1,h,T_{j+1}}\le \d_2(L,A)\|K_{1,j}\|_{1,h,T_j}
$$
with $\d_2(A,L)= C L^{-4\min\{|\h|,|\bar\h|\}}$: this quantity can be made small 
by taking $L$ large enough, given the choice of $\h$. 
\item {\it Norm of $\LL^{(3)}$.} By \pref{6.5} and \pref{6.62b}
\bal
&\|\LL^{(3)}_j (\f',V,x,\s)\|_{h,T_{j+1}(\f',V)}
\le L^2 \sum_{Y\in \SS_j(V)\atop Y\ni x}^{\bar Y=V}\sum_{q=0,-\s}
\|\Rem_{1,\d\f'}\EEE_j[\hK_{1,j}(q, \f,Y, x,\s)]\|_{h,T_{j+1}(\f',Y)}
\notag\\
&\qquad\qquad
\le  G_{j+1}(\f',V)A^{-|V|_{j+1}}\;\|K_{1,j}\|_{h,T_j}\; 
k^*_s(A,1/2)L^2\sum_{q=0,-\s} \r_1(q+\h\s,\a);
\eal
by \pref{6.90} we obtain 
$$
\|\LL^{(3)}_{j}\|_{1,h,T_{j+1}}\le \d_3(L,A)\|K_{1,j}\|_{1,h,T_j}
$$
with
$\d_3(L,A)= \frac{C}{\kappa_L}L^{-\min\{\h^2,\bar\h^2\}}$.
\ed
This proves the former of \pref{arkani} for $\r(L,A)=\d_1(L,A) +\d_2(L,A) + \d_3(L,A)$. 
The latter  of \pref{arkani} has a similar proof. 
Let us now consider \pref{goddar}. For $\LL^{(a,k)}_{2,j}$ we have 
$$
\LL^{(a,k)}_{2,j} (\f',V,x_1,\s_1,x_2,\s_2)=\sum_{p=1}^3\LL^{(a,p,k)}_{2,j} (\f',V,x_1,\s_1,x_2,\s_2)
$$
where
\bal
\LL^{(a,1,k)}_{2,j} (\f',V,x_1,\s_1,x_2,\s_2)
&:=
\sum_{Y\in\not\SS_j(V)\atop Y\ni x_1}^{\bar Y=V}
\EEE_j\lft[K^{(a,k)}_{2,j}(\f,Y,x_1,\s_1,x_2,\s_2)\rgt],
\lb{el1b}
\\
\LL^{(a,2,k)}_{2,j} (\f',V,x_1,\s_1,x_2,\s_2)
&:=
\sum_{Y\in \SS_j(V)\atop Y\ni x_1}^{\bar Y=V}
\sum_{q\in \ZZZ\atop
q\neq -\frac12(\s_1+\s_2)}
\EEE_j\lft[\hK^{(a,k)}_{2,j}(q,\f,Y,x_1,\s_1,x_2,\s_2)\rgt],
\lb{el2b}
\\
\LL^{(a,3,k)}_{2,j} (\f',V,x_1,\s_1,x_2,\s_2)
&:=\sum_{Y\in \SS_j(V)\atop Y\ni x_1}^{\bar Y=V}
\Rem_{0,\d\f'}\;\EEE_j
\lft[\hK^{(a,k)}_{2,j}(-\frac{\s_1+\s_2}{2}, \f,Y,x_1,\s_1,x_2,\s_2)\rgt].
\lb{el3b}
\eal
With estimates similar to the ones used in the previous discussions, 
we have a bound  
\be\lb{goddar3}
\|\LL^{(a, p,k)}_{2,j}\|_{2,h,T_{j+1}}\le \d_{a,p,k}(L,A,\h) \|K^{(a,k)}_{2,j}\|_{2,h,T_{j}}
\ee
where possible choices of the prefactors  $\d_{a,p,k}(L,A,\h)$'s are: 
using the third of  \pref{6.90}, $\d_{a,1,k}(L,A,\h)=A^{-\th}$ for a $\th>0$; 
using the second of \pref{6.90} and
\pref{lili},  $\d_{a,2,k}(L,A,\h)= CL^{-d(2\h)}$ for an $\h$ independent $C$ and
for $d(2\h)$ defined in Theorem \ref{dh}; using  the second of \pref{6.90} and
\pref{lili2},  $\d_{a,3,k}(L,A,\h)= C\lft(\sqrt{\kappa_L} L\rgt)^{-1}$. 
This proves \pref{goddar} for $\d=a$ and $\r(L,A,\h)= \sum_{p=1}^3\d_{a,p,k}(L,A,\h)$. 
The proof of \pref{goddar} for $\d=\bar a, b$ is similar. 

\subsection{Proof of Lemma \ref{bourgain}}\lb{s6.4b}
\begin{lemma}\lb{sc1} For $V_j$ given in \pref{vv},
\bal\lb{9:01}
&\EEE_j\lft[V_{j}(t_j,\F,B)\rgt]= 
(\tilde E_j- E_j)+
V_{j+1}(\tilde t_j,\F',B), 
\eal
where
\bal
\tilde E_j:=E_j-\frac{s_j}{2}|B|& \sum_{\m\in\hat u}(\dpr^{-\m}\dpr^\m\G_j)(0)
\eal
and $\tilde t_j$ is defined in \pref{5.25}.
\end{lemma}
\bpr 
From standard results on the correlations of Gaussian measures, 
\bal\lb{pt1}
&\EEE_j\lft[V_{0,j}(t_j,\F,B)\rgt]
= 
\frac{s_j}{2}\sum_{x\in B\atop \m\in\hat u}(\dpr^\m \f')_x^2
-|B|
\frac{s_j}{2}\sum_{\m\in\hat u}(\dpr^{-\m} \dpr^\m\G_j)(0)
\notag\\
&\qquad\qquad\qquad\qquad
+z_j L^{-2j} e^{-\frac{\a^2}{2}\G_j(0)}
\sum_{x\in B\atop \s=\pm1} e^{i\a\s\f'_x} 
\\\lb{pt1b}
&\EEE_j\lft[V_{1,j}(t_j,\F,B)\rgt]
=Z_j L^{-2j} e^{-\frac{\a^2}{2}\h^2 \G_j(0)} \sum_{x\in B\atop \s=\pm}J_{\s,x}
e^{i\a\h\s\f'_x} 
\notag\\
&\qquad\qquad\qquad\qquad
+\bar Z_j L^{-2j} e^{-\frac{\a^2}{2}\bar \h^2 \G_j(0)}  \sum_{x\in B\atop \s=\pm}J_{\s,x}
e^{i\a\bar \h\s\f'_x} 
\eal
These identities  give \pref{9:01}.
\epr
Via this Lemma, be obtain the following formulas 
for $\LL^{(c1)}_{1,j}$ and $\LL^{(c1)\dagger}_{1,j}$: 
\bal
&\LL^{(c1)}_{1,j}(\f',V,x,\s)
\notag\\
&=
\frac{Z_j}{Z_{j+1}}L^2 
\sum_{B\in \BB_j(V)\atop B\ni x}^{\bar B=V}\lft[e^{-\h^2\frac{\a^2}{2}\G_j(0)}\MM_{1,1,j}
- \sum_{X\in \SS_j}^{X\supset B}\Tay_{1,\d\f'} 
\EEE_j\lft[\hK_{1,j}(0, \d\f'+\z, X, x,\s)\rgt]\rgt]e^{i\h\a\s\f'_x}
\notag\\
&+\frac{Z_j}{Z_{j+1}}L^2 
\sum_{B\in \BB_j(V)\atop B\ni x}^{\bar B=V}\lft[e^{-\bar\h^2\frac{\a^2}{2}\G_j(0)}\MM_{2,1,j} 
- \sum_{X\in \SS_j}^{X\supset B}\Tay_{1,\d\f'} 
\EEE_j\lft[\hK_{1,j}(-\s, \d\f'+\z, X, x,\s)\rgt] \rgt]e^{i\bar\h\a\s\f'_x}
\lb{el4}
\eal
and
\bal
&\LL^{(c1)\dagger}_{1,j}(\f',V,x,\s)
\notag\\
&=
\frac{\bar Z_j}{\bar Z_{j+1}}L^2 
\sum_{B\in \BB_j(V)\atop B\ni x}^{\bar B=V}\lft[e^{-\bar \h^2\frac{\a^2}{2}\G_j(0)}\MM_{2,2,j}
- \sum_{X\in \SS_j}^{X\supset B}\Tay_{1,\d\f'} 
\EEE_j\lft[\hK_{1,j}^\dagger(0, \d\f'+\z, X, x,\s)\rgt]\rgt]e^{i\bar \h\a\s\f'_x}
\notag\\
&+\frac{\bar Z_j}{\bar Z_{j+1}}L^2 
\sum_{B\in \BB_j(V)\atop B\ni x}^{\bar B=V}\lft[e^{-\h^2\frac{\a^2}{2}\G_j(0)}\MM_{1,2,j} 
- \sum_{X\in \SS_j}^{X\supset B}\Tay_{1,\d\f'} 
\EEE_j\lft[\hK^\dagger_{1,j}(\s, \d\f'+\z, X, x,\s)\rgt] \rgt]e^{i\h\a\s\f'_x}.
\lb{el4bis}
\eal
Let us consider the two terms in the square brackets in the first line of \pref{el4}:
by \pref{conseq1} and \pref{conseq2} they  are equal to 
\bal\lb{el5}
\sum_{X\in \SS_j}^{X\ni x}\sum_{y\in X^*}
\EEE_j\lft[\frac{\dpr\hK_{1,j}}{\dpr \z_y}(0,\z,X,x,\s)\rgt]
\lft(\sum_{\m\in\hat u} (y-x)^\m \dpr^\m \f'_x-(\f'_y-\f'_x) \rgt).
\eal
Note that \pref{el5} depends on $\f'$ only via the factor 
 $u_x(y,\f'):=\sum_{\m\in\hat u} (y-x)^\m \dpr^\m \f'_x-(\f'_y-\f'_x)$ and 
that, with the notation of \pref{fdiff}, 
\be
D^n u_x(y,\f')\cdot (f_1, \ldots, f_n)=
\begin{cases}
u_x(y,f_1) \quad &\text{ if $n=1$}\\
0 \quad &\text{ if $n\ge 2$}\\
\end{cases}
.\ee
As $X\in S_j$, we have $X^*\subset V^*$ and $|y-x|\le CL^j$, so that
\be\lb{sofia}
\|u_x(\cdot, \f)\|_{\CC^2_j(X)}\le CL^{-2}\|\nabla_{j+1}^2 \f\|_{L^\io(V^*)}.
\ee
Finally  the $\|\cdot\|_{h,T_{j+1}(\f',X)}$ norm of \pref{el5} is bounded by 
\bal
&\sum_{X\in \SS_j}^{X\ni x}\|\EEE_j\lft[\hK_{1,j}(0,\z,X,x,\s)\rgt]\|_{h,T_{j+1}(0,X)}
\notag\\
&\qquad\times
\lft(\| u_x(\cdot,\f)\|_{\CC^2_j(X)}+\sup_{\|f\|_{\CC^2_{j+1}}=1}\| u_x(\cdot,f)\|_{\CC^2_j(X)}\rgt)
\notag\\
&\le C L^{-2(1+\h^2)} \|K_{1,j}\|_{1,h,T_j}
\lft(1+ \|\nabla_{j+1}^2 \f'\|_{L^\io(X^*)}\rgt)
\sum_{X\in \SS_j}^{X\ni x}(A/2)^{-|X|_j}
\notag\\
&\le C' \kappa_L^{-1 }L^{-2(1+\h^2)} \|K_{1,j}\|_{1,h,T_j}
G^{\rm str}_{j+1}(\f',V)
(A/2)^{-1}
\eal
where, to obtain the second line we used \pref{eqdh} at $\f'=0$ and \pref{sofia}.
The other lines in \pref{el4} and \pref{el4bis} can be dealt with exactly the same procedure. 
Finally, as $|V|_{j+1}=1$, $Z_j/Z_{j+1}\le 1$ and because of \pref{6.100b}, we obtain 
$$
\|\LL^{(c1)}_{1,j}(\f',V,x,\s)\|_{h,T_{j+1}(\f',V)}
\le C \kappa_L^{-1 }L^{-2\h^2} \|K_{1,j}\|_{1,h,T_j}
G_{j+1}(\f',V)
A^{-|V|_{j+1}}
$$
which proves the first of 
\pref{chaniotis2} for $\r(L,A,\h)=C \kappa_L^{-1 }L^{-2\min\{\h^2, \bar \h^2\}}$.
\subsection{Proof of Lemma \ref{bois2}}\lb{sc2}
This proof is a detailed calculation of the second order part of the 
RG map.
\begin{lemma} \lb{11:20}
If  the choice of the 
$w$'s functions is the one in \pref{lwn}, \pref{lw1n} and \pref{hofern},
and the choice for  $E_{j+1}$ and $t_{j+1}$, $\tilde t_j$, $t^*_j$ is the one 
in Section \ref{sc20},  then, 
for any $D\in \BB_{j+1}$ and $B\in \BB_j(D)$, 
\bal\lb{3sc2}
&{1\over 2}\EEE^T_j\lft[V_{1,j}(\tilde t_j,\F,B);V_{1,j}(\tilde t_j,\F,D^*)\rgt]
+\EEE^T_j\lft[V_{1,j}(\tilde t_j,\F,B);V_{0,j}(\tilde t_j,\f,D^*)\rgt] 
\notag\\
&+{1\over 2}\EEE^T_j\lft[V_{0,j}(\tilde t_j,\f,B);V_{0,j}(\tilde t_j,\f,D^*)\rgt]
+\sum_{X\in \SS_j}^{X\supset B}Q_{2,j}(\F',B,X)
\notag\\
&=W_{j+1}(\F',B)-\EEE_j\lft[W_{j}(\tilde t_j,\F,B)\rgt] 
+(E_{j+1}-E^*_j)|B|+ V_{j+1}(t_{j+1}-t^*_j,\F',B). 
\eal
\end{lemma}
\bpr
An explicit computation of Gaussian correlations (for $\a$ and $\a'$
any real parameter) yields:
\bal\lb{gaussian}
&\EEE^T_j\big[(\dpr^\m\z_x)^2;(\dpr^\n\z_{x+y})^2\big]
=2(\dpr^{-\m}\dpr^\n \G_j)(y)^2,
\notag\\
&\EEE^T_j\big[(\dpr^\m\z_x);(\dpr^\n\z_{x+y})\big]
=-(\dpr^{-\m}\dpr^\n \G_j)(y),
\notag\\
&\EEE^T_j\big[e^{i\a\z_x};(\dpr^\m\z_{x+y})^2\big]
=-\a^2e^{-\frac{\a^2}{2}\G_j(0)}(\dpr^\m \G_j)(y)^2,
\notag\\
&\EEE^T_j\big[e^{i\a\z_{x+y}};(\dpr^\m\z_{x})^2\big]
=-\a^2e^{-\frac{\a^2}{2}\G_j(0)}(\dpr^{-\m} \G_j)(y)^2,
\notag\\
&
\EEE^T_j\big[e^{i\a\z_{x}};(\dpr^\m\z_{x+y})\big]
=i\a e^{-\frac{\a^2}{2}\G_j(0)}(\dpr^\m \G_j)(y),
\notag\\
&
\EEE^T_j\big[e^{i\a\z_{x+y}};(\dpr^\m\z_{x})\big]
=-i\a e^{-\frac{\a^2}{2}\G_j(0)}(\dpr^{-\m} \G_j)(y),
\notag\\
&\EEE^T_j\big[e^{i\a\e\z_x};e^{i\a\e \e'\z_{x+y}}\big]= 
e^{-\frac{\a^2}2\G_j(0)}e^{-\frac{\a'^2}2\G_j(0)}
\lft( e^{-\a\a'\G_j(y)}-1\rgt).
\eal
Let $Y:=D^*\in \PP_{j+1}$. Let us separate 
the discussion of \pref{3sc2} into three parts.
\*
{\it\01. First part.}
Our goal is to determine the functions  $w_{0,\a, j}(y)$'s and the coefficients 
$t_{j+1}$  so to satisfy  Lemma \pref{bois2} for the part that doesn't depend 
on $J$:
\bal\lb{3sc2first}
&\frac12\EEE^T_j\lft[V_{0,j}(\tilde t_j,\F,B);V_{0,j}(\tilde t_j,\F, Y)\rgt]
=
W_{0,j+1}(\F',B)-\EEE_j\lft[W_{0,j}(\tilde t_j,\F,B)\rgt]
\notag\\
&\qquad+(E_{j+1}-E^*_j)|B|
+V_{0,j+1}(t_{j+1}-t^*_j,\F',B).
\eal
This identity was already verified in \citep{Fa12}. However, here we want to 
show how to re-derive it by means of an ansatz that can be generalized to the 
more sophisticated second an third parts.
We look for $w_{0,\a, j}(y)$, where $\a$ collects the various labels that appear in 
\pref{wj0}, into the form of sum of contribution gathered at each scale $n\le j-1$: 
$$
w_{0,\a, j}(y)=\sum_{n=1}^{j-1} R^{(j-1)}_{0,\a,n}(y).
$$
By use of \pref{gaussian},  
\bal\lb{tt}
&{1\over 2}\EEE^T_j\lft[V_{0,j}(\F,B);V_{0,j}(\f, Y)\rgt]=
s^2_j|B|\frac{1}{4}\sum_{\m,\n\in\hat u}
\sum_{y\in \ZZZ^2}(\dpr^{-\m}\dpr^\n \G_j)(y)^2
\notag\\
&-s^2_j\frac{1}{2}
\sum_{\m,\n\in\hat u}
\sum_{y\in \ZZZ^2}(\dpr^{-\m}\dpr^\n \G_j)(y)
\sum_{x\in B}(\dpr^\m\f'_x)(\dpr^\n\f'_{x+y})
\notag\\
&+z^2_j\frac{L^{-4j}}{2} 
\sum_{y\in \ZZZ^2}
e^{-{\a^2}\G_j(0)}\lft( e^{{\a^2}\G_j(y)}-1\rgt)
\sum_{x\in B\atop \s=\pm1}e^{i\a\s(\f'_x- \f'_{x+y})}
\notag\\
&+z^2_j\frac{ L^{-4j}}{2}
\sum_{y\in \ZZZ^2}
e^{-{\a^2}\G_j(0)}\lft( e^{-{\a^2}\G_j(y)}-1\rgt)
\sum_{x\in B\atop \s=\pm1}e^{i\a\s(\f'_x+\f'_{x+y})}
\notag\\
&+z_js_j \frac{i\a L^{-2j}}{2}
\sum_{y\in \ZZZ^2\atop\n\in\hat u}
e^{-\frac{\a^2}{2}\G_j(0)}(\dpr^\n \G_j)(y)
\sum_{x\in B\atop \s=\pm1}\s\lft[e^{i\a\s\f'_x} (\dpr^\n\f'_{x+y}) 
-e^{i\a\s\f'_{x+y}} (\dpr^{-\n}\f'_x)\rgt]
\notag\\
&-z_js_j \frac{\a^2L^{-2j}}{4}
\sum_{y\in \ZZZ^2\atop\n\in\hat u}
e^{-\frac{\a^2}{2}\G_j(0)}(\dpr^\n \G_j)^2(y)
\sum_{x\in B\atop \s=\pm1}\lft[e^{i\a\s\f'_x} 
+e^{i\a\s\f'_{x+y}}\rgt].
\eal
the above terms have to be re-arranged according to the 
following rule: 
each term is to be   
{\it either power-counting irrelevant} (see discussion after \pref{lili2}) {\it 
or local} (namely with all the fields $\f'$ dependent by a same point $x$), 
{\it or constant} (namely independent of $\f'$).  
Let us discuss each line of the right hand side member. 
The first line is a constant, which will be absorbed into $E_{j+1}$. 
The second line appears to be
marginal; in fact it is irrelevant, because one can plug in the identity
$$
(\dpr^\m\f'_x)(\dpr^\n\f'_{x+y})=(\dpr^\m\f'_x)
\lft[(\dpr^\n\f'_{x+y})-(\dpr^\n\f'_{x})\rgt]+
(\dpr^\m\f'_x)(\dpr^\n\f'_{x})
$$
and neglect the last term because of the  cancellation
\bal
&\sum_{\m,\n\in\hat u}\sum_{y\in \ZZZ^2}(\dpr^{-\m}\dpr^\n\G_j(y))=
-\sum_{i,j=0,1} \sin k_i \sin k_j\hat\G_j(k)\Big|_{k=0}=0.
\eal
The third line is relevant. To write it as the sum of an irrelevant term plus
a local one use the identity
\bal
e^{i\a\s(\f'_x- \f'_{x+y})}
&=
\lft[e^{i\a\s(\f'_x- \f'_{x+y})}-1+\frac{\a^2}4
|y|^2\sum_{\m\in\hat u} (\dpr^{\m}\f_x)^2\rgt]
\notag\\
&+1-\frac{\a^2}4 |y|^2
\sum_{\m\in\hat u} (\dpr^{\m}\f_x)^2.
\eal
Again, by symmetries, we have neglected a term, 
the linear order of 
the Taylor expansion in $y$: when plugged into \pref{tt}
this term cancels because it is odd in $\s$. 
Besides, the term proportional to  $|y|^2$ is chosen with a special 
form thanks to  the  partial  cancellation, 
for $m,n=0,1$, 
$$
\sum_{y\in \ZZZ^2} e^{-\a^2 \G_j(0)}\lft(e^{\a^2 \G_j(y)}-1\rgt)y_m y_n=
\frac{\d_{m,n}}{2}
\sum_{y\in \ZZZ^2} e^{-\a^2 \G_j(0)}\lft(e^{\a^2 \G_j(y)}-1\rgt)|y|^2,
$$
which makes irrelevant the sum of the terms in the square brackets.
The fourth and the fifth  lines of \pref{tt} are irrelevant. 
The only remaining relevant term is the sixth
line. To write it as the sum of an irrelevant term plus
a local one use the identity
$$
e^{i\a\s\f'_x} 
+e^{i\a\s\f'_{x+y}}=\lft[e^{i\a\s\f'_{x+y}}-e^{i\a\s\f'_x}\rgt]+2e^{i\a\s\f'_x}. 
$$
In conclusion, after all such operations, we obtain a new equivalent formula
for \pref{tt}:
\bal\lb{tt2}
&{1\over 2}\EEE^T_j\lft[V_{0,j}(\F,B);V_{0,j}(\f, Y)\rgt]
\notag\\
&=s^2_j |B|\frac14
\sum_{y\in \ZZZ^2\atop \m,\n\in\hat u}(\dpr^{-\m}\dpr^\n \G_j)^2(y)+z^2_j |B|L^{-4j}
\sum_{y\in \ZZZ^2}
e^{-{\a^2}\G_j(0)}
\lft( e^{{\a^2}\G_j(y)}-1\rgt)
\notag\\
&-s^2_j\frac12
\sum_{y\in \ZZZ^2\atop \m,\n\in\hat u}(\dpr^{-\m}\dpr^\n \G_j)(y)
\sum_{x\in B}(\dpr^\m\f'_x)\lft[(\dpr^\n\f'_{x+y})-(\dpr^\n\f'_{x})\rgt]
\notag\\
&+z^2_j\frac{L^{-4j}}2 
\sum_{y\in \ZZZ^2}
e^{-{\a^2}\G_j(0)}\lft( e^{{\a^2}\G_j(y)}-1\rgt)
\sum_{x\in B\atop \s=\pm1}
\lft[e^{i\a\s(\f'_x- \f'_{x+y})}-1+|y|^2\frac{\a^2}4
\sum_{\m\in\hat u} (\dpr^{\m}\f_x)^2\rgt]
\notag\\
&-z^2_j\frac{\a^2L^{-4j} }{2}
\sum_{y\in \ZZZ^2}
e^{-{\a^2}\G_j(0)}\lft( e^{{\a^2}\G_j(y)}-1\rgt)|y|^2
\frac12\sum_{x\in B\atop \m\in\hat u}
(\dpr^{\m}\f_x)^2
\notag\\
&+z^2_j\frac{L^{-4j}}{2}
\sum_{y\in \ZZZ^2}
e^{-{\a^2}\G_j(0)}\lft( e^{-{\a^2}\G_j(y)}-1\rgt)
\sum_{x\in B\atop \s=\pm1}e^{i\a\s(\f'_x+\f'_{x+y})}
\notag\\
&+ z_js_j\frac{\a L^{-2j}}{2}
\sum_{y\in \ZZZ^2\atop\n\in\hat u}
e^{-\frac{\a^2}{2}\G_j(0)}(\dpr^\n \G_j)(y)
\sum_{x\in B\atop \s=\pm1}i\s\lft[e^{i\a\s\f'_x} (\dpr^\n\f'_{x+y}) 
-e^{i\a\s\f'_{x+y}} (\dpr^{-\n}\f'_x)\rgt]
\notag\\
&-z_js_j\frac{ \a^2L^{-2j}}{4}
\sum_{y\in \ZZZ^2\atop\n\in\hat u}
e^{-\frac{\a^2}{2}\G_j(0)}(\dpr^\n \G_j)^2(y)
\sum_{x\in B\atop \s=\pm1}\lft[e^{i\a\s\f'_{x+y}}-e^{i\a\s\f'_x} \rgt]
\notag\\
&-z_js_j\frac{ \a^2L^{-2j}}{2}
\sum_{y\in \ZZZ^2\atop\n\in\hat u}
e^{-\frac{\a^2}{2}\G_j(0)}(\dpr^\n \G_j)^2(y)
\sum_{x\in B\atop \s=\pm1}e^{i\a\s\f'_x} .
\eal
The first part of our  ansatz is that the irrelevant terms 
which were  generated in the above integration
provide the $R^{(j)}_{0,\a,j}(y)$'s; more precisely,  plugging in \pref{wj0} 
$\tilde t_j$ instead of $t_j$
(namely  replacing  $s_j$ and $z_j$
with $s_{j+1}$ and $z_{j+1}$) and then comparing the irrelevant lines 
with \pref{wj0}, 
%\be\lb{tremaine2}
%s_j=s_{j+1} + O(z^2_j)
%\qquad 
%z_j=z_{j+1} + O(z^2_j)
%\ee
we set
\bal\lb{ansatz11}
&R^{(j)\m\n}_{0,a,j}(y)={1\over 2}
(\dpr^{-\m}\dpr^\n \G_{j})(y),
\notag\\
&R^{(j)}_{0,b,j}(y)={1\over 2}
e^{-\a^2\G_j(0)}\lft(e^{\a^2\G_j(y)}-1\rgt)L^{-4j},
\notag\\
&R^{(j)}_{0,c,j}(y)={1\over 2}
e^{-\a^2\G_j(0)}\lft(e^{-\a^2\G_j(y)}-1\rgt)L^{-4j},
\notag\\
&R^{(j)\m}_{0,d,j}(y)={\a\over 2}
e^{-{\a^2\over 2}\G_{j}(0)}
(\dpr^\m\G_{j})(y)
L^{-2n},
\notag\\
&R^{(j)}_{0,e,j}(y)={\a^2\over 4}
e^{-{\a^2\over 2}\G_{j}(0)}
\sum_{\m\in\hat u}(\dpr^\m\G_{j})^2(y)
L^{-2j}.
\eal
Next consider $\EEE_j\lft[W_{j,0}\rgt]$. As we have done in passing from \pref{tt} to 
\pref{tt2}, we write the result as a sum of terms each of which is either irrelevant or 
local, or constant. 
To do that, we need again some partial cancellations such as, 
for $m,n=0,1$,  
$$
\sum_{y\in \ZZZ^2}w_{0,b,j}(y)
\lft(e^{\a^2\G_j(y|0)}-1\rgt)y_m y_n=\d_{m,n}
\sum_{y\in \ZZZ^2}w_{0,b,j}(y)
\lft(e^{\a^2\G_j(y|0)}-1\rgt){|y|^2\over 2},
$$
that is a consequence of the invariance of $w_{0,b,j}(y)$ under the 
interchange of $y_0$ and $y_1$: this property will be apparent in the 
final choice of  $w_{0,b,j}(y)$ given in \pref{lwn}. 
The outcome the initial integration and subsequent re-arrangement is 
\bal\lb{tt3}
&\EEE_j\lft[W_{0,j}(t_j,\f',B)\rgt]
\notag\\
&=s_j^2|B|
\sum_{y\in \ZZZ^2\atop \m, \n\in\hat u} w_{0,a,j}^{\m\n}(y)
(\dpr^{-\m}\dpr^\n \G_j)(y)
+z_j^22 |B|\sum_{y\in \ZZZ^2}w_{0,b,j}(y)
\lft(e^{\a^2\G_j(y|0)}-1\rgt)
\notag\\
&-z_j^2 |B|\frac{\a^2}{2}
\sum_{y\in \ZZZ^2\atop \m\in\hat u}w_{0,b,j}(y)
|y|^2 
(\dpr^{-\m}\dpr^{\m}\G_j)(0)
\notag\\
&-s_j^2
\sum_{y\in \ZZZ^2\atop \m, \n\in\hat u} w_{0,a,j}^{\m\n}(y)
\sum_{x\in B}(\dpr^\m\f'_x)
\Big[(\dpr^\n\f'_{x+y})- (\dpr^\n\f'_x)\Big]
\notag\\
&+z_j^2\sum_{y\in \ZZZ^2}w_{0,b,j}(y)e^{\a^2\G_j(y|0)}
\sum_{x\in B\atop\s=\pm} 
\Bigg[ e^{i\s\a(\f'_{x}-\f'_{x+y})}
-1
+|y|^2\frac{\a^2} 4\sum_{\m\in\hat u}(\dpr^{\m}\f'_x)^2
\Bigg]
\notag\\
&-z_j^2 \a^2
\sum_{y\in \ZZZ^2}w_{0,b,j}(y)
\lft(e^{\a^2\G_j(y|0)}-1\rgt)|y|^2
\frac12
\sum_{x\in B\atop\m\in\hat u} (\dpr^{\m}\f'_x)^2
\notag\\
&+z_j^2\sum_{y\in \ZZZ^2} 
w_{0,c,j}(y)e^{-\a^2 (\G_j(0)+\G_j(y))}
\sum_{x\in B\atop\s=\pm}
e^{i\s\a(\f'_{x}+\f'_{x+y})}
\notag\\
&+z_js_j
\sum_{y\in \ZZZ^2\atop \m\in\hat u} 
w^\m_{0,d,j}(y)e^{-\frac{\a^2}{2} \G_j(0)}
\sum_{x\in B\atop\s=\pm}
i\s\lft[
e^{i\s\a\f'_{x}}(\dpr^\m\f'_{x+y})-
e^{i\s\a\f'_{x+y}}(\dpr^{-\m}\f'_x)\rgt]
\notag\\
&-z_j s_j
\sum_{y\in \ZZZ^2\atop \m\in\hat u} 
\lft[w_{0,e,j}(y)+\a w^\m_{0,d,j}(y)\dpr^\m \G_j(y)\rgt]
e^{-\frac{\a^2}{2} \G_j(0)}
\sum_{x\in B\atop\s=\pm}
\lft[e^{i\s\a\f'_{x+y}}-e^{i\s\a\f'_{x}}\rgt]
\notag\\
&-z_js_j 2\a
\sum_{y\in \ZZZ^2\atop \m\in\hat u} 
w^\m_{0,d,j}(y)e^{-\frac{\a^2}{2} \G_j(0)}\dpr^\m \G_j(y)
\sum_{x\in B\atop\s=\pm}e^{i\s\a\f'_{x}}.
\eal
The second part of our ansatz is that the factors produced in the above integration
transform $R^{(j-1)}_{0,\a,n}(y)$ into $R^{(j)}_{0,\a,n}(y)$; more precisely, 
plugging in \pref{tt3} $\tilde t_j$ instead of $t_j$
and then comparing the irrelevant terms with \pref{wj0}, 
\bal\lb{ansatz21}
&R_{0,a,n}^{(j)\m\n}(y)=R_{0,a,n}^{(j-1)\m\n}(y),
\notag\\
&R^{(j)}_{0,b,n}(y)=
R^{(j-1)}_{0,b,n}(y)e^{\a^2\G_{j}(y|0)},
\notag\\
&R^{(j)}_{0,c,n}(y)=
R^{(j-1)}_{0,c,n}(y)e^{-\a^2[\G_{j}(0)+\G_{j}(y)]},
\notag\\
&R^{(j)\m}_{0,d,n}(y)=R^{(j-1)\m}_{0,d,n}(y)
e^{-{\a^2\over 2}\G_{j}(0)},
\notag\\
&R^{(j)}_{0,e,n}(y)=R^{(j-1)}_{0,e,n}(y)e^{-{\a^2\over 2}\G_{j}(0)}+R^{(j-1)\m}_{0,d,n}(y)
e^{-{\a^2\over 2}\G_{j}(0)}\a \dpr^\m\G_{j}(y).
\eal
Finally, it is straightforward to solve \pref{ansatz21} with boundary data 
\pref{ansatz11}; the result is 
\bal\lb{lw}
&R_{0,a,n}^{(j-1)\m\n}(y)={1\over 2}
(\dpr^{-\m}\dpr^\n \G_{n})(y),
\notag\\
&R^{(j-1)}_{0,b,n}(y)=
\frac12
e^{-\a^2\G_{j-1,n+1}(0|y)}
e^{-\a^2\G_n(0)}\lft(e^{\a^2\G_n(y)}-1\rgt)L^{-4n},
\notag\\
&R^{(j-1)}_{0,c,n}(y)={1\over 2}
e^{-\a^2[\G_{j-1,n+1}(0)+\G_{j-1, n+1}(y)]}
e^{-\a^2\G_n(0)}\lft(e^{-\a^2\G_n(y)}-1\rgt)L^{-4n},
\notag\\
&R^{(j-1)\m}_{0,d,n}(y)={\a\over 2}
e^{-{\a^2\over 2}\G_{j-1, n}(0)}
(\dpr^\m\G_{n})(y)
L^{-2n},
\notag\\
&R^{(j-1)}_{0,e,n}(y)={\a^2\over 4}
e^{-{\a^2\over 2}\G_{j-1,n}(0)}
\sum_\m \lft[\lft(\dpr^\m\G_{j-1,n}\rgt)^2(y)-
\lft(\dpr^\m\G_{j-1,n+1}\rgt)^2(y)\rgt]
L^{-2n}.
\eal
Besides, collecting the marginal and relevant terms from \pref{tt2} and \pref{tt3}
we obtain 
\bal\lb{dfm2} 
&a_j:=\a^2\sum_{y\in \ZZZ}|y|^2
\lft[\sum_{n=0}^{j}R^{(j)}_{0,b,n}(y)-\sum_{n=0}^{j-1}R^{(j-1)}_{0,b,n}(y)\rgt],
\notag\\
&b_j:=\sum_{y\in \ZZZ\atop\m\in \hat e}
\lft[2\a L^{2j}
w^\m_{0,d,j}(y)\dpr^\m \G_j(y)+\frac{\a^2}2\lft(\dpr^\m\G_{j}\rgt)^2(y)\rgt].
\eal
This proves that \pref{lwn} and \pref{dfm3} yield \pref{3sc2first}.
\vskip1em
{\it\02. Second term.}
This term  contains one factor of external field $J_{x,\s}$.
We look for $w_{1,\a,j}(y)$ into the form 
$$
w_{1,\a,j}(y)=\sum_{n=0}^{j-1} R^{(j-1)}_{1,\a,n}(y)
$$
where $R^{(j-1)}_{1,\a,n}(y)$ will be determined by means of an ansatz to  obtain
\bal\lb{3sc2second}
\EEE^T_j\lft[V_{1,j}(\tilde t_j,\F,B);V_{0,j}(\tilde t_j,\F, Y)\rgt]
&=W_{1,j+1}(\F',B)-\EEE_j\lft[W_{1,j}(\tilde t_j, \F,B)\rgt]
\notag\\
&+V_{1,j+1}(t_{j+1}-t^*_j,\F',B).  
\eal
We find
\bal\lb{taylor}
&\EEE^T_j\lft[V_{1,j}(\F,B);V_{0,j}(\f, Y)\rgt]
\notag\\
&=Z_jz_j L^{-4j}
\sum_{y\in \ZZZ^2}
e^{-(1+\h^2)\frac{\a^2}{2}\G_j(0)}\lft(e^{-\a^2\h\G_j(y)}-1\rgt)
\sum_{x\in B\atop \s=\pm1}
J_{x,\s} e^{i\a\s(\h\f'_x+\f'_{x+y})}
\notag\\
&+\bar Z_jz_j L^{-4j}
\sum_{y\in \ZZZ^2}
e^{-(1+\bar\h^2)\frac{\a^2}{2}\G_j(0)}\lft(e^{\a^2\bar\h\G_j(y)}-1\rgt)
\sum_{x\in B\atop \s=\pm1}
J_{x,\s} e^{i\a\s(\bar \h\f'_x- \f'_{x+y})}
\notag\\
&+Z_jz_j L^{-4j}
\sum_{y\in \ZZZ^2}
e^{-(1+\h^2)\frac{\a^2}{2}\G_j(0)}\lft(e^{\a^2\h\G_j(y)}-1\rgt)
\sum_{x\in B\atop \s=\pm1}
J_{x,\s} e^{i\a\s(\h\f'_x-\f'_{x+y})}
\notag\\
&+\bar Z_jz_j L^{-4j}
\sum_{y\in \ZZZ^2}
e^{-(1+\bar\h^2)\frac{\a^2}{2}\G_j(0)}\lft(e^{-\a^2\bar \h\G_j(y)}-1\rgt)
\sum_{x\in B\atop \s=\pm1}
J_{x,\s} e^{i\a\s(\bar \h\f'_x+\f'_{x+y})} 
\notag\\
&
+Z_j s_j  i\a \h L^{-2j}
\sum_{y\in \ZZZ^2\atop\n\in\hat u}
 e^{-\h^2\frac{\a^2}{2}\G_j(0)}(\dpr^\n \G_j)(y)
\sum_{x\in B\atop \s=\pm1}J_{x,\s} \s e^{i\h\a\s\f'_x} 
(\dpr^\n\f'_{x+y}) 
\notag\\
&
+\bar Z_j s_j  i\a \bar \h  L^{-2j}
\sum_{y\in \ZZZ^2\atop \n\in\hat u} 
e^{-\bar \h^2\frac{\a^2}{2}\G_j(0)}(\dpr^\n \G_j)(y)
\sum_{x\in B\atop \s=\pm1}J_{x,\s} \s e^{i\bar\h\a\s\f'_x} 
(\dpr^\n\f'_{x+y})
\notag\\
&-Z_js_j \frac{\a^2\h^2L^{-2j}}{2}
 \sum_{y\in \ZZZ^2\atop\n\in\hat u} 
e^{-\h^2\frac{\a^2}{2}\G_j(0)}(\dpr^\n \G_j)^2(y)
\sum_{x\in B\atop \s=\pm1}J_{x,\s} e^{i\h\a\s\f'_x} 
\notag\\
&-\bar Z_js_j
\frac{\a^2\bar \h^2L^{-2j}}{2}
\sum_{y\in \ZZZ^2\atop\n\in\hat u} e^{-\bar \h^2\frac{\a^2}{2}\G_j(0)}(\dpr^\n \G_j)^2(y)
\sum_{x\in B\atop \s=\pm1}J_{x,\s} e^{i\bar \h\a\s\f'_x}.
\eal
We want to reorganize the summation \pref{taylor} 
so that every term is  either irrelevant or local. 
The first two lines are irrelevant, because the absolute value
of their total charge is $|\h+1|>1$ or $|\bar\h-1|>1 $. 
The third  and fourth lines are relevant; to write it as sum of an irrelevant term and 
a local one we extract the Taylor expansion in $y^\m$ up 
to the first order: for example,  for the third
line   this means that 
we plug in the identity
\bal 
e^{i\a\s(\h\f'_x-\f'_{x+y})}
&=e^{i\a\s \bar \h\f'_x}
e^{i\a\s(\f'_x-\f'_{x+y})}
\notag\\
&=e^{i\a\s \bar \h\f'_x}
\lft[e^{i\a\s(\f'_x-\f'_{x+y})}-1-i\a\s\sum_{\m\in\hat u} y^\m(\dpr^\m\f'_x)\rgt]
\notag\\
&+e^{i\a\s \bar\h\f'_x}
+i\a\s e^{i\a\s \bar \h\f'_x}\sum_{\m\in\hat u} y^\m(\dpr^\m\f'_x).
\eal
However, since $\sum_{y\in \ZZZ^2}\lft(e^{\a^2\h\G_j(y)}-1\rgt)y^\m=0$, 
the last term (once
replaced into the third line) cancels.
The fifth and sixth lines are apparently relevant; 
in fact, they are irrelevant as one can see 
by plugging in the identity 
$$
(\dpr^\n\f'_{x+y})=\lft[(\dpr^\n\f'_{x+y})-(\dpr^\n\f'_{x})\rgt]+(\dpr^\n\f'_{x})
$$
and observing that 
$\sum_{y\in \RRR^2} (\dpr^\n \G_j)(y)=0$ so that, the last term, which
is $y$-independent, give vanishing contribution.
Finally, the seventh and eighth lines are relevant; however, they are 
already local. 
In conclusion, an equivalent formulation of \pref{taylor} is 
\bal\lb{taylor2}
&\EEE^T_j\lft[V_{1,j}(\F,B);V_{0,j}(\f, Y)\rgt]
\notag\\
&=Z_jz_j L^{-4j}
\sum_{y\in \ZZZ^2}
e^{-(1+\h^2)\frac{\a^2}{2}\G_j(0)}\lft(e^{-\a^2\h\G_j(y)}-1\rgt)
\sum_{x\in B\atop \s=\pm1}
J_{x,\s} e^{i\a\s(\h\f'_x+\f'_{x+y})}
\notag\\
&+\bar Z_jz_j L^{-4j}
\sum_{y\in \ZZZ^2}
e^{-(1+\bar\h^2)\frac{\a^2}{2}\G_j(0)}\lft(e^{\a^2\bar\h\G_j(y)}-1\rgt)
\sum_{x\in B\atop \s=\pm1}
J_{x,\s} e^{i\a\s(\bar \h\f'_x- \f'_{x+y})}
\notag\\
&+Z_jz_j L^{-4j}
\sum_{y\in \ZZZ^2}
e^{-(1+\h^2)\frac{\a^2}{2}\G_j(0)}\lft(e^{\a^2\h\G_j(y)}-1\rgt)
\sum_{x\in B\atop \s=\pm1}
J_{x,\s} e^{i\a\s \bar \h\f'_x}
\notag\\
&\qquad\times \lft[e^{i\a\s(\f'_x-\f'_{x+y})}-1-i\a\s\sum_\m y^\m\dpr^\m\f'_x\rgt]
\notag\\
&+Z_jz_j L^{-4j}
\sum_{y\in \ZZZ^2}
e^{-(1+\h^2)\frac{\a^2}{2}\G_j(0)}\lft(e^{\a^2\h\G_j(y)}-1\rgt)
\sum_{x\in B\atop \s=\pm1}
J_{x,\s} e^{i\a\s \bar \h\f'_x}
\notag\\
&+\bar Z_jz_j L^{-4j}
\sum_{y\in \ZZZ^2}
e^{-(1+\bar\h^2)\frac{\a^2}{2}\G_j(0)}\lft(e^{-\a^2\bar \h\G_j(y)}-1\rgt)
\sum_{x\in B\atop \s=\pm1}
J_{x,\s} e^{i\a\s \h\f'_x}
\notag\\
&\qquad\times\lft[e^{-i\a\s(\f'_x-\f'_{x+y})}-1+i\a\s\sum_\m y^\m\dpr^\m\f'_x\rgt] 
\notag\\
&+\bar Z_jz_j L^{-4j}
\sum_{y\in \ZZZ^2}
e^{-(1+\bar\h^2)\frac{\a^2}{2}\G_j(0)}\lft(e^{-\a^2\bar \h\G_j(y)}-1\rgt)
\sum_{x\in B\atop \s=\pm1}
J_{x,\s} e^{i\a\s \h\f'_x}
\notag\\
&
+Z_j s_j  i\h \a L^{-2j}
\sum_{y\in \ZZZ^2\atop\n\in\hat u}
 e^{-\h^2\frac{\a^2}{2}\G_j(0)}(\dpr^\n \G_j)(y)
\sum_{x\in B\atop \s=\pm1}J_{x,\s} e^{i\h\a\s\f'_x}
\s \lft[(\dpr^\n\f'_{x+y}) -(\dpr^\n\f'_{x})\rgt]
\notag\\
&
+\bar Z_j s_j  i \bar \h  \a L^{-2j}
\sum_{y\in \ZZZ^2\atop\n\in\hat u} 
e^{-\bar \h^2\frac{\a^2}{2}\G_j(0)}(\dpr^\n \G_j)(y)
 \sum_{x\in B\atop \s=\pm1} J_{x,\s} e^{i\bar\h\a\s\f'_x} 
\s\lft[(\dpr^\n\f'_{x+y})-(\dpr^\n\f'_{x})\rgt]
\notag\\
&-Z_js_j \frac{\a^2\h^2L^{-2j}}{2}
\sum_{y\in \ZZZ^2\atop\n\in\hat u} 
e^{-\h^2\frac{\a^2}{2}\G_j(0)}(\dpr^\n \G_j)^2(y)
\sum_{x\in B\atop \s=\pm1}J_{x,\s} e^{i\h\a\s\f'_x} 
\notag\\
&-\bar Z_js_j
\frac{\a^2\bar \h^2L^{-2j}}{2}
\sum_{y\in \ZZZ^2\atop\n\in\hat u} 
e^{-\bar \h^2\frac{\a^2}{2}\G_j(0)}(\dpr^\n \G_j)^2(y)
\sum_{x\in B\atop \s=\pm1}J_{x,\s} e^{i\bar \h\a\s\f'_x}.
\eal
Replacing $t_j$ with $\tilde t_j$ and then 
comparing \pref{taylor2} with \pref{wj1}, 
%\be\lb{tremaine}
%\frac{Z_j}{Z_{j+1}}= L^{-2} e^{\h^2 \frac{\a^2}2 \G_j(0)}+O(z_j)\;
%\qquad 
%\frac{\bar Z_j}{\bar Z_{j+1}}= L^{-2} e^{{\bar \h}^2 \frac{\a^2}2 \G_j(0)}+O(z_j),
%\ee
%
we formulate the  ansatz
\bal\lb{ansatz12}
R^{(j)}_{1,b,j}(y)&=L^{-2j}
e^{-\frac{\a^2}{2}\G_{j}(0)}
\lft(e^{-\a^2\h\G_j(y)}-1\rgt),
\notag\\
\bar R^{(j)}_{1,b,j}(y)&=
 L^{-2j}
e^{-\frac{\a^2}{2}\G_{j}(0)}
\lft(e^{\a^2\bar \h\G_j(y)}-1\rgt),
\notag\\
R^{(j)}_{1,c,j}(y)&=
L^{-2j} 
e^{-\frac{\a^2}{2}\G_{j}(0)}
\lft(e^{\a^2\h\G_{j}(y)}-1\rgt),
\notag\\
\bar R^{(j)}_{1,c,j}(y)&=
L^{-2j} 
e^{-\frac{\a^2}{2}\G_{j}(0)}
\lft(e^{-\a^2\bar \h\G_j(y)}-1\rgt),
\notag\\
R^{(j)\n}_{1,d,j}(y)&=
i\a\h 
(\dpr^\n \G_j)(y)
\notag\\
\bar R^{(j)\n}_{1,d,j}(y)&=
i\a\bar \h 
(\dpr^\n \G_j)(y).
\eal
Next consider  $\EEE_j [W_{1,j}]$: the outcome 
of the integration and the of the rearrangement  into  
terms that are either irrelevant or local is
\bal \lb{taylor3}
&\EEE_j\lft[W_{1,j}(\F, B)\rgt]
\notag\\
&=
z_j Z_j L^{-2j}\sum_{y\in \ZZZ^2} w_{1,b,j}(y)
e^{-\frac{\a^2}2 (1+\h^2)\G_j(0)}
e^{-\a^2\h\G_j(y)}
\sum_{x\in B\atop \s=\pm}
J_{x,\s} e^{i\a\s(\h\f'_x+\f'_{x+y})}
\notag\\
&+z_j \bar Z_j L^{-2j}\sum_{y\in \ZZZ^2} \bar w_{1,b,j}(y)
e^{-\frac{\a^2}2(1+\bar \h^2)\G_j(0)}
e^{\a^2\bar \h\G_j(y)}\sum_{x\in B\atop \s=\pm}
J_{x,\s} e^{i\a\s(\bar \h\f_x-\f_{x+y})} 
\notag\\
&+z_j Z_j L^{-2j}\sum_{y\in \ZZZ^2} w_{1,c,j}(y)
e^{-\frac{\a^2}2\bar\h^2\G_j(0)}
e^{\a^2 \h \G_j(y|0)}
\sum_{x\in B\atop \s=\pm}
J_{x,\s} e^{i\a\s\bar \h\f_x} 
\notag\\
&\qquad\times
\lft[e^{-i\a\s(\f_{x+y}-\f_x)}-1+i\a\s y^\m\sum_{\m\in\hat u}(\dpr^\m \f_{x})\rgt]
\notag\\
&+z_j Z_j L^{-2j}\sum_{y\in \ZZZ^2} w_{1,c,j}(y)
e^{-\frac{\a^2}2\bar\h^2\G_j(0)}
\lft(e^{\a^2 \h \G_j(y|0)}-1\rgt)
\sum_{x\in B\atop \s=\pm}
J_{x,\s} e^{i\a\s\bar \h\f_x}
\notag\\
&
+z_j \bar Z_j L^{-2j}\sum_{y\in \ZZZ^2} \bar w_{1,c,j}(y)
e^{-\frac{\a^2}2 \h^2\G_j(0)}
e^{-\a^2 \bar \h \G_j(y|0)}
\sum_{x\in B\atop \s=\pm}
J_{x,\s}e^{i\a\s\h\f_x} 
\notag\\
&\qquad\times\lft[ e^{i\a\s(\f_{x+y}-\f_x)} -
1-i\a\s y^\m\sum_{\m\in\hat u}(\dpr^\m \f_{x})\rgt]
\notag\\
&+z_j Z_j L^{-2j}\sum_{y\in \ZZZ^2} \bar w_{1,c,j}(y)
e^{-\frac{\a^2}2 \h^2\G_j(0)}
\lft(e^{-\a^2 \bar \h \G_j(y|0)}-1\rgt)
\sum_{x\in B\atop \s=\pm}
J_{x,\s} e^{i\a\s \h\f_x}
\notag\\
&
+
s_j Z_j L^{-2j}
\sum_{y\in \ZZZ^2\atop\n\in\hat u}w^\n_{1,d,j} (y)
e^{-\frac{\a^2}2\h^2\G_j(0)} 
\sum_{x\in B\atop  \s=\pm}J_{x,\s} e^{i\h\a\s\f_x} 
\s \lft[(\dpr^\n\f_{x+y})- (\dpr^\n\f_{x})\rgt]
\notag\\
&
+
s_j Z_j L^{-2j}i\h\a 
\sum_{y\in \ZZZ^2\atop\n\in\hat u}w^\n_{1,d,j} (y)
\dpr^\n \G_j(y)
e^{-\frac{\a^2}2\h^2\G_j(0)} 
\sum_{x\in B\atop  \s=\pm}J_{x,\s} e^{i\h\a\s\f_x} 
\notag\\
&
+
s_j \bar Z_j L^{-2j}
\sum_{y\in \ZZZ^2\atop\n\in\hat u}\bar w^\n _{1,d,j} (y)
e^{-\frac{\a^2}2\bar\h^2\G_j(0)} 
\sum_{x\in B\atop  \s=\pm}J_{x,\s} e^{i\bar\h\a\s\f_x} 
\s \lft[(\dpr^\n\f_{x+y})- (\dpr^\n\f_{x})\rgt]
\notag\\
&
+
s_j \bar Z_j L^{-2j}i\bar \h\a 
\sum_{y\in \ZZZ^2\atop\n\in\hat u} 
\bar w^\n_{1,d,j} (y)
\dpr^\n \G_j(y)
e^{-\frac{\a^2}2\bar \h^2\G_j(0)} 
\sum_{x\in B\atop  \s=\pm}J_{x,\s} e^{i\bar \h\a\s\f_x}.
\eal 
Note that, as we did to derive \pref{taylor2}, 
we used some cancellations, which in this case are
consequence of the parity of $w_{1,c,j}(y)$ and 
$\bar w_{1,c,j}(y)$ in $y$ as seen from \pref{lw1n}: 
$$
\sum_{y\in \ZZZ^2} w_{1,c,j}(y)
e^{-\frac{\a^2}2\bar\h^2\G_j(0)}
e^{\a^2 \h \G_j(y|0)} y^\m
=
\sum_{y\in \ZZZ^2} \bar w_{1,c,j}(y)
e^{-\frac{\a^2}2 \h^2\G_j(0)}
e^{\a^2 \bar \h \G_j(y|0)} y^\m=0.
$$
Therefore, replacing $t_j$ with $\tilde t_j$, we formulate the  
second part of the  ansatz
\bal\lb{ansatz22}
R^{(j)}_{1,b,n}(y)&=R^{(j-1)}_{1,b,n}(y)
e^{-\frac{\a^2}{2}\G_{j}(0)}
e^{-\h\a^2\G_{j}(y)},
\notag\\
\bar R^{(j)}_{1,b,n}(y)&=\bar R^{(j-1)}_{1,b,n}(y)
e^{-\frac{\a^2}{2}\G_{j}(0)}
e^{\bar \h\a^2\G_{j}(y)},
\notag\\
R^{(j)}_{1,c,n}(y)&=R^{(j-1)}_{1,c,n}(y)
e^{-\frac{\a^2}{2}\G_{j}(0)}
e^{\h\a^2\G_{j}(y)},
\notag\\
\bar R^{(j)}_{1,c,n}(y)&=\bar R^{(j-1)}_{1,c,n}(y)
e^{-\frac{\a^2}{2}\G_{j}(0)}
e^{-\bar\h\a^2\G_{j}(y)},
\notag\\
R^{(j)\n}_{1,d,n}(y)&=R^{(j-1)\n}_{1,d,n}(y),
\notag\\
\bar R^{(j)\n}_{1,d,n}(y)&=\bar R^{(j-1)\n}_{1,d,n}(y).
\eal
Finally it is easy to solve \pref{ansatz22} 
with  initial data \pref{ansatz12}: we obtain
\bal\lb{lw1}
R^{(j-1)}_{1,b,n}(y)&=L^{-2n} 
e^{-\frac{\a^2}{2}\G_{j-1,n}(0)}
e^{-\h\a^2\G_{j-1,n+1}(y)}\lft(e^{-\h\a^2\G_n(y)}-1\rgt),
\notag\\
\bar R^{(j-1)}_{1,b,n}(y)&=
L^{-2n} 
e^{-\frac{\a^2}{2}\G_{j-1,n}(0)}
e^{\bar \h\a^2\G_{j-1,n+1}(y)}\lft(e^{\bar \h\a^2\G_n(y)}-1\rgt),
\notag\\
R^{(j-1)}_{1,c,n}(y)&=
L^{-2n} 
e^{-\frac{\a^2}{2}\G_{j-1,n}(0)}
e^{\h\a^2\G_{j-1,n+1}(y)}\lft(e^{\h\a^2\G_n(y)}-1\rgt),
\notag\\
\bar R^{(j-1)}_{1,c,n}(y)&=
L^{-2n} 
e^{-\frac{\a^2}{2}\G_{j-1,n}(0)}
e^{-\bar\h\a^2\G_{j-1,n+1}(y)}\lft(e^{-\bar \h\a^2\G_n(y)}-1\rgt),
\notag\\
R^{(j-1)\n}_{1,d,n}(y)&=
i\a\h
(\dpr^\n \G_n)(y),
\notag\\
\bar R^{(j-1)\n}_{1,d,n}(y)&=
i\a\bar \h
(\dpr^\n \G_n)(y).
\eal
Besides, comparing with \pref{lk4}, the marginal and relevant terms of
\pref{taylor2} and \pref{taylor3} give
\bal\lb{lw2}
m_{2,1, j}&=
\sum_{y\in \ZZZ^2}
\lft[\sum_{n=0}^{j} R^{(j)}_{1,c,n}(y) e^{-\frac{\a^2}2 (2\h-1) \G_j(0)}
-\sum_{n=0}^{j-1} R^{(j-1)}_{1,c,n}(y)\rgt],
\notag\\
m_{1,2, j}&=
\sum_{y\in \ZZZ^2}
\lft[\sum_{n=0}^{j} \bar R^{(j)}_{1,c,n}(y) e^{\frac{\a^2}2 (2\bar\h+1) \G_j(0)}
-\sum_{n=0}^{j-1} \bar R^{(j-1)}_{1,c,n}(y)\rgt],
\notag\\
m_{1,1, j}&= \frac{\a^2\h^2}2\sum_{y\in \ZZZ^2 \atop \n\in\hat u} 
\lft[(\dpr^\n \G_j)^2(y)+
2(\dpr^\n \G_{j-1,0})(y) 
(\dpr^\m \G_j)(y)\rgt],
\notag\\
m_{2,2, j}&=\frac{\a^2\bar\h^2}2\sum_{y\in \ZZZ^2 \atop\n\in\hat u} 
\lft[(\dpr^\n \G_j)^2(y)+
2(\dpr^\n \G_{j-1,0})(y) 
(\dpr^\m \G_j)(y)\rgt].
\eal
This proves that \pref{lw1n} and \pref{dfm4} yield \pref{3sc2second}.
\vskip1em
{\it\03. Third term.} This term is quadratic in $J$.  We look for $w_{2,\a,j}(y)$
(where again $\a$ is the collections of various labels, compare with \pref{wj2}) 
into the form 
$$
w_{2,\a,j}(y)=\sum_{n=1}^{j-1} R^{(j-1)}_{2,\a,n}(y);
$$
then $ R^{(j-1)}_{2,\a,n}(y)$ will be determined by an ansatz to obtain 
\bal\lb{3sc2third}
&{1\over 2}\EEE^T_j\lft[V_{1,j}(\tilde t_j,\F,B);V_{1,j}(\tilde t_j,\F, Y)\rgt]
+\sum_{X\in \SS_j}^{X\supset B}Q_{2,j}(\F',B,X)
\notag\\
&=W_{2,j+1}(\F',B)-\EEE_j\lft[W_{2,j}(\tilde t_j\F,B)\rgt].
\eal
The first term in \pref{3sc2third} is 
\bal\lb{sec1}
&{1\over 2}\EEE^T_j\lft[V_{1,j}(\F,B);V_{1,j}(\F, Y)\rgt]
\notag\\
&=Z_j^2\frac{L^{-4j}}{2}\sum_{y\in\ZZZ^2\atop \e=\pm1} 
e^{-\h^2\a^2\G_j(0)}\lft(e^{-\h^2\a^2\e \G_j(y)}-1\rgt)
\sum_{x\in B\atop \s=\pm1}J_{x,\s}J_{\s\e,x+y} e^{i\h\a\s(\f'_x+\e \f'_{x+y})}
\notag\\
&+Z_j\bar Z_j\frac{L^{-4j}}{2}\sum_{y\in\ZZZ^2\atop \e=\pm1} 
e^{-(\h^2+\bar \h^2)\frac{\a^2}{2}\G_j(0)}
\lft(e^{-\h\bar \h\a^2\e \G_j(y)}-1\rgt)
\sum_{x\in B\atop \s=\pm1}\lft[J_{x,\s}J_{\s\e,x+y} 
e^{i\a\s(\h\f'_x+\e\bar\h \f'_{x+y})}\rgt]
\notag\\
&+Z_j\bar Z_j\frac{L^{-4j}}{2}\sum_{y\in\ZZZ^2\atop \e=\pm1} 
e^{-(\h^2+\bar \h^2)\frac{\a^2}{2}\G_j(0)}
\lft(e^{-\h\bar \h\a^2\e \G_j(y)}-1\rgt)
\sum_{x\in B\atop \s=\pm1}\lft[J_{x,\s}J_{\s\e,x+y} 
e^{i\a\s(\bar \h\f'_{x}+\e \h \f'_{x+y})}\rgt]
\notag\\
&+\bar Z_j^2\frac{L^{-4j}}{2}\sum_{y\in\ZZZ^2\atop \e=\pm1} 
e^{-\bar \h^2\a^2\G_j(0)}\lft(e^{-\bar\h^2\a^2\e \G_j(y)}-1\rgt)
\sum_{x\in B\atop \s=\pm1}J_{x,\s}J_{\s\e,x+y} e^{i\bar\h\a\s(\f'_x+\e \f'_{x+y})},
\eal
where the parameters $t_j$ have to be replaced with $\tilde t_j$;
taking into account also 
the second term in \pref{3sc2third},  we set
\bal\lb{ansatz13}
R^{(j)\e}_{2,a,j}(y)&:=\frac12 Z_j^2L^{-4j}
e^{-\h^2\a^2\G_j(0)}\lft(e^{-\h^2\a^2\e \G_j(y)}-1\rgt),
\notag\\
\bar R^{(j)\e}_{2,a,j}(y)&:=\frac12 \bar Z_j^2L^{-4j}
e^{-\bar \h^2\a^2\G_j(0)}\lft(e^{-\bar \h^2\a^2\e \G_j(y)}-1\rgt),
\notag\\
R^{(j)\e}_{2,b,j}(y)&:=\frac12 Z_j\bar Z_j L^{-4j}
e^{-(\h^2+\bar \h^2)\frac{\a^2}{2}\G_j(0)}\lft(e^{-\h\bar \h\a^2\e
  \G_j(y)}-1\rgt),
\notag\\
R^{(j)\e}_{2,c,j}(y)&:=\sum_{k=0}^j 2^{-(j-k)} L^{-4k} e^{-L^{-k} |y|}
\notag\\
&\times \lft\{Z_k^2 \frac12\sum_{\s=\pm1}\sum_{X\in \SS_j}^{X\ni 0}
\EEE_j\lft[ \hat K_{2,j}^{(a,k)}\lft(-\s\frac{1+\e}2,\z, X,0, \s, y, \s\e\rgt)\rgt]
\rgt.
\notag\\
&+\lft.
\bar Z^2_k \frac12\sum_{\s=\pm1}\sum_{X\in \SS_j}^{X\ni 0}\EEE_j\lft[ 
\hat K_{2,j}^{(\bar a,k)}\lft(\s\frac{1+\e}2,\z, X,0, \s, y, \s\e\rgt)
\rgt]
\rgt.
\notag\\
&+\lft.
Z_k\bar Z_k \frac12\sum_{\s=\pm1}\sum_{X\in \SS_j}^{X\ni 0}
\EEE_j\lft[ 
\hat K_{2,j}^{(b,k)}\lft(-\s\frac{1-\e}2,\z, X, 0, \s, y,\s\e\rgt)
\rgt]\rgt\}.
\eal
Next, we find 
\bal
&\EEE_j\lft[W_{2,j}(\F,B)\rgt]
\notag\\
&
=\sum_{y\in\ZZZ^2\atop\e=\pm}w^\e_{2,a,j}(y)
e^{-\h^2 \a^2 (1+\e)\G_j(0)}e^{-\h^2 \a^2 \e\G_j(y|0)}
\sum_{x\in B\atop \s=\pm}J_{x,\s}J_{\s\e,x+y} 
e^{i\h\a\s(\f'_x+\e\f'_{x+y})}
\notag\\
&+\sum_{y\in\ZZZ^2\atop\e=\pm}\bar w^\e_{2,a,j}(y)
e^{-\bar \h^2 \a^2 (1+\e)\G_j(0)}e^{-\bar \h^2 \a^2 \e\G_j(y|0)}
\sum_{x\in B\atop \s=\pm}J_{x,\s}J_{\s\e,x+y} 
e^{i\bar\h\a\s(\f'_x+\e \f'_{x+y})}
\notag\\
&+\sum_{y\in\ZZZ^2\atop\e=\pm}w^\e_{2,b,j}(y)
e^{-(\h+\e \bar \h)^2 \frac{\a^2}2 \G_j(0)}e^{-\h \bar \h \a^2 \e\G_j(y|0)}
\sum_{x\in B\atop \s=\pm}
J_{x,\s}J_{\s\e,x+y} 
\notag\\
&\qquad\qquad\qquad\qquad
\times\lft[e^{i\a\s(\h\f'_x+\e\bar\h \f'_{x+y})}+
e^{i\a\s(\bar\h\f'_x+\e\h \f'_{x+y})}\rgt]
\notag\\
&+\sum_{y\in\ZZZ^2\atop\e=\pm1}w^\e_{2,c,j}(y)
e^{-\frac{\a^2}2(1+\e)^2(\h-\frac12)^2 \G_j(0)}
\sum_{x\in B\atop \s=\pm}
J_{x,\s}J_{\e\s,x+y} e^{i\a\s(1+\e)(\h-\frac12)\f'_x}.
\eal
Hence the second part of the ansatz is 
\bal\lb{ansatz23}
R^{(j)\e}_{2,a,n}(y)&=
R^{(j-1)\e}_{2,a,n}(y)e^{-\h^2 \a^2 (1+\e)\G_j(0)}e^{-\h^2 \a^2 \e\G_j(y|0)},
\notag\\
\bar R^{(j)\e}_{2,a,n}(y)&=\bar R^{(j-1)\e}_{2,a,n}(y)
e^{-\bar \h^2 \a^2 (1+\e)\G_j(0)}e^{-\bar \h^2 \a^2 \e\G_j(y|0)},
\notag\\
R^{(j)\e}_{2,b,n}(y)&=R^{(j-1)\e}_{2,b,n}(y)
e^{-(\h+\e \bar \h)^2 \frac{\a^2}2 \G_j(0)}e^{-\h \bar \h \a^2 \e\G_j(y|0)},
\notag\\
R^{(j)\e}_{2,c,n}(y)&=R^{(j-1)\e}_{2,c,n}(y)
e^{-\frac{\a^2}2(1+\e)^2(\h-\frac12)^2 \G_j(0)}.
\eal
Solving \pref{ansatz23} with initial data  \pref{ansatz13} we obtain 
\bal\lb{hofer}
R^{(j-1)\e}_{2,a,n}(y)&=\frac12 Z_n^2L^{-4n}
e^{-\h^2(1+\e)\a^2\G_{j-1,n+1}(0)}
\notag\\
&\qquad\qquad
\times e^{-\h^2\a^2\e\G_{j-1,n+1}(y|0)}
e^{-\h^2\a^2\G_n(0)}\lft(e^{-\h^2\a^2\e \G_n(y)}-1\rgt)
\notag\\
\bar R^{(j-1)\e}_{2,a,n}(y)&=\frac12 \bar Z_n^2L^{-4n}
e^{-\bar \h^2(1+\e)\a^2\G_{j-1,n+1}(0)}
\notag\\
&\qquad\qquad
\times 
e^{-\bar \h^2\a^2\e\G_{j-1,n+1}(y|0)}
e^{-\bar \h^2\a^2\G_n(0)}\lft(e^{-\bar \h^2\a^2\e \G_n(y)}-1\rgt)
\notag\\
R^{(j-1)\e}_{2,b,n}(y)&=\frac12 Z_n\bar Z_n L^{-4n}
e^{-(\h+\e\bar\h)^2\frac{\a^2}{2}\G_{j-1,n+1}(0)}
\notag\\
&\qquad\qquad
\times 
e^{-\h\bar \h\a^2\e\G_{j-1,n+1}(y|0)}
e^{-(\h^2+\bar \h^2)\frac{\a^2}{2}\G_n(0)}
\lft(e^{-\h\bar \h\a^2\e
  \G_n(y)}-1\rgt),
\notag\\
R^{(j-1)\e}_{2,c,n}(y)&:=
e^{-\frac{\a^2}2(1+\e)^2(\h-\frac12)^2 \G_{j-1,n+1}(0)}\sum_{k=0}^n 2^{-(n-k)} L^{-4k} e^{-L^{-k} |y|}
\notag\\
&\!\times \lft\{Z_k^2 \frac12\sum_{\s=\pm1}\sum_{X\in \SS_n}^{X\ni 0}
\EEE_j\lft[ \hat K_{2,n}^{(a,k)}\lft(-\s\frac{1+\e}2,\z, X,0, \s, y, \s\e\rgt)\rgt]
\rgt.
\notag\\
&+\lft.
\bar Z^2_k \frac12\sum_{\s=\pm1}\sum_{X\in \SS_n}^{X\ni 0}\EEE_j\lft[ 
\hat K_{2,n}^{(\bar a,k)}\lft(\s\frac{1+\e}2,\z, X,0, \s, y, \s\e\rgt)
\rgt]
\rgt.
\notag\\
&+\lft.
Z_k\bar Z_k \frac12\sum_{\s=\pm1}\sum_{X\in \SS_n}^{X\ni 0}
\EEE_j\lft[ 
\hat K_{2,n}^{(b,k)}\lft(-\s\frac{1-\e}2,\z, X, 0, \s, y,\s\e\rgt)
\rgt]\rgt\}.
\eal
In conclusion, \pref{hofern} yield \pref{3sc2third}.
This concludes the proof of Lemma \ref{11:20}.
\epr

\section{Remainder part of the RG map}
\begin{lemma}\lb{crone}
If $z>0$ is small enough and  $|s_j|,|z_j|\le c_0|q_j|$, 
$\|K_{0,j}\|_{h,T_j}\le c_0|q_j|^2$, there exists $C\=C(A, L, \a)$ such that, 
\be\lb{cosmo}
\|\RR_{1,j}\|_{1, h, T_{j+1}}\le C \lft[|q_j|^2
+|q_j|\|K_{1,j}\|_{1,h,T_j}+|q_j|\|K^\dagger_{1,j}\|_{1,h,T_j}\rgt]; 
\ee
besides the same bound is valid for $\|\RR^\dagger_{1,j}\|_{1, h, T_{j+1}}$.
\end{lemma}
\bpr 
We begin with an exact formula for $\RR_{1,j}$. 
From \pref{4.19bis} we have
$$
P_j(D)= P_{0,j}(D)+ P_{1,j}(D)
+ P_{2,j}(D)+ P_{\ge 3,j}(D)
$$ 
where, if $\tilde V_{0,j}(D):=V_{0,j}(D)-\EEE_j\lft[ V_{0,j}(D)\rgt]$
and $\tilde V_{1,j}(D):=V_{1,j}(D)-\EEE_j\lft[ V_{1,j}(D)\rgt]$, 
\bal
P_{0,j}(D)
&= \tilde V_{0,j}(D)-\Big(V_{0,j+1}+\d E_j|D|-\EEE_j\lft[ V_{0,j}\rgt]\Big)
+\lft(e^{U_{0,j}(D)}-1-V_{0,j}(D)\rgt)
\notag\\
&- 
\lft(e^{U_{0,j+1}(D)+\d E_{j}|D|}-1-(V_{0,j+1}(D)+\d E_j|D|)\rgt),
\notag\\
P_{1,j}(D)
&= \tilde V_{1,j}(D)+ \lft(e^{U_{0,j}(D)}-1\rgt)
\tilde V_{1,j}(D)
\notag\\
&+\lft(e^{U_{0,j}(D)}-e^{U_{0,j+1}(D)+\d E_{j}|D|}\rgt) \EEE_j[V_{1,j}(D)]
\notag\\
&
-e^{U_{0,j+1}(D)+\d E_{j}|D|}\Big(V_{1,j+1}(D)-\EEE_j[V_{1,j}(D)]\Big)
\notag\\
&+e^{U_{0,j}(D)}W_{1,j}(D)-e^{U_{0,j+1}(D)+\d E_{j}|D|} W_{1,j+1}(D),
\notag\\
P_{2,j}(D)
&=\frac12e^{U_{0,j}(D)}\lft(V_{1,j}(D)^2+W_{1,j}(D)^2+2 W_{2,j}(D)\rgt)
\notag\\
&\quad- 
\frac12e^{U_{0,j+1}(D)+\d E_{j}|D|} \lft(V_{1,j+1}(D)^2+W_{1,j+1}(D)^2+2 W_{2,j+1}(D)\rgt);
\eal
while $P_{\ge 3,j}(D)$ contains the rest of $P_{j}(D)$.
Therefore we find
\bal\lb{cosmo:57}
\RR^{(1)}_{1,j}(Y')&=
\sum_{D\in \BB_{j+1}}^{D=Y'}
\EEE_j\lft[\lft(e^{U_{0,j}(D)}-1-V_{0,j}(D)\rgt)V_{1,j}(D)\rgt]
\notag\\
&-
\sum_{D\in \BB_{j+1}}^{D=Y'}
\lft(e^{U_{0,j+1}(D)+\d E_j|D|}-1-V_{0,j+1}(D)-\d E_j|D|\rgt)\EEE_j[V_{1,j}(D)]
\notag\\
&-
\sum_{D\in \BB_{j+1}}^{D=Y'}
\lft(e^{U_{0,j+1}(D)+\d E_j|D|}-1\rgt)\Big(V_{1,j+1}(D)-\EEE_j[V_{1,j}(D)]\Big)
\notag\\
&-
\sum_{D\in \BB_{j+1}}^{D=Y'}
\Big(V_{0,j+1}(D)+\d E_j|D|-\EEE_j[V_{0,j}(D)]\Big) V_{1,j+1}(D)
\notag\\
&-
\sum_{D\in \BB_{j+1}}^{D=Y'}
\lft(e^{U_{0,j+1}(D)+\d E_j|D|}-1\rgt) W_{1,j+1}(D)
\notag\\
&+
\sum_{D\in \BB_{j+1}}^{D=Y'}
\EEE_j\lft[\lft(e^{U_{0,j}(D)}-1\rgt) W_{1,j}(D)\rgt],
\notag\\
\RR^{(2)}_{1,j}(Y')&=
\sum_{D_1, D_2\in \BB_{j+1}\atop D_1\neq D_2}^{D_1\cup D_2=Y'}
\EEE_j\lft[\Big(P_{0,j}(D_1)-\tilde V_{0,j}(D_1)\Big)\tilde V_{1,j}(D_2)\rgt]
\notag\\
&+
\sum_{D_1, D_2\in \BB_{j+1}\atop D_1\neq D_2}^{D_1\cup D_2=Y'}
\EEE_j\lft[P_{0,j}(D_1)\Big(P_{1,j}(D_2)-\tilde V_{1,j}(D_2)\Big)\rgt],
\notag\\
\RR^{(3)}_{1,j}(Y')
&=
\sum_{D\in \BB_{j+1}}^{D=Y'}\Big[\EEE_j[W_{1,j}(D)]-\EEE_j[W_{1,j}(\tilde t_j, D)]\Big],
\notag\\
\RR^{(4)}_{1,j}(Y')
&=
\sum_{D_1, D_2\in \BB_{j+1}}^{D_1\cup D_2=Y'}
\Big[\EEE^T_j\lft[V_{0,j}(D_1); V_{1,j}(D_2)\rgt]- 
\EEE^T_j\lft[V_{0,j}(\tilde t_j, D_1); V_{1,j}(\tilde t_j, D_2)\rgt]\Big],
\notag\\
\RR^{(5)}_{1,j}(Y')&=
\sum_{Y_0\in \CC_{j+1}(X_0)\ge 1\atop |Z|_{j+1}+\CC_{j+1}(X_0\cup X_1)\ge 2}^{\to Y'} 
\EEE_j\lft[P_{0,j}^Z R_{0,j}^{X_1}\rgt] J_{0,j}^{X_0\bs Y_0, (D)}J_{1,j}(D_{Y_0}, Y_0)
\notag\\
&+
\sum_{Y_1\in \CC_{j+1}(X_1)\ge 1\atop |Z|_{j+1}+\CC_{j+1}(X_0\cup X_1)\ge 2}^{\to Y'} 
\EEE_j\lft[P_{j,0}^Z R_{0,j}^{X_1\bs Y_1}R_{1,j}(Y_1) \rgt] J_{0,j}^{X_0, (D)}
\notag\\
&+
\sum_{\CC_{j+1}(X_0\cup X_1)\ge 1\atop B\in \BB_{j+1}(Z)\neq\emptyset}^{\to Y'} 
\EEE_j\lft[P_{1,j}(B)P_{0,j}^{Z\bs B} R_{0,j}^{X_1}\rgt] J_{0,j}^{X_0, (D)},
\notag\\
\RR^{(6)}_{1,j}(Y')&=
\sum_{\CC_{j+1}(X_0\cup X_1)\ge 1\atop B\in \BB_{j+1}(Y'\bs W)}^{\to Y'} 
e^{-\d E_j|Y'|+U_{0,j+1}(Y'\bs W)}\Big(V_{1,j+1}(B)+W_{1,j+1}(B)\Big)
\EEE_j\lft[P_{0,j}^Z R_{0,j}^{X_1}\rgt] J_{0,j}^{X_0, (D)}
\notag\\
&+
\sum_{Y_0\in \CC_{j+1}(X_0)}^{\to Y'} 
\lft(e^{-\d E_j|Y'|+U_{0,j+1}(Y'\bs W)}-1\rgt)\EEE_j\lft[P_{0,j}^Z R_{0,j}^{X_1}\rgt] 
J_{0,j}^{X_0\bs Y_0, (D)} J_{1,j}(D_{Y_0}, Y_0)
\notag\\
&+
\sum_{Y_1\in \CC_{j+1}(X_1)}^{\to Y'} 
\lft(e^{-\d E_j|Y'|+U_{0,j+1}(Y'\bs W)}-1\rgt)\EEE_j
\lft[P_{0,j}^Z R_{0,j}^{X_1\bs Y_1}R_{1,j}(Y_1)\rgt] J_{0,j}^{X_0, (D)}
\notag\\
&+
\sum_{\CC_{j+1}(X_0\cup X_1)\ge 1\atop B\in \BB_{j+1}(Z)}^{\to Y'} 
\lft(e^{-\d E_j|Y'|+U_{0,j+1}(Y'\bs W)}-1\rgt)\EEE_j\lft[P_{0,j}^{Z\bs B}
P_{1,j}(B) R_{0,j}^{X_1}\rgt] J_{0,j}^{X_0, (D)},
\notag\\
\RR^{(7)}_{1,j}(Y')&=
\sum_{Z\in \PP_{j+1}\atop |Z|_{j+1}\ge 3 }^{Z=Y'}\sum_{B\in \BB_{j+1}(Z)}
\EEE_j\lft[P_{0,j}^{Z\bs B} P_{1,j}(B)\rgt]
\notag\\
&+ \lft(e^{-\d E_j|Y'|}-1\rgt)
\sum_{Z\in \PP_{j+1}\atop |Z|_{j+1}\ge 2 }^{Z=Y'}\sum_{B\in \BB_{j+1}(Z)}
\EEE_j\lft[P_j^{Y'}\rgt]
\notag\\
&+ \lft(e^{-\d E_j|Y'|}-1\rgt)
\sum_{B\in \BB_{j+1}}^{B=Y'}
\EEE_j\lft[P_{1,j}(B)- \tilde V_{1,j}(B)\rgt],
\notag\\
\RR^{(8)}_j(Y')&=
\sum_{X\in \PP_j\atop |\CC_j(X)|\ge 2}^{\bar X=Y'}
\sum_{Y_0\in\CC_j(X) }\EEE_{j}\lft[K_{1,j}(Y_0)\prod_{Y\in \CC_j(X\bs Y_0)} K_{0,j}(Y)\rgt],
\notag\\
\RR^{(9)}_j(Y')&=\sum_{X\in \PP_j\atop B\in \BB_{j+1}(Y'\bs X)}^{\bar X=Y'} 
\EEE_j\lft[\Big(V_{1,j}(B)+ W_{1,j}(B)\Big)
e^{U_{0,j}(Y'\bs X)}\prod_{Y\in \CC_j(X)} K_{0,j}(Y)\rgt]
\notag\\
&+\sum_{X\in \PP_j\atop Y_0\in \CCC_{j+1}(X)}^{\bar X=Y'} 
\EEE_j\lft[\lft(e^{U_{0,j}(Y'\bs X)}-1\rgt)K_{1,j}(Y_0)\prod_{Y\in \CC_j(X\bs Y_0)} K_{0,j}(Y)\rgt].
\eal
The reason of \pref{cosmo} is that each of the above terms falls into 
one of two classes: a) those terms which, besides containing a factor of $V_{1,j}$ or $W_{1,j}$, 
also contain at least two factors  of $s_j$, $z_j$, or  one factor of  $K_{0,j}$;
b) those terms which contain one factor of either $K_{1,j}$ or $K^\dagger_{1,j}$
and at least one factor of $s_j,z_j$ or $K_{0,j}$
To proceed more formally, we need formula (6.74) of \citep{Br07}
and a  simpler version of the bounds 
in  Lemma 14 of \citep{Fa12}.
\begin{lemma}[\citep{Br07}]
There exists a $\th >0$ such that, for any $X\in \PP_j$, 
\be\lb{6:25}
(1+2\th) |\bar X|_{j+1}\le |X|_j + 8 (1+2\th ) |\CC_j(X)|.
\ee 
\end{lemma}
\begin{lemma}[\citep{Fa12}] Under the hypothesis of Lemma \ref{crone}, 
for a $\th >0$, there exists a $C\=C(A,L,\a)$ such that 
\bal
\lb{6:27}
&\|e^{U_{0,j}(\f,D)}-1\|_{h,T_j(\f,D)}\le C  |q_j| G_j^{\rm str}(\f,D),
\\
\lb{6:28}
&\|e^{U_{0,j+1}(\f',D)+\d E_j|D|}-1\|_{h,T_j(\f',D)}\le C |q_j|
G^{\rm str}_{j+1}(\f',D),
\\
&\|P_{0,j}(\f',\z,D)\|_{h,T_j(\f',D)}
\le C A^{-(1+\th)} |q_j|\lft[G^{\rm str}_j(\f,D)+G^{\rm str}_{j+1}(\f',D)\rgt],
\lb{6:29}
\\
&\|J_{0,j}(\f',D,Y)\|_{h,T_j(\f',Y)} 
\le C A^{-(1+\th)|D^*|_{j+1}} |q_j|^2
G^{\rm str}_{j+1}(\f',D), 
\lb{6:30}
\\
\lb{6:31}
&\|R_{0,j}(\f',\z,Y)\|_{h,T_j(\f',Y)}
\le C A^{-(1+\th)|Y|_{j+1}}|q_j|^2
\lft[G^{\rm str}_{j+1}(\f',Y)+ G_{j}(\f,Y)\rgt]. 
\eal
\end{lemma}
Let us consider, for example, $\RR^{(8)}_j(Y)$. Extracting the dependence in $J$
be obtain, as usual, two terms:
\bal
\RR^{(8)}_j(\f',Y, x,\s)&= 
\sum_{X\in \PP_j\atop |\CC_j(X)|\ge 2}^{\bar X=Y}
\sum_{Y_0\in\CC_j(X)\atop Y_0\ni x }
\EEE_{j}\lft[K_{1,j}(\f,Y_0, x,\s)\prod_{Y\in \CC_j(X\bs Y_0)} K_{0,j}(Y)\rgt],
\eal 
and a similar one proportional to $K^\dagger_{1,j}(\f,Y_0, x,\s)$. Using \pref{6.51}, 
\pref{6.54} and the inequality $A^{-|X|_j}\le A^{-(1+2\th)|Y|_{j+1}} A^{8(1+2\th)|\CC_j(X)|}$
which is a consequence of \pref{6:25},  
a bound for $\|\RR^{(8)}_j(\f',Y,x,\s)\|_{h,T_j(\f',Y)}$ is, for a $C\=C(A,L,\a)$, 
\bal
&G_{j+1}(\f',Y)\|K_{1,j}\|_{1,h,T_j} 
\sum_{X\in \PP_j\atop |\CC_j(X)|\ge 2}^{\bar X=Y}  A^{-|X|_{j}} 2^{|X|_j} (C |q_j|)^{|C_j(X)|-1} 
\notag\\
&\le G_{j+1}(\f',Y)\|K_{1,j}\|_{1,h,T_j} 4^{L^2|Y|_{j+1}} A^{-(1+2\th)|Y|_{j+1}}
\sum_{p\ge 2}  A^{8(1+2\th)p} (C |q_j|)^{p-1} 
\notag\\
&\le G_{j+1}(\f',Y)\|K_{1,j}\|_{1,h,T_j} A^{-|Y|_{j+1}}
C_1 |q_j|, 
\eal
where the last inequality holds if one first chooses $A$ large enough so that 
$4^{L^2} A^{-2\th}\le 1$, and then chooses $|q_1|$ small enough so that the series in $p$
is convergent. To obtain the second line we also used that in the sum in the first 
line there are no more than $2^{|Y|_j}\le 2^{L^2|Y|_{j+1}}$ terms. 

As a second sample case, consider one  
of the terms in $\RR^{(6)}_{1,j}(Y)$, which, after the extraction of  
$Z_j L^{2j}J_{x,\s}$, is
\bal\lb{test}
\frac{Z_{j+1} }{Z_{j}}L^{-2} e^{i\a\h\s\f'_x}
\sum_{\CC_{j+1}(X_0\cup X_1)\ge 1\atop B\in \BB_{j+1}(Y\bs W), B\ni x}^{\to Y} 
e^{-\d E_j|Y|+U_{0,j+1}(Y\bs W)} 
\EEE_j\lft[P_{0,j}^Z R_{0,j}^{X_1}\rgt] J_{0,j}^{X_0, (D)}.
\eal
A bound for the norm $\|\cdot\|_{h,T(\f',Y)}$ of this term is made of three kinds of 
factors: a product of field regulators, a product of factors of $A^{-1}$, and
a product of factors of $|q_j|$.  
Collecting all the factors of field regulators we obtain
\bal
&G^{\rm str}_{j+1}(\f',Y\bs W)\prod_{D\in \BB_{j+1}(Z)} 
\Big[G^{\rm str}_j(\f,D)+G^{\rm str}_{j+1}(\f',D)\Big]
\notag\\
&\times\prod_{Y\in \CC_{j+1}(X_1)}\Big[G_j(\f,Y)+ G^{\rm str}_{j+1}(\f',Y)\Big]
G^{\rm str}_{j+1}(\f',X_0)
\notag\\
&\le
\sum_{W_1\in P_{j+1}(Z)\atop W_2\in ((X_1))_{j+1}}
G_j(\f,W_1\cup W_2)G^{\rm str}_{j+1}(\f',Y\bs (W_1\cup W_2)), 
\eal
where $((X_1))_{j+1}$ is the collection of all the possible unions 
of connected parts of $X_1$.  Since the number of terms in the sum
over $W_1$ and $W_2$ 
is not larger than $2^{|Z|_{j+1} +|\CC_{j+1}(X_1)|}$, 
by \pref{6.54} the expectation of such 
factors is bounded by 
$$
2^{|Z|_{j+1} +|\CC_{j+1}(X_1)|}2^{L^2|Z|_{j+1} + L^2|X_1|_{j+1} }G^{\rm str}_{j+1}(\f',Y). 
$$
Next, collecting the $A^{-1}$ factors coming from \pref{6:29}, \pref{6:30}, \pref{6:31}, 
we obtain a factor not larger than
$A^{-(1+\th)|Y|_{j+1}}$. 
In conclusion, a bound for \pref{test} is, for a $C\=C(A,L,\a)$,  
\bal
&G^{\rm str}_{j+1}(\f',Y) 2^{(1+L^2)|Y|_{j+1} }A^{-(1+\th)|Y|_{j+1}} 
\notag\\
&\times \sum_{\CC_{j+1}(X_0\cup X_1)\ge 1}^{\to Y} 
(C|q_j|)^{|Z|_{j+1}+ 2|\CC_{j+1}(X_0\cup X_1)|} 
\notag\\
&\le G^{\rm str}_{j+1}(\f',Y) 2^{(5+L^2)|Y|_{j+1} }A^{-(1+\th)|Y|_{j+1}} 
\sum_{p\ge 1}
(C|q_j|)^{2p} 
\notag\\
&  \le 
C_1 G^{\rm str}_{j+1}(\f',Y) A^{-|Y|_{j+1}} 
|q_j|^{2},
\eal
where we used that the sum in the second line has no more than 
$4^{|Y|_{j+1}+|\CC_{j+1}(X_0)|}\le 2^{4|Y|_{j+1}}$ (indeed each connected component of $X_0$ 
has to be a small polymer); besides we assumed $A$ large enough so that 
$2^{(5+L^2)}A^{-\th}\le 1$, as well as $|q_1|$ small enough so that the series in $p$ is
convergent.
The other terms of \pref{cosmo:57} can be studied in a similar manner. 
\epr
\begin{lemma}\lb{fassin}
If $z>0$ is small enough and  $|s_j|,|z_j|\le c_0|q_j|$, 
$\|K_{0,j}\|_{h,T_j}\le c_0|q_j|^2$ and  
$\|K_{1,j}\|_{1,h,T_j}\le  c_0|q_j|^2$,
there exists $C\=C(A, L, \a)$ such that,
\bal\lb{geary}
\|\RR^{\d,k}_{2,j}\|_{2, h, T_{j+1}} \le
\begin{cases} 
C |q_k| &\text{for $k=j$} \\
C |q_j| \|K^{\d,k}_{2,j}\|_{2,h,T_j}\qquad&\text{for $0\le k\le j-1$} 
\end{cases}
\eal
\end{lemma}
\bpr 
When $k=j$ the term $\RR^{\d,k}_{2,j}$ is generated by a term in 
$K_{2,j+1}$ which contain at least two factors of $V_{1,j}$, $W_{1,j+1}$ 
or $K_{1,j}$ and at least one factor of $s_{j}$, $z_j$ or $K_{0,j}$. 
When $k\le j+1$, 
term $\RR^{\d,k}_{2,j}$ is generated by terms in $K_{2,j+1}$ which contain at least 
one factor of $K^{\d,k}_{2,j}$ and one factor of 
 $s_{j}$, $z_j$ or $K_{0,j}$. Note that in the case $k=j$ 
the norms on the right hand side are $\|\cdot\|_{h,T_j}$ or $\|\cdot\|_{1,h,T_j}$, in which the 
size of the sets are weighed with a factor $A$, whereas on the norm on the left hand side
is $\|\cdot\|_{2,h,T_j}$ in which the 
size of the sets are weighed with a factor $\sqrt A$: 
this provides the factor $e^{-L^{-k}|x_1-x_2|}$ in \pref{maldacena}.
\epr
\subsection{Proof of Theorem \ref{leibler}}\lb{7.1}
Let us consider the first of \pref{leibler0}. We have 
\bal
&K_{1,j+1}(\F',Y,x,\s)=\LL_{1,j}(\F',Y,x,\s)+ \RR_{1,j}(\F',Y,x,\s).
\eal
From \pref{slk} we see that the assumption of Lemma \ref{crone} is satisfied; therefore, 
using also \pref{arkani2}, we obtain
\bal
\|K_{1,j+1}\|_{1,h,T_{j+1}}&\le \r(L,A,\h) \|K_{1,j}\|_{1,h,T_j}
\notag\\
&+ C \lft[|q_j|^2
+|q_j|\|K_{1,j}\|_{1,h,T_j}+|q_j|\|K^\dagger_{1,j}\|_{1,h,T_j}\rgt]  
\eal 
Assuming by induction that $\|K_{1,j}\|_{1,h,T_j}\le 2C|q_j|^2$ and that 
$\|K^\dagger_{1,j}\|_{1,h,T_j}\le 2C|q_j|^2$, 
we obtain that $\|K_{1,j+1}\|_{1,h,T_{j+1}}\le 2C|q_{j+1}|$. 
(We also used that  $\r(L,A,\h)\le 1/4$ for $L$ and $A$ large enough; that
$|q_j|/|q_{j+1}|\le 1+ \sqrt{ab}|z|\le 2$ for $|z|$ small enough; and that 
$|q_j|\le c_0|z|\le \frac18$ for $|z|$ small enough.)
This proves  the first of \pref{leibler0}. The second can be obtained in a similar way.  
\subsection{Proof of Theorem \ref{pherson}}
Let us consider the bound for $K^{(a,k)}_{2,j+1}$.
From the previous definitions, suppressing the dependence in 
$(\F',Y, x_1, \s_1, x_2, \s_2)$,  we have
\bal
K^{(a,k)}_{2,j+1}=2\LL^{(a,k)}_{2,j}
+ 2\RR^{(a,k)}_{2,j}
\eal
where the factors $2$
stem from the prefactor $2^{-(j-k)}$ 
in \pref{maldacena}.
Because of  \pref{slk} the assumptions in Lemma \ref{fassin} 
are satisfied; therefore, with the aid of \pref{goddar2}, we have
\bal
\|K^{(a,k)}_{2,j+1}\|_{2,h,T_{j+1}}&\le 2\r(L,A,\h)\|K^{(a,k)}_{2,j}\|_{2,h,T_{j}}
\notag\\
&+
\begin{cases} 
2C |q_k| &\text{for $k=j$} \\
2C |q_j| \|K^{(a,k)}_{2,j}\|_{2,h,T_j}\qquad&\text{for $0\le k\le j-1$} 
\end{cases}
.
\eal
Therefore, for $L$ and $A$ large enough and $|z|$ small enough 
it is easy to show inductively  a bound  such as 
$\|K^{(a,k)}_{2,j}\|_{2,h,T_{j}}\le 4C|q_k|$. $K^{(\bar a,k)}_{2,j+1}$  and 
$K^{(b,k)}_{2,j+1}$ can be dealt with  in a similar way. 

\section{Flow of the fractional charge renormalization}
Merging \pref{lk4} and \pref{lk2b} we obtain \pref{sk}, namely 
the equation that 
describe the flow of the renormalization parameters $Z_j$ and $\bar Z_j$. 
To study such flows we need an explicit computation of some of the coefficients. 
In this section we set $\a^2=8\p$.  
%Let us recapitulate 
%some of the statements in  Lemma \ref{scott} and Lemma \ref{t3.1b}.
%\begin{lemma}\mp{questo si puo' togliere}
%\lb{scott}
%There exists a $j$-independent 
%$C\=C(L)$ such that
%%, for $\th
%%:=\min\{\frac14, \h, |\bar \h|\}$, 
%%
%\bal
%&|a_j-a|\le C L^{-\frac14 j}
%&&|b_j-b|\le C L^{-\frac14 j}
%\lb{scott1}\\
%&|m_{1,1,j}-\h^2  b|\le C L^{-\frac j4}
%&&|m_{2,2,j}-\bar \h^2 b|\le C L^{-\frac j4}
%\lb{scott2}
%\eal
%where $a=8\p^2 e^{-8\p \tilde c_E} \ln L$ and $b=2\ln L$.
%Besides, if $\h=-\bar \h= \frac12$,
%\bal
%&|m_{2,1,j}-\frac{\sqrt{ab}}2|\le C L^{-\frac j4}
%&&|m_{1,2,j}-\frac{\sqrt{ab}}2|\le C L^{-\frac j4};
%\lb{scott3}\eal 
%if instead $\h\neq \frac12$, then  $|m_{2,1,j}|, |m_{1,2,j}|\le C$.
%\end{lemma}
%The proof is in Appendix \ref{pscott}.
The calculation of \pref{scott1},
was already done  in \citep{Fa12}. Note that \pref{scott3} is only valid 
for $\h=\frac12$; for other values of $\h$ in $(0,1)$ 
we just need that $|m_{2,1,j}|, |m_{1,2,j}|$
are bounded, see below.  
Using \pref{slk} and \pref{scott1}, \pref{scott2}, 
the equation for the fractional charge renormalization constants \pref{sk} becomes 
\bal\lb{matrix}
\begin{pmatrix}
Z_{j+1}\\\\\bar Z_{j+1}
\end{pmatrix}
=&\begin{pmatrix} 
L^2e^{-4\p\h^2\G_j(0)}& 0\\\\
0&L^2e^{-4\p\bar \h^2\G_j(0)}
\end{pmatrix}
\notag\\
&\qquad\times
\begin{pmatrix} 
1 -\h^2 |q_j|+\tilde \MM_{1,1,j} & m_{1,2,j} z_j+
\MM_{1,2,j}
\\\\
m_{2,1,j} z_j+ \MM_{2,1,j}& 
1 -\bar \h^2 |q_j|+\tilde  \MM_{2,2,j} 
\end{pmatrix}
\begin{pmatrix}
Z_j \\\\ \bar Z_j
\end{pmatrix}
\eal
where 
\bal
&\tilde \MM_{1,1,j}=-(m_{1,1,j}s_j-\h^2|q_j|)+
\MM_{1,1,j},
\notag\\
&\tilde \MM_{2,2,j}=-(m_{2,2,j}s_j-\bar \h^2|q_j|)+
\MM_{2,2,j}.
\eal
Because of \pref{slk},  \pref{leibler0} and \pref{scott2},
for a $C\=C(L)$ and $m=1,2$,
\be\lb{pacm}
|\tilde \MM_{m,m, j}|\le  C \lft[|q_j| L^{-\frac14j} +   
\frac{\t |q_1|}{[1+|q_1| (j-1)]^{\frac32}}\rgt].
\ee
Let us consider two different cases,   
$|\h|=|\bar \h|$, or  $|\h|<|\bar \h|$; the  case 
 $|\bar\h|<|\h|$ gives the same formulas after interchanging 
$Z_j$ and $\h$ with $\bar Z_j$ and $-\bar\h$. 
\subsection{Case $|\h|=|\bar \h|$} 
In this case  $\h=-\bar \h=\frac12$ and \pref{scott3} holds.  
Therefore, if we introduce $Z^+_j:= Z_j+\bar Z_j$ 
and $Z^-_j=Z_j-\bar Z_j$: then
\bal\lb{iteration}
Z^+_{j+1}
&=L^{2}e^{-\p\G_j(0)}
\lft(1 -\frac14|q_j|+\frac12 q_j +\MM_{+,j}\rgt) 
Z^+_{j},
\notag\\
Z^-_{j+1}
&=L^{2}e^{-\p\G_j(0)}
\lft(1 -\frac14|q_j|-\frac12 q_j +\MM_{-,j}\rgt) 
Z^-_{j},
\eal
where 
\bal
&\MM_{+,j}:= \lft(m_{1,2,j} z_j - \frac 12 q_j\rgt)+\tilde \MM_{1,1,j} + \MM_{2,1,j}, 
\notag\\
&\MM_{-,j}:= -\lft(m_{1,2,j} z_j - \frac 12 q_j\rgt)+\tilde \MM_{1,1,j} - \MM_{2,1,j}.
\eal
It is easy to see that also $\MM_{+,j}$ and $\MM_{-,j}$
satisfy the bound \pref{pacm}.
In the physical case $z>0$, one has  
$|q_j|=q_j$ and then
\bal\lb{4:58}
Z^+_{j+1}
&=Z^+_{1} e^{2j\ln L-\p\G_{j,1}(0)+\frac14 \sum_{k=1}^j q_k +\sum_{k=1}^j  m_{+,k}},
\notag\\
Z^-_{j+1}
&=Z^-_{1} e^{2j\ln L-\p\G_{j,1}(0)-\frac34\sum_{k=1}^jq_k+\sum_{k=1}^j m_{-,k}}, 
\eal
where
\bal
&m_{+,j}:=\log\lft(1 +\frac14 q_j+\MM_{+,j}\rgt)-\frac14 q_j,
\notag\\
&m_{-,j}:=\log\lft(1 -\frac34 q_j+\MM_{-,j}\rgt)+\frac34 q_j.
\eal
Hence $|m_{+,j}|$ and $|m_{-,j}|$ satisfy 
a bound like \pref{pacm}. Therefore, for a $C\=C(L)$ and 
for three constants $\{\tilde c_{\s}:\s=0,\pm\}$ that are vanishing for $\t, |q_1|\to 0$,
one has:  for $\s=\pm$, $m_{\s,k}$ is summable and 
$$
\lft|\sum_{k=1}^j  m_{\s,k}-\tilde c_{\s}\rgt|
\le C \frac{|q_1| +\t}{\sqrt{1+|q_1|(j-1)}}; 
$$
while $q_k$ is not summable but 
$$
\lft|\sum_{k=1}^j q_k-\ln (1+ |q_1| j)- \tilde c_0\rgt|\le C|q_j|.
$$
Setting  $c_+:=\tilde c_++\tilde c_0$ and $c_-:=\tilde c_-+\tilde c_0$, 
from \pref{4:58} one finds the  explicit formula for $Z^+_j$ and $Z^-_j$
in  \pref{leibler1}.
\subsection{Case  $|\h|<|\bar \h|$}  
When $0<\h<\frac12$, we expect that 
the sequence  $(Z_j)$ dominates  $(\bar Z_j)$; therefore we recast \pref{matrix} 
 as
\be\lb{pezzini}
\begin{pmatrix}
Z_{j+1}\\\\\bar Z_{j+1}
\end{pmatrix}
=L^2e^{-4\p\h^2\G_j(0)-\h^2 |q_j|+m_{j}} 
\lft[\begin{pmatrix} 
1& 0\\\\
0&\ell_j 
\end{pmatrix}
+\begin{pmatrix} 
0
& m_{-,j}
\\\\
m_{+,j}&0 
\end{pmatrix}\rgt]
\begin{pmatrix}
Z_j \\\\ \bar Z_j
\end{pmatrix}
\ee
where
\bal
&m_{j}:=\ln\lft(1-\h^2 |q_j| +\tilde M_{1,1,j}\rgt)+\h^2 |q_j|,
\notag\\
&\ell_j:= e^{-4\p(\bar\h^2-\h^2)\G_j(0)+\h^2|q_j|-m_{j}} 
\lft(1-\bar \h^2|q_j| +\tilde M_{2,2,j}\rgt),
\notag\\
&m_{-,j}:=e^{\h^2|q_j|-m_{j}} \lft(m_{1,2,j} z_j +M_{1,2,j}\rgt),
\notag\\
&m_{+,j}:=e^{-4\p(\bar\h^2-\h^2)\G_j(0)+\h^2|q_j|-m_{j}} 
\lft(m_{2,1,j} z_j + M_{2,1,j}\rgt).
\eal
For a $C\=C(L)$ and $\s=\pm$ one has 
\bal\lb{rhea}
&|m_j|\le C \lft[|q_1| L^{-\frac14j} +\t \frac{|q_1|}{[1+|q_1|(j-1)]^{\frac32}}\rgt],
\notag\\
&|\ell_j|\le L^{-2(\bar \h^2-\h^2)}\lft[1+C|q_j| +CL^{-\frac j4}\rgt],
\notag\\
&|m_{\s,j}|\le C |q_j|.
\eal
The difference with the case $\h=-\bar\h$ is that 
in \pref{rhea} the coefficient $m_{j}$ is absolutely  summable in $j$,  
while $m_{+,j}$ and $m_{-,j}$ are not. This will be compensated
by the presence of several factors of $\ell_j<1$. 
For $z>0$, the solution of \pref{pezzini} is 
\bal\lb{pezzini2}
\begin{pmatrix}
Z_{j+1}\\\\\bar Z_{j+1}
\end{pmatrix}
&= L^{2j} e^{-4\p \h^2\G_{j,1}(0)-\h^2\sum_{k=1}^j  q_k +\sum_{k=1}^j m_k} Q(j,1)
\begin{pmatrix}
Z_1\\\\ 
\bar Z_1
\end{pmatrix}
\eal
where $Q(f,i)$ is a two-by-two matrix parametrized by two integers $f\ge i$:  
\be\lb{daniela}
Q(f,i)=\prod_{n=i}^f
\lft[\begin{pmatrix} 
1& 0\\\\
0&\ell_n
\end{pmatrix}
+\begin{pmatrix} 
0
& m_{-,n}
\\\\
m_{+,n}&0 
\end{pmatrix}\rgt].
\ee
From the definition \pref{daniela}, for any $1\le j_0\le j$,  we have the factorization   
$Q(j,1)=Q(j,j_0) Q(j_0-1,1)$. We will take advantage of it by choosing
a $j_0$ that is large when the difference $\h^2-\bar \h^2$ is small; and 
estimating  $Q(j,j_0)$ and  $Q(j_0-1,1)$ in  different ways. This will avoid that 
$L^{-1}$ (and hence $z$) be vanishing in the limit $\h\to\frac12$.     
\begin{lemma}
If $0\le z\le \frac14$, for every $0\le \h<\frac12$ 
there exist a scale $j_0\=j_0(\h)$ and 
a constant $C_0\=C_0(\h)$ such that:  
\bd
\item Estimates for the entries of $Q(j_0-1,1)$  are
\bal\lb{melyssa1}
&|Q(j_0-1,1)_{1,1}-1|\le C_0 |q_1|^2,
\notag\\
&|Q(j_0-1,1)_{1,2}-\sum_{d=1}^{j_0-1} \ell_{1}\cdots\ell_{d-1}m_{-,d}|\le C_0 |q_1|^3,
\notag\\
&|Q(j_0-1,1)_{2,1}|\le C_0 |q_1|,
\notag\\ 
&|Q(j_0-1,1)_{2,2}- \ell_1\cdots \ell_{j_0-1}|\le C_0 |q_1|^2.
\eal
\item For $m=1,2$ the limits $\bar Q_{1,m}(j_0):=\lim_{j\to \io}Q(j,j_0)_{1,m}$ exist
and 
\bal\lb{stat2}
&|\bar Q_{1,1}(j_0)-1|\le C_0 \sqrt{|q_1|},
\notag\\
&|\bar Q(j_0)_{1,2}-\sum_{d\ge j_0}\ell_{j_0}\cdots\ell_{d-1}m_{-,d}|\le C_0 |q_1|^{\frac32}; 
\eal
besides, estimates for the speed of convergence of the above limits are
\bal\lb{stat1}
&|Q(j,j_0)_{1,1}-\bar Q(j_0)_{1,1}|\le C_0 |q_j|,
&|Q(j,j_0)_{1,2}-\bar Q(j_0)_{1,2}|\le C_0 |q_j|,
\notag\\
&|Q(j,j_0)_{2,1}|\le C_0 |q_j|,\qquad 
&|Q(j,j_0)_{2,2}-\ell_{j_0}\cdots\ell_{j}|\le C_0  |q_j||q_1|.
\eal
\ed
In the limit $\h\to\frac12$ the constant $C_0(\h)$ is divergent.
\end{lemma}
\bpr
Consider \pref{daniela} and expand the product of the sum.
The interpretation of the result can be given in terms of the process 
of two ``states'', 
$A_1$ and $A_2$, 
and one  ``particle'': at each time $n=i, i+1,i+2,\ldots, f$ the particle 
can either 
hold in one of the two states or jump to the other.   The ``cost'' of staying in state 
$A_1$ or $A_2$ at time $n$ is $1$ and $\ell_n$, respectively. The cost of jumping 
form $A_1$ to $A_2$  or from $A_2$ to $A_1$ at time $n$ is 
$m_{+,n}$ and $m_{-,n}$ respectively. Let us denote 
$$
\sum_{u_1,d_1,\ldots,u_n, d_n}^{*[i,f]}
$$
the sum with constraint $i\le u_1< d_1<\cdots <u_n< d_n \le f$.
\bd
\item The entry $Q(f,i)_{1,1}$ is the sum of the  cost of all the patterns 
 that start and end at $A_1$, 
$$
Q(f,i)_{1,1}=1+ \sum_{n\ge 1}\sum_{u_1,d_1,\ldots,u_n, d_n}^{*[i,f]} \prod_{s=1}^n m_{+,u_s} 
\ell_{u_s+1}\cdots\ell_{d_s-1} m_{-,d_s}, 
$$
where:
$n$ is the number of intervals of time that the particle has spent in $A_2$; 
$u_s$'s are the times in which the particle jumps from $A_1$ to $A_2$, and 
$d_s$'s are the times in which the particle jumps from $A_2$ to $A_1$. 
As it is easy to see from \pref{rhea}, there exists a constant $C$ such that, 
if $0\le z\le \frac14$,  
then $\ell_n\le C$ and $|m_{+,u_s} m_{-,d_s}|\le C |q_1|^2$. 
Hence, for a $C_0\=C_0(j_0)$ (and divergent in the limit 
$j_0\to \io$) 
\be\lb{urra1a}
|Q(j_0-1,1)_{1,1}-1|\le C_0 |q_1|^2.
\ee
However,  if $j_0$ is larger that a $j'_0\=j'_0(\h)$, 
then one has the better bound 
$\ell_n\le L^{-(\h^2-\bar\h^2)}$ for every $n\ge j_0$.  
Therefore, if $d_s-1\ge u_s\ge j_0$, then  
$|m_{+,u_s} m_{-,d_s}|\le C |q_{u_s}|^2 $;  if also $d_s-u_s\ge 2$, 
then  $\ell_{u_s+1}\cdots\ell_{d_s-1}\le 
L^{-(\bar\h^2 -\h^2)(d_{s}-u_s-2)}$. In this way we obtain 
that $\lim_{j\to \io}Q(j,j_0)_{1,1}$ exists and 
\bal
\lft|Q(j,j_0)_{1,1}-1\rgt|
&\le \sum_{n\ge 1} C^n 
\lft(\sum_{u\ge j_0} q^2_{u} \sum_{w\ge0 }
L^{-(\bar\h^2 -\h^2)w}\rgt)^n
\notag\\
&\le \sum_{n\ge 1} \lft(\frac{|q_1|}{1+|q_1| j_0}
\frac{\tilde C}{1-L^{-(\bar\h^2 -\h^2)}}\rgt)^n .
\eal
Since $\frac{|q_1|}{1+|q_1| j_0}\le \sqrt{|q_1|} j_0^{-\frac12}$, 
if $j_0$ is larger than a $j''_{0}(\h)$ the term of the last series is 
bounded by $|q_1|^{\frac n2}$. Therefore, for $|q_1|<\frac14$,  such  series 
is summable and 
\be\lb{urra1b}
\lft|Q(j,j_0)_{1,1}-1\rgt|\le 2\sqrt{|q_1|}.
\ee
To study the speed of convergence of $\lim_{j\to \io}Q(j,j_0)_{1,1}$, 
consider an $f>j$ and the difference
\bal
Q(f,j_0)_{1,1}-Q(j,j_0)_{1,1}=\sum_{n\ge 1}\sum_{u_1,d_1,\ldots,u_n, d_n\atop d_n\ge j+1}^{*[j_0,f]} 
\prod_{s=1}^n m_{+,u_s} 
\ell_{u_s+1}\cdots\ell_{d_s-1} m_{-,d_s}.
\eal
As $|m_{+,u_n} \ell_{u_n+1}\cdots\ell_{d_n-1} m_{-,d_n}|\le C_0 |q_{d_n}|^2 
L^{-\frac12(\bar\h^2 -\h^2)(d_n-u_n-2)}$ for a $C_0\=C_0(\h)$, using a similar argument and 
the constraint $d_n\ge j+1$, 
$$
|Q(f,j_0)_{1,1}-Q(j,j_0)_{1,1}|\le C_0 |q_j|.
$$
\item The entry $Q(f,i)_{1,2}$ is the sum of the cost of all the patterns that start 
at $A_2$ and end at $A_1$, 
\bal
Q(f,i)_{1,2}
&=\sum_{d=i}^f \ell_{i}\cdots\ell_{d-1}m_{-,d}
\lft[1+ 
\sum_{n\ge 1}\sum_{u_1,d_1,\ldots,u_n,d_n}^{*[d+1,f]} \prod_{s=1}^{n-1} m_{+,u_s} 
\ell_{u_s+1}\cdots\ell_{d_s-1} m_{-,d_s}\rgt]
\notag\\
&=\sum_{d=i}^f \ell_{i}\cdots\ell_{d-1}m_{-,d} Q(f,d+1)_{1,1}.
\eal
Therefore, for constants $C$ and  $C_0\=C_0(j_0)$, we have
\bal\lb{urrah2a}
&\lft|Q(j_0-1,1)_{1,2}-\sum_{d=1}^{j_0-1} \ell_{1}\cdots\ell_{d-1}m_{-,d}\rgt|
\notag\\
&\le j_0 C^{j_0}|q_1|\sup_{d\le j_0-1} |Q(j_0-1,d)_{1,1}-1|\le C_0 |q_1|^3.
\eal
Besides, $\lim_{j\to \io}Q(j,j_0)_{1,2}$ exists and for a constant $C_1\=C_1(\h,j_0)$
\bal\lb{urrah2b}
&\lft|Q(j,j_0)_{1,2}-\sum_{d=j_0}^{j}
\ell_{j_0}\cdots\ell_{d-1}m_{-,d}\rgt|
\notag\\
&\le \frac{C|q_1|}{1-L^{-(\bar\h^2 -\h^2)}} 
\sup_{d\ge j_0} |Q(j,d)_{1,1}-1|\le C_1 |q_1|^{\frac32}.
\eal
To study the speed of convergence of the limit consider an $f\ge j+1$ and the difference 
\bal
Q(f,j_0)_{1,2}-Q(j,j_0)_{1,2}
&=\sum_{d=j_0}^j \ell_{i}\cdots\ell_{d-1}m_{-,d} \lft[Q(f,d+1)_{1,1}-Q(j,d+1)_{1,1}\rgt]
\notag\\
&+\sum_{d=j+1}^f \ell_{i}\cdots\ell_{d-1}m_{-,d} Q(f,d+1)_{1,1}.
\eal
Then 
\bal
\lft|Q(f,j_0)_{1,2}-Q(j,j_0)_{1,2}\rgt|
&\le \frac{C|q_1|}{1-L^{-(\bar\h^2 -\h^2)}} 
\sup_{j_0\le d\le j}\lft|Q(f,d+1)_{1,1}-Q(j,d+1)_{1,1}\rgt|
\notag\\
&+\frac{C|q_j|}{1-L^{-(\bar\h^2 -\h^2)}} \sup_{j+1\le d\le f}\lft| Q(f,d+1)_{1,1}\rgt|
\le C_0|q_j|.
\eal
\item The entry $Q(f,i)_{2,1}$ is the sum of the cost of all the patterns that start 
at $A_1$ and end at $A_2$,
\bal
Q(f,i)_{2,1}
&=\sum_{u=i}^f 
\lft[1+ 
\sum_{n\ge 1}\sum_{u_1,d_1,\ldots,u_n,d_n}^{*[i,u-1]} \prod_{s=1}^{n-1} m_{+,u_s} 
\ell_{u_s+1}\cdots\ell_{d_s-1} m_{-,d_s}\rgt]m_{+,u}
\ell_{u+1}\cdots\ell_{f}
\notag\\
&=\sum_{u=i}^f Q(u-1,i)_{1,1} \;m_{+,u}
\ell_{u+1}\cdots\ell_{f}.
\eal
For constants $C$ and  $C_0\=C_0(j_0)$, we have 
\be\lb{urrah3a}
|Q(j_0-1,i)_{2,1}|\le j_0C^{j_0} |q_1| \sup_{u\le j_0-1} |Q(u,i)_{1,1}|
\le C_0 |q_1|.
\ee
If $j-1\ge u\ge j_0$,  for a constant $C_1\=C_1(\h)$ we have 
$|m_{+,u}\ell_{u+1}\cdots\ell_{j}|\le C_1 |q_j| L^{-\frac12(\bar \h^2-\h^2)(j-u)}$; hence
\be\lb{urrah3}
|Q(j,j_0)_{2,1}|\le 2C_1|q_j|.
\ee
\item The entry $Q(f,i)_{2,2}$ is the total cost of all the patterns that start 
and end at $A_2$: 
\bal
Q(f,i)_{2,2}
&=\ell_{i}\cdots\ell_{f}
\notag\\
&+\sum_{d<u=i}^f \ell_{i}\cdots\ell_{d-1}m_{-,d}\; Q(u-1, d+1)_{1,1}\;
m_{+,u}
\ell_{u+1}\cdots\ell_{f}.
\eal
For $C_0\=C_0(j_0)$, 
\be\lb{urrah4a}
|Q(j_0-1,i)_{2,2}-\ell_{1}\cdots\ell_{j_0-1}|\le C_0|q_1|^2\;;
\ee
besides
\be\lb{urrah4b}
|Q(j,j_0)_{2,2}-\ell_{j_0}\cdots\ell_{j}|\le C_0|q_1||q_j|.
\ee
\ed
From these formulas, one obtains
\pref{melyssa1}, \pref{stat2} and \pref{stat1}. 
\epr
Combining the  bounds of this Lemma, we obtain:
\bal\lb{am0}
&Q(j,1)_{1,1}=e^{\tilde c_1+ \tilde r_{1,j}},  
\notag\\
&Q(j,1)_{1,2}=\sum_{d\ge 1} \ell_1\cdots \ell_{d-1} m_{-,d}e^{\tilde c_2+\tilde s_{1,j}},
\notag\\
&Q(j,1)_{2,1}=\tilde r_{2,j,},
\notag\\
&Q(j,1)_{2,2}=e^{-(\bar\h^2-\h^2)[4\p\G_{j,1}(0)+\sum_{k=1}^j q_k]+ \tilde c_3 + \tilde r_{3,j}}
+ \tilde r_{4,j},
\eal
where: $|\tilde c_{1}|\le C_0 \sqrt{|q_1|}$;  
$|\tilde c_{2}|, |\tilde c_{3}|\le C_0 |q_1|$;  
$|\tilde r_{1}|, |\tilde r_{2}|, |\tilde r_{3}|\le C_0 |q_j|$; 
$|\tilde s_{1,j}|\le C_0 (1+|q_1|j)^{-1}$.  
Plugging \pref{am0} into \pref{pezzini2}, one 
obtains point 2. of Theorem \ref{leiblerbis}.

Finally, note that the  lowest order in $z$ of $\lim_{j\to \io}Q(j,1)_{1,2}$ is 
\bal\lb{3:13pm}
&z\sum_{j\ge 1} e^{-4\p (\bar\h^2-\h^2)\G_{j-1,1}(0)} m_{1,2,j}
\notag\\
&=z e^{4\p (\bar\h^2-\h^2)\G_{0}(0)}
\lft(\sum_{j\ge 0} e^{-4\p (\bar\h^2-\h^2)\G_{j-1,0}(0)} m_{1,2,j}
- m_{1,2,0}\rgt)
\notag\\
&:=z e^{4\p (\bar\h^2-\h^2)\G_{0}(0)}
\lft(c(\h)
- m_{1,2,0}\rgt)
\eal
From the  definition of $m_{1,2,j}$ in 
\pref{lw2} we find 
\bal
&c(\h)=\sum_{n\ge 0}e^{-4\p (\bar\h^2-\h^2)\G_{\io,0}(0)}\sum_{y\in \ZZZ^2}\bar R^{(\io)}_{1,c,n}(y)
\notag\\
&=
\sum_{n\ge 0}L^{-2n}e^{-4\p (\bar\h^2-\h^2)\G_{n-1,0}(0)}\sum_{y\in \ZZZ^2}
e^{-\bar \h \a^2 \G_{\io, n+1}(y|0)}e^{\bar \h\a^2\G_n(0)} \lft(e^{-\bar \h \a^2\G_n(y)}-1\rgt)
\eal
where the series in $n$ is summable and strictly positive, as one can verify by the 
inequality  $e^x-1\ge x+\frac12 x^2 e^{-x_0}$,  valid for any $x:|x|\le x_0$, 
and by the fact that $-\bar \h> 0$,  $|\G_n(y)|\le \G_n(0)$, 
$$
\sum_{y\in \ZZZ^2}\G_n(y)\ge 0, \qquad \sum_{y\in \ZZZ^2}\G_n(y)^2>0.
$$
\section{Exact asymptotic formulas}

\subsection{Proof of Lemma \ref{scott} and Lemma \ref{t3.1b}}
\lb{pscott}
A key result is the following Lemma, in which we introduce a 
continuous approximation of the covariances $\G_j$ which has a simpler
scaling transformation. 
\begin{lemma}
\lb{seiberg} 
Consider the  set of  ``continuous covariances'' 
$\tilde \G_j(x)$, $j=0,1, \ldots R-1$,  
defined for $x\in \RRR^2$ as 
\be\lb{seiberg0}
\tilde \G_j(x):=\int\frac{d^2p}{(2\p)^2}\; e^{i x p}\;  \frac{u(L^j p)-u(L^{j+1} p)}{p^2}
\ee
where $u(p)$ is a differentiable even function such that
$u(L^j p)-u(L^{j+1} p)\ge 0$ for every $j$ and 
\be\lb{seiberg00}
\qquad
u(0)=1,\qquad |u(p)|\le \frac{C}{1+|p|^4} .
\ee 
There exists a special choice of $u$ and a constant $C>0$ such that,  
for every $x\in \ZZZ^2$,
\mp{forse ci vuole la derivata discreta anche per $\tilde \G_j$} 
\be\lb{seiberg3}
|\G_j(x)-\tilde \G_j(x)|\le C L^{-\frac14j},
\qquad
|\dpr^\m \G_j(x)-\tilde \G_j^{,\m}(x)|
\le C L^{-\frac54j},
\ee
where the upper label $^{,\m}$ indicates the continuous derivative (as opposed
to $\dpr^\m$ that is the lattice one).
\end{lemma}
The proof is in Appendix A.3. of \citep{Fa12}.
\pref{seiberg0} and \pref{seiberg0} have
important consequences: first, 
for every $x\in \RRR^2$
\be\lb{seiberg1}
\tilde \G_j(x)=\tilde \G_0(L^{-j}x);
\ee
second, 
\be\lb{seiberg4}
\tilde \G_j(0)= \int\frac{d^2p}{(2\p)^2}\; \;  \frac{u(L^j p)-u(L^{j+1} p)}{p^2}
=\frac{1}{2\p} \ln L;
\ee
finally, $\tilde \G_{\io,0}(x|0)$ is a differentiable function and, 
asymptotically for large $|x|$,
\be\lb{seiberg2}
\tilde \G_{\io,0}(x|0):=\sum_{j=0}^\io
\lft[\tilde \G_j(x)-\tilde \G_j(0)\rgt]=-\frac1{2\p} \ln |x| + \tilde c_E + o(1)
\ee
where $o(1)$ is a vanishing term in the limit $|x|\to \io$ and $\tilde c_E$ is a constant.

Consider  the coefficient $a_j$ in \pref{dfm2}.
Let $\tilde R^{(j-1)}_{0,b,n}$ and $\tilde \G_j(0|y)$ be the same function as
$R^{(j-1)}_{0,b,n}$ and $\G_j(0|y)$, respectively, 
but with $\tilde \G_j(x)$ in place of $\G_j(x)$ for any $j$. 
Using \pref{seiberg3} we have
\bal
\sum_{y\in \ZZZ^2} |y|^2 
\lft|R^{(j)}_{0,b,j}-\tilde R^{(j)}_{0,b,j}(y)\rgt|
&\le C(L) L^{-\frac14 j}
\notag\\
\sum_{y\in \ZZZ^2} |y|^2 \lft|R^{(j)}_{0,b,n}(y)-R^{(j-1)}_{0,b,n}(y)-\tilde R^{(j)}_{0,b,n}(y)+
\tilde R^{(j-1)}_{0,b,n}(y)\rgt|
&\le C(L) L^{-\frac j4 }L^{-\frac34 (j-n)}
\eal
therefore in the definition of $a_j$ we can replace $R^{(j-1)}_{0,b,n}$ with 
$\tilde R^{(j-1)}_{0,b,n}$ up to an error $C L^{-\frac j4}$. Besides, 
$$
\lft|\sum_{y\in \ZZZ^2} |y|^2 
\tilde R^{(j)}_{0,b,n}(y)-\int\! d^2y \;  |y|^2 
\tilde R^{(j)}_{0,b,n}(y)\rgt|\le C(L) L^{-j};
$$
therefore in the formula for $a_j$ replacing the sum with an integral 
generates and error not larger than $CL^{-\th j}$ for a $\th<1$. 
In conclusion, an equivalent formula for 
$a_j$ is, up to  an $O(L^{-\frac{j}{4}})$ error, 
\bal\lb{aup}
{\a^2\over 2} \int\!d^2y\; y^2
\lft[\sum_{n=0}^j \tilde R^{(j)}_{0,b,n}(y)
-\sum_{n=0}^{j-1} \tilde R^{(j-1)}_{0,b,n}(y)\rgt].
\eal
We now take advantage of the exact scale transformation \pref{seiberg1}.
We have $\tilde R^{(j-1)}_{n}(y)=L^4 \tilde R^{(j)}_{n+1}(yL)$; hence 
the two sums in \pref{aup}  cancel each others 
almost completely, and \pref{aup} becomes
\bal
{\a^2\over 2} \int\!d^2y\; y^2
\tilde R^{(j)}_{0,b,0}(y)
&={\a^2\over 2} \int\!d^2y\; y^2
e^{-\a^2\tilde \G_{\io,1}(0|y)}
e^{-\a^2\tilde \G_{0}(0)}
\lft(e^{\a^2\tilde \G_{0}(y)}-1\rgt)
+O(L^{-j})
\notag\\
&=\frac{\a^2}{2} \int\!\frac{d^2y}{y^2}\; \lft[w(y)-
w(y L^{-1})L^{4-\frac{\a^2}{2\p}} \rgt]
+O(L^{-j})
\eal
for 
$w(y)=y^4e^{-\a^2\tilde \G_{\io,0}(0|y)}$; a new
$O(L^{-j})$  error in the first line 
is due to the replacement of $e^{-\a^2\tilde \G_{j,1}(0|y)}$ with 
$e^{-\a^2\tilde \G_{\io,1}(0|y)}$.  At $\a^2=8 \p$, the last integral can 
be exactly computed only using the differentiability of $w(y)$
and the boundary values $w(0)=0$ and
$\lim_{y\to\io}w(y)=e^{-8\p \tilde c_E}$, see \pref{seiberg2}. 
This proves the first of \pref{scott1}.

Now consider the coefficient $m_{2,1,j}$ in \pref{lw2}. 
Arguing as done for $a_j$, 
up to $O(L^{-\th j})$ corrections, 
it is given by 
\be\lb{aup2}
\int\!d^2 y \; 
\lft[\sum_{n=0}^{j} \tilde R^{(j)}_{1,c,n}(y)e^{- \frac{\a^2}{2} (2\h-1)\tilde \G_j(0)}-
\sum_{n=0}^{j-1} \tilde R^{(j-1)}_{1,c,n}(y) \rgt]
\ee 
where the formula for $\tilde R^{(j-1)}_{1,c,n}(y)$ is obtained from the formula for 
$R^{(j-1)}_{1,c,n}(y)$ by replacing $\G_j(x)$ with $\tilde \G_j(x)$. 
In the case $\h=-\bar \h =\frac12$, by means of the exact scaling 
$\tilde R^{(j-1)}_{1,c,n}(y)=L^2 \tilde R^{(j)}_{1,c,n+1}(yL)$, \pref{aup2} becomes
\bal
\int\!d^2 y \; 
\tilde R^{(j)}_{1,c,0}(y)
&=
\int\!d^2 y \;e^{-\frac{\a^2}2 \tilde \G_{\io,1}(0|y)} e^{-\frac{\a^2}2 \tilde \G_{0}(0)} 
\lft(e^{\frac{\a^2}2 \tilde \G_{0}(y)}-1\rgt)+ O(L^{-j}) 
\notag\\
&=\int\!\frac{d^2y}{y^2}\; \lft[\sqrt{w(y)}-
\sqrt{w(y L^{-1})}L^{2-\frac{\a^2}{4\p}} \rgt]
+O(L^{-j})
\eal
where $w(y)$ is the same function introduced for $a_j$. At $\a^2=8\p$ the last integral 
can be exactly computed only using the differentiability of $\sqrt{w(y)}$ for $y\neq 0$ 
and the boundary values $\sqrt{w(0)}=0$ and
$\lim_{y\to\io}\sqrt{w(y)}=e^{-4\p \tilde c_E}$. This proves the first of \pref{scott3}; the 
second is also proven because at $\h=-\h=\frac12$ one has $m_{1,2,j}=m_{2,1,j}$.    
 
Finally, consider the coefficient $b_j$ in \pref{dfm2} and $m_{1,1,j}$, $m_{2,2,j}$ in \pref{lw2}. 
With the same argument used for $a_j$, an equivalent formula the last two 
coefficients, up to an $\h^2$ or $\bar \h^2$  prefactor, is   
\bal\lb{aba}
&\frac{\a^2}{2}\sum_{\m=\hat e}\int\! d^2y\;
\lft[\lft(\tilde \G_{j}^{,\m}(y)\rgt)^2 + 2\tilde \G_{j-1,0}^{,\m}(y)\tilde \G_{j}^{,\m}(y)\rgt]
+O(L^{-\frac {j}{4}})
\notag\\
&=\frac{\a^2}{2}\sum_{\m=\hat e}\int\! d^2y\;
\lft[\lft(\tilde \G_{j,0}^{,\m}(y)\rgt)^2 -
\lft(\tilde \G_{j-1,0}^{,\m}(y)\rgt)^2)\rgt]
+O(L^{-\frac {j}{4}})
\eal
At $\a^2=8\p$, \pref{aba} is also an equivalent formula for $b_j$. 
As $\tilde \G_{j-1,0}^{,\m}(y)=L \tilde \G_{j,1}^{,\m}(yL)$,
the last integral in \pref{aba} becomes
\bal
&\frac{\a^2}{2}\sum_{\m=\hat e}\int\! d^2y\;
\lft[\lft(\tilde \G_{j,0}^{,\m}(y)\rgt)^2 -
\lft(\tilde \G_{j,1}^{,\m}(y)\rgt)^2\rgt]
%=\frac{\a^2}{2}\sum_{\m=\hat e}
%\int\! d^2y\;
%\lft[2\tilde \G_{j,1}^{,\m}(y)\tilde \G_{0}^{,\m}(y)
%+(\tilde\G_{0}^{,\m})^2(y)\rgt]
\notag\\
&=
\frac{\a^2}{2}\int\!\frac{d^2p}{(2\p)^2}
\frac{[u( p)]^2-[u(L p)]^2}{p^2}
+
\a^2\int\!\frac{d^2p}{(2\p)^2}
\frac{[u(p)]-[u(L p)]}{p^2} u(L^{j+1}p)
\eal
In the last line, the former integral can be exactly computed while the latter, 
using the boundedness of the derivatives of $u$ is $O(L^{-j})$. 
This proves \pref{scott2} and the second of \pref{scott1}.

\subsection{Proof of Theorem \ref{finale}}
From \pref{hofern} we have
\be\lb{albanese}
w^-_{2,a,R}(y)
=\frac12 \sum_{n=0}^{R-1}
Z_n^2L^{-4n}
e^{\h^2\a^2\G_{R,n+1}(y|0)}
e^{-\h^2\a^2\G_n(0)}\lft(e^{\h^2\a^2 \G_n(y)}-1\rgt).
\ee 
Use the inequality 
$Z_n^2 L^{-4n}\le C_\th L^{-\th n}$ for a $\th< \min\{4\h^2,4\bar \h^2, 1\}$, 
to replace  the 
function $\G_{R,n+1}(y|0)$ with $\G_{\io,n+1}(y|0)$ up to an $O(L^{-R\th})$ 
error term. Note that 
each $y$ can be uniquely written as  
$y=L^{n_0}\t$ for $|\t|\in [1,L)$ and an integer $n_0$;
then $e^{\h^2\a^2 \G_n(y)}-1=0$ every time 
$n\le n_0-1$ so that, in \pref{albanese}, one can actually start the sum 
from $n=n_0$. Accordingly,  a formula for  $\lim_{R\to\io}w^-_{2,a,R}(y)$ is 
\bal\lb{albanese1}
w^-_{2,a}(y)=\frac12 \sum_{n\ge n_0}
Z_n^2L^{-4n}
e^{\h^2\a^2 \G_{\io,n+1}(y|0)}
e^{-\h^2\a^2 \G_n(0)}\lft(e^{\h^2\a^2 \G_n(y)}-1\rgt) .
\eal
Using the same  argument, a formula for  $w^-_{2,\bar a}(y)=\lim_{R\to\io}\bar w^-_{2,a,R}(y)$ 
is given by \pref{albanese1} after replacing $Z_j$ and $\h$ with $\bar Z_j$ and $-\bar \h$.
Let us consider three different cases.

\subsubsection{Case $0<\h<\frac12$.}
It is convenient to write  \pref{albanese1} 
as
the difference of two convergent series
\bal
\lb{epifanio}
&\frac12 \sum_{n\ge n_0}
Z_n^2L^{-4n}e^{\h^2\a^2 \G_{\io,n}(y|0)}-\frac12 \sum_{n\ge n_0}
Z_n^2L^{-4n}e^{-\h^2\a^2 \G_n(0)}e^{\h^2\a^2  \G_{\io,n+1}(y|0)}.
\eal
By replacing in the second series the factor $Z_n^2L^{-4n}e^{-\h^2\a^2 \G_n(0)}$ 
with the almost identical
factor $Z_{n+1}^2L^{-4(n+1)}$, each term of the latter series cancels a term in the former, 
so that only the term for $n=n_0$ in the first series survives; besides, by definition of $n_0$
we have $\G_{\io,n_0}(y|0)= \G_{\io,0}(y|0)+ \G_{n_0-1,0}(0)$. Hence %\pref{epifanio} becomes
\bal\lb{zalone0}
w^-_{2,a}(y)=&\frac12 
Z_{n_0}^2L^{-4n_0}e^{\h^2\a^2 \G_{n_0-1,0}(0)}e^{\h^2\a^2 \G_{\io,0}(y|0)}
\notag\\
&-\frac12 \sum_{n=n_0}^{\io}
\lft(Z_n^2L^{-4n}e^{-\h^2\a^2 \G_n(0)}-Z_{n+1}^2L^{-4(n+1)}\rgt)e^{\h^2\a^2 \G_{\io,n+1}(y|0)}.
\eal 
The first term in \pref{zalone0} is the leading one. Indeed, 
from  \pref{leibler2} we have 
\bal\lb{zalone}
&Z^2_{j}L^{-4j}
=e^{-\h^2\a^2 \G_{j-1,0}(0)}[1+|q_1| (j-1)]^{-2\h^2}e^{\tilde c_1+\tilde r_{1,j}},
\notag\\
&\bar Z^2_{j}L^{-4j}
=e^{-\h^2\a^2 \G_{j-1,0}(0)}[1+|q_1| (j-1)]^{-2\h^2} r_{2,j}, 
\eal
where  $|r_{m,j}|\le C(1+|q_1|j|)^{-\frac12}$ for any $m=1,2$ and 
$\tilde c_1$, $\tilde c_2$ are vanishing in the limit $z\to 0$.
Therefore   the first term of \pref{zalone0} is 
\be\lb{zalone0C}
\frac{e^{8\p \h^2 c_E}}{2|y|^{4\h^2} } 
(1+ |q_1|\log_L |y|)^{-2\h^2}
e^{\tilde c}(1+o(1)),
\ee
for $o(1)$ a term bounded by $C(\log|y|)^{-\frac12}$.
The second term in \pref{zalone0}
is  subleading by a factor $(\log_L|y|)^{-\frac12}$ at least. 
Indeed from \pref{6:52pm} we also  find 
\bal
&|Z^2_n L^{-4n} e^{-\h^2\a^2 \G_n(0)}-Z^2_{n+1} L^{-4(n+1)}|
\notag\\
&\le 
Z^2_{n+1} L^{-4(n+1)}\lft[\lft(\frac{1+|q_1|n}{1+|q_1| (n-1)}\rgt)^{2\h^2}
\lft(\frac{1+\tilde s_{1,n}}{1+\tilde s_{1,n+1}}\rgt)-1\rgt]
\notag\\
&\le 
C \frac{L^{-4\h^2 n_0}}{(1+|q_1| n_0)^{2\h^2+\frac12}}L^{-4\h^2(n-n_0)}.
\eal
Summing  over $n\ge n_0$, one obtains the bound
$ C|y|^{-4\h^2}(1+|q_1| n_0)^{-2\h^2-\frac12}$, which is subleading
with respect to \pref{zalone0C}. 
Instead, for  $w_{2,\bar a}(y)$
the method used above does not work; however,  we can provide 
an upper bound that shows that $w_{2,\bar a}(y)$ is subleading 
with respect to $w_{2,a}(y)$: as 
$$
e^{\bar \h^2\a^2 \G_{\io,n+1}(y|0)}
e^{-\bar \h^2\a^2 \G_n(0)}\lft|e^{\bar \h^2\a^2 \G_n(y)}-1\rgt|\le C \bar \h^2\a^2 |\G_n(y)|, 
$$
 $w_{2,a}(y)$ is bounded by 
$$
C \bar \h^2\a^2 \sum_{n\ge n_0}
\bar Z^2_{n} L^{-4n}|\G_n(y)|\le C \frac{L^{-4\h^2 n_0}}{(1+|q_1| n_0)^{2\h^2+\frac12}}. 
$$
Next consider $w^-_{2,b,R}$ in \pref{hofern}.
For any $0\le \th <1$ and a corresponding constant $C_\th\=C_\th(\h)$,  
we have the bound
\bal
|w^-_{2,b,R}(y)| 
&\le \h\bar \h\a^2 C\sum_{n=1}^{R-1} Z_n L^{-2n} \bar Z_n L^{-2n}
e^{-\frac{\a^2}{2}\G_{R-1,n+1}(0)}
\notag\\
&\le \h\bar \h\a^2 C_\th L^{-4\h^2\th R}.
\eal
Hence $\lim_{R\to\io}w^-_{2,b,R}(y)=0$. 
Finally, consider $w^{-}_{2,c}(y)$.  From \pref{hofern} we find
\bal
|w^{-}_{2,c}(y)|&\le
\sum_{k\ge 0} L^{-4k} e^{-L^{-k} |y|}
\sum_{n\ge k}  2^{-(n-k)} 
\notag\\
&\times \lft\{Z_k^2 \frac12\sum_{\s=\pm1}\sum_{X\in \SS_n}^{X\ni 0}
\EEE_j\lft[ |\hat K_{2,n}^{(a,k)}\lft(0,\z, X,0, \s, y, -\s\rgt)|\rgt]
\rgt.
\notag\\
&\quad+\lft.
\bar Z^2_k \frac12\sum_{\s=\pm1}\sum_{X\in \SS_n}^{X\ni 0}
\EEE_j\lft[ 
|\hat K_{2,n}^{(\bar a,k)}\lft(0,\z, X,0, \s, y, -\s\rgt)|
\rgt]
\rgt.
\notag\\
&\hskip-2em+\lft.
Z_k\bar Z_k \frac12\sum_{\s=\pm1}\sum_{X\in \SS_n}^{X\ni 0}
\EEE_j\lft[ 
|\hat K_{2,n}^{(b,k)}\lft(\s,\z, X, 0, \s, y,-\s\rgt)|
\rgt]\rgt\}
\notag\\
&\hskip-2em\le SA^{-\frac12}k^*_s(\sqrt A,1/2)
\sum_{k\ge 0} L^{-4k} e^{-L^{-k} |y|}
\sum_{n\ge k}  2^{-(n-k)}   
\notag\\
&\hskip-2em\times \lft\{Z_k^2 
\|K_{2,n}^{(a,k)}\|_{2,h,T_j}+
\bar Z^2_k  
\|K_{2,n}^{(\bar a,k)}\|_{2,h,T_j}+
Z_k \bar Z_k  
\|K_{2,n}^{(b,k)}\|_{2,h,T_j}\rgt\}.
\eal
Hence, from \pref{maldacena1},  we obtain
\bal
|w^{-}_{2,c}(y)|&\le
\frac{|q_1|C}{|y|^{4\h^2}(1+|q_1|\log_L |y|)^{2\h^2+1}}. 
\eal

\subsubsection{Case $\frac12<\h<1$.}
The fundamental difference with the previous case is in the formula for the 
renormalization constants. 
From  \pref{leibler2} we have 
\bal\lb{checco}
&\bar Z^2_{j}L^{-4j}e^{\bar \h^2\a^2 \G_{j-1,0}(0)}
=[1+|q_1| (j-1)]^{-2\bar \h^2}c(\h)^2z^2e^{\tilde c_1+\tilde r_{1,j}},
\notag\\
&Z^2_{j}L^{-4j}e^{\bar \h^2\a^2 \G_{j-1,0}(0)}
=[1+|q_1| (j-1)]^{-2\bar \h^2}e^{\tilde c_2}\tilde r_{2,j}, 
\eal 
where $c(\h)$ is the positive constant in Theorem \ref{leiblerbis}. 
Now, proceeding with $w_{2,\bar a}(y)$ with the same method that 
in the previous section was used for $w_{2, a}(y)$ we obtain the formula 
\be\lb{zalone0D}
z^2 c(\h)^2\frac{e^{8\p \bar \h^2 c_E}}{2|y|^{4\bar \h^2} } 
(1+ |q_1|\log_L |y|)^{-2\bar \h^2}
e^{\tilde c}(1+o(1)),
\ee
for $o(1)$ a term bounded by $C(\log|y|)^{-\frac12}$. 
Conversely, proceeding with $w_{2,a}(y)$ with the same method that 
in the previous section was used for $w_{2,\bar a}(y)$ we obtain the bound 
$$
C \h^2\a^2 \sum_{n\ge n_0}
Z^2_{n} L^{-4n}|\G_n(y)|\le C \frac{L^{-4\bar \h^2 n_0}}{(1+|q_1| n_0)^{2\bar \h^2+\frac12}}, 
$$
which is subleading with respect to \pref{zalone0D}.
Finally, with the same arguments of the previous section, $w_{2,a}(y)=0$ and
\bal
|w^{-}_{2,c}(y)|&\le
\frac{|q_1|C}{|y|^{4\bar \h^2}(1+|q_1|\log_L |y|)^{2\bar \h^2+1}}. 
\eal
\subsubsection{Case $\h=\frac12$} 
From  \pref{leibler1},  we have
\bal\lb{6:52pm}
&Z_{j}=\frac{Z_{j}^++Z_{j}^-}{2}=\frac12 L^{2j} e^{-\p \G_{j-1,0}(0)}(1+|q_1|(j-1))^{\frac14} 
e^{\tilde c+\tilde s_{j}}
\eal
where $\tilde c$ vanishes for $z\to0$ and 
$|\tilde s_{j}|\le \frac{C}{\sqrt{1+|q_1|j}}$. Hence
\bal\lb{zalone2}
Z^2_{n_0}L^{-4n_0}e^{\h^2\a^2 \G_{n_0-1,0}(0)}
=\frac14 (1+|q_1| (n_0-1))^{\frac12}e^{2\tilde c+2\tilde s_{n_0}}
\eal
and the formula for $w^-_{2,a}(y)$ is 
\be\lb{zalone0D}
\frac{e^{2\p c_E}}{8|y| } 
(1+ |q_1|\log_L |y|)^{\frac12}
e^{\tilde c}(1+o(1)).
\ee
Now consider $w^-_{2,\bar a}(y)$. 
Since a formula for $\bar Z_j=(Z^+_j-Z^-_j)/2$ is again given by \pref{6:52pm}, 
--but for numerically different $\tilde c$ and $\tilde s_j$--  
also  the formula for $w^-_{2,\bar a}(y)$
is \pref{zalone0D}.
Finally, consider $w^{-}_{2,c}(y)$.
\bal
|w^{-}_{2,c}(y)|&\le
\frac{|q_1|C}{|y|(1+|q_1|\log_L |y|)^{\frac12}}.
\eal
This completes the proof of point 1 of Theorem \ref{finalevero}. 
\subsection{Proof of Theorem \ref{finale}}
For the sake of brevity in this section we denote $\tilde \EEE_R$ the limiting
expectation $\lim_{m\to 0}\EEE_R$. 
Let us consider the last term of \pref{finalcorr}
\bal\lb{farmers}
&e^{-\d E_R|\L|}
\tilde \EEE_R\lft[\frac{\dpr^2 K_{2,R}(\F)}{\dpr J_{x,+}\dpr J_{0,-}}\rgt]_{J=0}
=e^{-\d E_R|\L|}
\sum_{k=0}^R 2^{-(R-k)} e^{-L^{-k}|x|} L^{-4k}
\notag\\
&\qquad\times
\Bigg\{Z^2_k\tilde\EEE_R\lft[K_{2,R}^{(a,k)}\rgt]
+\bar Z^2_k\tilde\EEE_R\lft[K_{2,R}^{(\bar a,k)}\rgt]
+Z_k\bar Z_k\tilde\EEE_R\lft[K_{2,R}^{(b,k)}\rgt]\Bigg\}
\eal
where we suppressed in  $K_{2,R}^{(\d,k)}$ the dependence in  $(\z,\L, x,+,0,-)$. 
Using \pref{r2},\pref{ex6.53bis} and \pref{maldacena1}, 
an upper bound for the absolute value of \pref{farmers} is 
$$
C'A^{-1}e^{C|q_R|} |q_R|
\sum_{k=0}^R 2^{-(R-k)} e^{-L^{-k}|x|} \lft(L^{-4k}Z_k^2 + L^{-4k}\bar Z_k^2\rgt) \frac{|q_k|}{|q_R|}.
$$
In the limit $R\to \io$, this bound 
is vanishing: indeed $|q_R|\to 0$ while the sum remains bounded by  the fact that
$Z_k L^{-2k}, \bar Z_k L^{-2k}\le C$. 
Next, consider the first term in \pref{finalcorr}: expanding the product inside the 
square brackets, one obtains four terms. Since they can all be studied in similar way, 
let us consider one of them:
\bal\lb{farmers1}
&e^{-\d E_R|\L|}\tilde\EEE_R\lft[e^{V_{0,R}(\z) + W_{0,R}(\z)} \frac{\dpr V_{1,R}(\F)}{\dpr J_{x,+}}
\frac{\dpr V_{1,R}(\F)}{\dpr J_{0,-}}\rgt]_{J=0}
\notag\\
&=
e^{-\d E_R|\L|}L^{-4R}\tilde\EEE_R\lft[e^{V_{0,R}(\z) + W_{0,R}(\z)} 
\lft(Z_R^2 e^{i\h \a(\z_x-\z_0)}+\bar Z_R^2 e^{i\bar \h \a(\z_x-\z_0)}\rgt)\rgt]
\notag\\
&+
e^{-\d E_R|\L|}L^{-4R}\tilde\EEE_R\lft[e^{V_{0,R}(\z) + W_{0,R}(\z)} 
Z_R \bar Z_R \lft(e^{i \a (\h\z_x-\bar\h\z_0)}+e^{i \a (\bar\h\z_x-\h\z_0)}\rgt)\rgt]
\eal
It is easy  to see that, for a $C\=C(\a)$ and for $z$ smaller than a $z(L,\a)$, 
$$
\|V_{0,R}(\z,\L)\|_{h,T_R(\L)}\le C |q_R|\lft(1+\max_{n=1,2}\|\nabla_R^n \z\|^2_{L^\io(\L)}\rgt)
\le C |q_R|+\frac12\ln G^{\rm str}(\z,\L),
$$
$$
\|W_{0,R}(\z,\L)\|_{h,T_R(\L)}\le C |q_R|^2\lft(1+\max_{n=1,2}\|\nabla_R^n \z\|^2_{L^\io(\L)}\rgt)
\le C |q_R|^2+\frac12\ln G^{\rm str}(\z,\L),
$$
$$
\|e^{i \a_1\z_x}\|_{h,T_R(\L)}\le e^{h|\a|};
$$
therefore, for any $\a_1, \a_2\in \RRR$,  
\bal
&\lft|e^{V_{0,R}(\z) + W_{0,R}(\z)} e^{i (\a_1\z_x+\a_2\z_0)}\rgt|
\notag\\
&\le\|e^{V_{0,R}(\z) + W_{0,R}(\z)}\|_{h,T_R(\L)} \;\|e^{i \a_1\z_x}\|_{h,T_R(\L)}\; 
\|e^{i \a_2\z_0}\|_{h,T_R(\L)} 
\notag\\
&\le e^{h|\a_1|+h|\a_2|}e^{2C |q_R|}G^{\rm str}(\z,\L).
\eal
In conclusion, the absolute value of \pref{farmers1} can be bounded by 
$$
C(\a) e^{C|q_R|}\lft(L^{-4R}Z_R^2 + L^{-4R}\bar Z_R^2\rgt).
$$
In the limit $R\to \io$ this  bound is vanishing since  
$|q_R|, L^{-2R} Z_R,L^{-2R} \bar Z_R \to 0$. 
The remaining term of \pref{farmers1} is the one that gives the 
right hand side of \pref{canaletto}. To prove this fact, we need 
to study the 
difference
\bal\lb{am}
&e^{-\d E_R|\L|}
\tilde\EEE_R\lft[e^{V_{0,R}(\z) + W_{0,R}(\z)} 
\frac{\dpr^2 W_{2,R}(\F)}{\dpr J_{x,+}\dpr J_{0,-}}\rgt]_{J=0}
%\notag\\
%&\qquad
-2\lft[w_{2,a,R}^-(x)+ \bar w_{2,a,R}^-(x) +w_{2,c,R}^-(x)\rgt]
\notag\\
&=2w_{2,a,R}^-(x)
\lft\{e^{-\d E_R|\L|}
\tilde\EEE_R\lft[e^{V_{0,R}(\z) + W_{0,R}(\z)} 
e^{i\h \a(\z_x-\z_0)}\rgt]-1\rgt\}
\notag\\
&+2\bar w_{2,a,R}^-(x)
\lft\{e^{-\d E_R|\L|}
\tilde\EEE_R\lft[e^{V_{0,R}(\z) + W_{0,R}(\z)} 
e^{i\bar \h \a(\z_x-\z_0)}\rgt]-1\rgt\}
\notag\\
&+2\bar w_{2,b,R}^-(x)
e^{-\d E_R|\L|}
\tilde\EEE_R\lft[e^{V_{0,R}(\z) + W_{0,R}(\z)} 
e^{i\a(\h \z_x-\bar \h\z_0)}\rgt]
\notag\\
&+2\bar w_{2,b,R}^-(x)
e^{-\d E_R|\L|}
\tilde\EEE_R\lft[e^{V_{0,R}(\z) + W_{0,R}(\z)} 
e^{i\a(\bar \h \z_x-\h\z_0)}\rgt].
\notag\\
&+2 w_{2,c,R}^-(x)
\lft\{e^{-\d E_R|\L|}
\tilde\EEE_R\lft[e^{V_{0,R}(\z) + W_{0,R}(\z)}\rgt]-1\rgt\}.
\eal
Observe that, by \pref{p4a} and \pref{p4b}  
$$
\lim_{R\to \io}\tilde\EEE_R\lft[
e^{i\h \a(\z_x-\z_0)}\rgt]=1, \qquad 
\tilde\EEE_R\lft[e^{i\a(\h \z_x-\bar \h\z_0)}\rgt]=0.
$$
From them it is  easy to show that \pref{am} is vanishing in the limit 
$R\to \io$. 
\appendix
\section{Functional Integral Formulation}
\lb{appA}
\subsection{Sine-Gordon transformation}
It has been long known that free-energy and  correlations of the Coulomb gas 
can be formulated as expectations with respect to a Gaussian measure 
\citep{Kac59,Sie60}.
Since  the Yukawa potential $W_\L(x,m)$ in \pref{yuk} is strictly positive definite, 
a finite dimensional Gaussian field $\{\f_x:x\in \L\}$ 
is defined by assigning zero mean and 
covariance \pref{wruble}. Therefore, for real $\s_1, \ldots, \s_n$  and $x_1\ldots, x_n\in \L$, 
we have
\bal\lb{white}
&\EEE_{m,\b}\lft[\exp\Big(\sum_{j=1}^n \s_j \f_{x_j}\Big)\rgt]
\notag\\
&= 
e^{-\frac\b2Q^2 W_\L(0;m)}\exp\Big\{-\frac\b2\sum_{i,j=1}^n \s_i\s_j 
\lft[W_\L(x_i-x_j;m)-W_\L(0;m)\rgt]\Big\}
\eal
where $Q:=\sum_{j=1}^n \s_j$. 
Now note that in the limit $m\to 0$ the coefficient $W_\L(0;m)$ is positively 
divergent; whereas under the same limit $W_\L(x;m)-W_\L(0;m)$ converges to 
$W_\L(x|0)$ in \pref{elp}; hence
\bal\lb{white2}
&\lim_{m\to 0}\EEE_{m,\b}\lft[\exp\Big(\sum_{j=1}^n \s_j \f_{x_j}\Big)\rgt]
\notag\\
&= 
\begin{cases}
\exp\lft\{-\frac\b2\sum_{i,j=1}^n \s_i\s_j 
W_\L(x_i-x_j|0)\rgt\}& \quad  \text{if $\sum_j \s_j=0$}
\\
0 & \quad  \text{otherwise}.
\end{cases}
\eal
Using Taylor expansion  in $z$ and \pref{white2} we have
\bal\lb{white3}
&\lim_{m\to 0}\EEE_{m,\b}\lft[e^{2z\sum_{x\in \L}\cos\f_x}\rgt]=Z_\L(\b,z),
\notag\\
&\lim_{m\to 0}\EEE_{m,\b}\lft[e^{i\h (\f_x-\f_y)} e^{2z\sum_{x\in \L}\cos\f_x}\rgt]=Z^{p_1,p_2}_\L(\b,z);
\eal
from these the functional integral formulation \pref{whitman} follows. 
\subsection{Multiscale decomposition of the Gaussian measure}
We review the construction of the multiscale representation; for details 
the reader can consult  Appendix A of \citep{Fa12}. 
In that paper we gave an explicit procedure to obtain  the decomposition 
\be\lb{dec}
W_\L(x;m)=\sum_{j=0}^{R-1}\G_{j}(x;m)+ \G'_{R}(x;m)
:=\G_{R-1,0}(x;m)+ \G'_{R}(x;m),
\ee
where the terms involved 
are such that $\G_j(x)\=\G_{j}(x;0)$ and $\G'_{R}(x)\=\G'_{R}(x;m)$ satisfy 
the properties discussed after \pref{p4}. 
For any $s\in (0,\frac12)$, consider the
non-negative definite, $s$--dependent  potential  
$$
\tilde W_\L(x;m):=\frac{1-s}{|\L|}\sum_{p\in \L^*}
\frac{\hat\G_{R-1,0}(p;0)+ \hat \G'_{R}(p;m)}{1+s \hat\D(p)
\lft[\hat\G_{R-1,0}(p;0)+ \hat \G'_{R}(p;m)\rgt]} e^{i xp}
$$
where $\hat\G_{R-1,0}(p;m)$ and $\hat \G'_{R}(p;m)$ are the Fourier transforms
of $\G_{R-1,0}(x;m)$ and $\G'_{R}(x;m)$. Call $\tilde \EEE_{m,\b}$ the associated 
Gaussian expectation. In the limit $m\to 0$, regardless of $s$, 
$\tilde W_\L(0;m)$ is positively divergent, while 
$\tilde W_\L(x;m)- \tilde W_\L(0;m)$ converges to $W_\L(x|0)$ --indeed, 
since $W_\L(x;m)$ is the inverse of $-\D +m^2$,  one has $\hat\D(p)
\lft[\hat\G_{R-1,0}(p;0)+ \hat \G'_{R}(p;0)\rgt]\=-1$. Therefore,
using  \pref{white} and \pref{white2}, 
it is easy to see that \pref{white3} are still valid 
if we replace $\EEE_{m,\b}$ with $\tilde \EEE_{m,\b}$. 

Now let us consider \pref{white3} with the latter Gaussian expectation.
For reason related to the RG procedure, $\tilde W_\L$ has been chosen so to be able to
extract from the measure and to add to the interaction 
a counterterm proportional to $\frac s2 (\dpr^\m \f_x)^2$. 
However  note that it is not known whether
$\tilde W_\L(x;m)$ is strictly positive definite;
therefore, to have a Gaussian measure with a density, define   $g(x;m)$ such that 
$\b\tilde W_\L(x;m)=\sum_{y\in \L} g(x-y;m)g(y;m)$; then, for any integrable  function 
$F(\f)$, such as the ones in \pref{white3},  we have 
$$
\tilde \EEE_{m,\b}\lft[F(\f)\rgt]= 
\EEE_I \lft[F (g^\f)\rgt]
$$
where $\EEE_I$ is the Gaussian expectation such that  
$\EEE_I[\f_x \f_y]=\d_{x,y}$, 
and $g^\f_x:=\sum_{y\in \L} g(x-y;m)\f_y$.
If $\a^2:=\b(1-s)$ and
$m_s:=\frac{m}{\sqrt{1-s}}$, we have
\bal\lb{A4}
\tilde \EEE_{m_s,\b}\big\{F(\f)\big\}
&=\EEE_A\Bigg\{ F(g^\f)\;
\exp\Bigg[\frac s{2\a^2}\sum_{x\in \L\atop \m\in\hat u}(\dpr^\m g^\f_x)^2\Bigg]\Bigg\}
\NN_\L(s;m) 
\notag\\
&=\EEE_B\Bigg\{ F(\a \f)\;
\exp\Bigg[\frac s{2}\sum_{x\in \L\atop \m\in\hat u}(\dpr^\m \f_x)^2\Bigg]\Bigg\}
\NN_\L(s;m) 
\eal
where $\EEE_A$ and $\EEE_B$ are the expectations with respect to the Gaussian measure 
with covariances 
$$
\EEE_A[\f_x \f_y]= \frac{1}{|\L|}\sum_{p\in \L^*}
\lft\{1+s \hat\D(p)\lft[\hat\G_{R-1,0}(p;0)+ \hat \G'_{R}(p;m)\rgt]\rgt\},
$$ 
$$
\EEE_B[\f_x \f_y]= \frac{1}{|\L|}\sum_{p\in \L^*}
\lft[\hat\G_{R-1,0}(p;0)+ \hat \G'_{R}(p;m)\rgt];
$$ 
and $\NN_\L(s;m)$ takes into account the different normalization of two
measures,
\be
\NN_\L(s;m)=\prod_{k\in \L^*}
\lft[{m^2-(1-s)\hat\D(k)\over m^2-\hat\D(k)}\rgt]^{\frac12}.
\ee
Finally  \pref{pot0} follows from the identity
\be\lb{A5}
\EEE_{B}\big[F(\f)\big] = \EEE_{R}\EEE_{R-1}\cdots\EEE_{1}\EEE_{0}
\big[F(\z^{(R)}+\z^{(R-1)}+\cdots+\z^{(0)} )\big] 
\ee
where $\z^{(R)},\z^{(R-1)},\ldots,\z^{(0)}$ are two-by-two independent Gaussian fields with 
covariances 
\bal
\EEE_{j}[\z^{(j)}_x\z^{(j)}_y]=
\begin{cases}
\G_j(x-y;0)\=\G_j(x-y)& \text{for $j=0,1,\ldots, R-1$},
\notag\\
\G'_R(x-y;m)\=\G'_R(x-y) &\text{for $j=R$}.
\end{cases}
\eal
\section{Proof of the Power Counting Theorems}
\lb{A6}
\subsection{Some Preliminary Inequalities}
\begin{lemma}\lb{lf1}
Let   $F\in \NN_j(X)$ with $X\in \SS_j$. 
For any $x_0\in X$, if $(\d\f)_x:=\f_x-\f_{x_0}$ and $\r:=5L^{-1}$,  
\be\lb{f1}
\|F(\d\f)\|_{h,T_{j+1}(\f,X)}\le 
\|F(\x)\|_{\r h,T_j(\x,X)}\Big|_{\x_x=\d\f_x};
\ee
\end{lemma}
\bpr
\pref{f1} follows from 
the identity 
$$
\sum_{x\in X^*} f_x {\dpr F(\d\f)\over \dpr\f_x}=
\lft.\sum_{x\in X^*} (\d f)_x {\dpr F\over \dpr\f_x}(\x)\rgt|_{\x=\d\f}
$$
and the fact that, for  $X$ small,   
$\|\d f\|_{\CC^2_j(X)}\le 5 L^{-1} \|f\|_{\CC^2_{j+1}(X)}$.
\epr
\begin{lemma}\lb{lf2}
Let   $F\in \NN_j(X)$
and $X\in \PP_j$. 
Given $\ps\in \CC^2_j(X)$, if $\D:=\|\ps\|_{\CC^2_j(X)}$,
\be\lb{f2}
\|F(\f+\ps)\|_{h,T_{j}(\f,X)}\le \|F(\f)\|_{h+\D,T_{j}(\f,X)};
\ee
\end{lemma}
\bpr
From the definition of the norm $T^n_j(\f,X)$,  
$$
\|D^n F(\f+\ps)||_{T^n_j(\f,X)}\le \sum_{m\ge0} \frac{\D^m}{m!} \|D^{m+n}F(\f)\|_{T^{m+n}_j(\f,X)}.
$$
From this \pref{f2} follows. 
\epr
\begin{lemma}\lb{lf3}
Let   $F\in \NN_j(X)$ with $X\in \PP_j$. 
For $h>0$ and $m\in \NNN$ 
\be\lb{f3}
\|\Rem_{m,\x} F(\x)\|_{h,T_{j}(\x,X)}
\le 
2(1+h^{-1}\|\x\|_{\CC^2_j(X)})^{m+1}
\sup_{t\in [0,1]} \|F(t\x)\|_{h,T^{\ge m+1}_{j}(t\x,X)}.
\ee
where
$\|F(\f)\|_{h,T^{\ge m}_j(\f,X)}:=\sum_{n\ge m}{h^n\over n!}\|D^n F(\f)\|_{T^n_j(\f,X)}$
\end{lemma}
\bpr
For $n\ge m+1$, obviously 
\be\lb{f3a}
\|D^n\Rem_{m,\z} F(\x)\|_{T^n_j(\x,X)}=\|D^n F(\x)\|_{T^n_j(\x,X)}.
\ee
For $0\le n\le m$, 
\bal
&D^n \Rem_{m,\x}(\x)\cdot (f_1,\ldots,f_n)
\notag\\
&=D^n F(\x)\cdot (f_1, \ldots, f_n)
-\Tay_{m-n,\x} \lft[D^n F(\x)\cdot (f_1, \ldots, f_n)\rgt]
\notag\\
&=\Rem_{m-n,\x} \lft[D^n F(\x)\cdot (f_1, \ldots, f_n)\rgt]
\notag\\
&=\int_0^1dt\; {(1-t)^{m-n}\over (m-n)!} 
D^{m+1}_{\x} F(t\x)\cdot (f_1,\ldots,f_n,\x,\ldots,\x);
\eal
then, 
\be\lb{f3b}
\|D^n\Rem_{m,\z} F(\x)\|_{T^n_j(\f,X)}\le\frac{\|\x\|^{m+1-n}_{\CC^2_j(X)}}{(m+1-n)!}
\sup_{t\in [0,1]}\|D^{m+1}F(t\x)\|_{T^{m+1}_j(t\x,X)}.
\ee
From \pref{f3a} and \pref{f3b} we obtain 
\bal
&\sum_{n=0}^{m+1} \frac{h^n}{n!}\|D^n\Rem_{m,\z} F(\x)\|_{T^n_j(\f,X)}
\notag\\
&\le 
(1+h^{-1}\|\x\|_{\CC^2_j(X)})^{m+1}\frac{h^{m+1}}{(m+1)!}
\sup_{t\in [0,1]}\|D^{m+1}F(t\x)\|_{T^{m+1}_j(t\x,X)}
\notag\\  
&\le 
(1+h^{-1}\|\x\|_{\CC^2_j(X)})^{m+1}
\sup_{t\in [0,1]}\|F(t\x)\|_{h,T^{\ge m+1}_j(t\x,X)}  
\eal
From that, \pref{f3} follows. 
\epr
\subsection{Charged Components Decomposition}\lb{abCG}
By induction on the scale $j$, the polymer activities $K_{0,j}(\f,X)$ are invariant 
under the global 
translations $\f_y\to \f_y+\frac{2m\p}{\a}$ for any $m\in \ZZZ$.
Define
the function of real variable  
$F(t):=K_{0,j}(\f+t,X)$, which is smooth and 
periodic of period $2\p/a$. Expanding 
$F(t)$ in (absolutely convergent) Fourier series and setting $t=0$,  
one obtains the first of \pref{dec3} with charged components  
$$
\hK_{0,j}(q,\f,X):={\a\over 2\p}\int_{0}^{2\p\over\a}\!ds\ 
K_{0,j}(\f+s,X) e^{-iq\a s }.
$$
Besides, 
since $G_j(\f,X)$ only depends 
upon the derivatives of $\f$,  
\be\lb{cin}
\|\hK_{0,j}(q,\f,X)\|_{h,T_j(\f,X)}
\le \|K_{0,j}(X)\|_{h,T_j(X)}  G_j(\f,X),
\ee
which proves \pref{cin0}.
To obtain the other two of \pref{dec3}, one can verify by inspection 
of \pref{5.25} and \pref{exprem} and inductively that   
$e^{-i\h \a\s \f_x}K_{1,j}(\f,X,x,\s)$ and 
$e^{-i\bar\h \a\s \f_x}K^\dagger_{1,j}(\f,X,x,\s)$ are invariant under 
the transformation $\f_y\to \f_y+\frac{2m\p}{\a}$ for any $m\in \ZZZ$.
Therefore the charged components in these cases are
$$
\hK_{1,j}(q,\f,X, x, \s):={\a\over 2\p}\int_{0}^{2\p\over\a}\!ds\ 
K_{1,j}(\f+s,X, x,\s) e^{-i(q+\h\s)\a s },
$$
$$
\hK^\dagger_{1,j}(q,\f,X, x, \s):={\a\over 2\p}\int_{0}^{2\p\over\a}\!ds\ 
K_{1,j}(\f+s,X, x, \s) e^{-i(q+\bar\h\s)\a s }.
$$
Again it is not difficult to see that 
\be
\|\hK_{1,j}(q,\f,X, x, \s)\|_{h,T_j(\f,X)}
\le \|K_{1,j}(X, x, \s)\|_{h,T_j(X)}  G_j(\f,X),
\ee
\be
\|\hK^\dagger_{1,j}(q,\f,X, x, \s)\|_{h,T_j(\f,X)}
\le \|K^\dagger_{1,j}(X, x, \s)\|_{h,T_j(X)}  G_j(\f,X),
\ee
which proves \pref{cin1}. The proof of \pref{dec3.0} and \pref{cin2} follows from 
similar arguments. 
\subsection{Proof of the first dimensional bound}\lb{ptdh}
Here we  prove Theorem \ref{dh}, which provides the first type of 
dimensional bound. 
We begin with setting up some notations. 
Consider the Gaussian expectation  $\EEE_j$ with covariance $\G_j$ 
and also the Gaussian expectation $\EEE_I$ with covariance $I=(\d_{i,j})$. 
Decompose $\G_j$ as $\G_j= g_j\circ
g_j$ and call  $(g_j f)_x:= \sum_{y\in \L} g_j(x-y) f_y$ 
and likewise for $(\G_j f)_x$. 
Consider an integrable charge $p$ activity $F(\f)\=F(\f,X)$. 
Under the imaginary translation $\z_x\to \z_x+i(g_jf)_x$ where 
$f$ is any test function  with finite support, 
$$
\EEE_j[F(\f)]=\EEE_I[F(\f'+(g_j\z))]=
e^{{1\over 2} (f,\G_jf)}\;
\EEE_j\lft[e^{-i(\z,f)} F(\f+i (\G_j f))\rgt].
$$
(The measure $\EEE_I$ is involved in the identity to avoid to make the imaginary 
translation in a  {\it degenerate} Gaussian measure, as in principle $\EEE_j$ could  be.)
Now use the identity $F(\f)=e^{i\a p\th }F(\f-\th)$ for any constant complex 
field $\th$:   
calling $\ps_x:=(\G_j f)_x$ and, 
for $x_0\in X$,  setting $\d\ps_x:=(\G_j f)_x-(\G_j f)_{x_0}$, we have 
\bal\lb{op}
\EEE_j\lft[F(\f)\rgt]
&= e^{{1\over 2}(f, \G_j f)-\a p (\d_{x_0}, \G_j f)}\;
\EEE_j\lft[e^{-i(\z,f)} F(\f+i \d\ps,)\rgt]
\eal
where $(\d_{x_0})_x:=\d_{x,x_0}$. 
In order to minimize  the prefactor in  the r.h.s. of 
\pref{op}, one can set 
$f_x=\a p\d_{x,x_0}$. However, the size of such an $f_x$ grows in $p$, 
and this conflicts with the assumption of finite radius of analyticity
for all the activities $F(\f)$. To avoid this problem, 
we consider two cases:
\bd
\item if $|p|\le 1$, we make the optimal choice $f_x=\a p\d_{x,x_0}$ 
\item if $|p|>1$, we follow \citep{DH00} and set $f_x=\a {\rm\; sgn}(p)\d_{x,x_0}$
(for $ {\rm sgn}(x):=x/|x|$).
\ed 
Therefore, from \pref{op} we obtain
\bal\lb{ct1}
&\|\EEE_j\lft[F(\f)\rgt]\|_{h,T_{j+1}(\f',X)}
\le e^{-d(p)\frac{\a^2}{2}\G_j(0)}\;
\EEE_j\lft[\|F(\f+i \d\ps)\|_{h,T_{j+1}(\f',X)}\rgt]
\eal
where $d(p):=p^2$ for $|p|\le 1$ and $d(p):=2|p|-1$ otherwise. 
Note that according to definition  \pref{hut} 
for any value of $p$ we have 
$$
\D:=\|\d\ps\|_{\CC^2_j(X)}\le \frac h2.
$$
Now consider the expectation on the r.h.s. of \pref{ct1}; 
and set  $\r:=5 L^{-1}$, $H_x:=\z_x+i(\d\ps)_x$, .
Since $\|e^{ip \f_{x_0}}\|_{h,T_{j+1}(\f',X)}$ is less than
$e^{h|p|\a}$ (which is  $L$-independent), by
 \pref{f1}, \pref{f2}, \pref{cin.0},
and for $L$ so large that $\r\le \frac 12$, (hence $\r h+\D\le h$) 
\bal\lb{ct2}
&\|F(\f+i\d\ps)\|_{h,T_{j+1}(\f',X)}
\le e^{h|p|\a}
\|F(\d\f'+H)\|_{h,T_{j+1}(\f',X)} 
\cr
&\qquad\qquad\qquad\qquad\qquad
\le e^{h|p|\a}\|F(\x+H)\|_{\r h,T_j(\x, X)}\Big|_{\x:=\d\f'}
\cr
&\qquad\qquad\qquad\qquad\qquad
\le e^{h|p|\a}\|F(\x+\z)\|_{\r h+\D,T_j(\x,X)}\Big|_{\x:=\d\f'}
\cr
&\qquad\qquad\qquad\qquad\qquad
\le e^{h|p|\a}\|F\|_{h,T_j} A^{-|X|_j} G_j(\f,X).
\eal
The last inequality is due to the fact that
$G_j(\f,X)$ depends on the derivatives of $\f$, and then 
$G_j(\d\f+\z,X)=G_j(\f,X)$.
Finally,  \pref{eqdh} is obtained by 
plugging \pref{ct2} into \pref{ct1} and using \pref{6.54} for the 
integration $\EEE_j$. 
This completes the proof of Theorem \ref{dh}. 
\subsection{Proof of the second dimensional bound}\lb{callan}
We want to prove Theorem \ref{by}, which gives the 
second dimensional bound. 
From \pref{ct1} and the inequality 
$\|e^{ip \f_{x_0}}\|_{h,T_{j+1}(\f',X)}\le e^{h|p|\a}$,
we find 
\bal\lb{ct5a}
&\|\Rem_{m,\d\f'} \EEE_j\lft[F(\f)\rgt]\|_{h,T_{j+1}(\f', X)}
\le e^{-d(p)\frac{\a^2}{2}\G_j(0)}\;
\|\Rem_{m,\d\f'} \EEE_j\lft[F(\f+i\d\ps)\rgt]\|_{h,T_{j+1}(\f', X)}
\notag\\
&\le e^{h|p|\a} e^{-d(p)\frac{\a^2}{2}\G_j(0)}\;
\EEE_j\lft[\|\Rem_{m,\d\f'}F(\d\f'+H)\|_{h,T_{j+1}(\f',X)}\rgt]
\eal
where $H_x:=\z_x+i\d\ps_x$.
As in the previous proof,  $\D:=\|\d\ps\|_{C^2_j(X)}\le \frac h2$ and 
$\r:= 5 L^{-1}$ is small for large enough $L$. 
Now use  \pref{f1}, \pref{f3} and \pref{f2} to obtain 
(the definition of the seminorm 
$\|\cdot\|_{\r h,T^{\ge m+1}_j(\f,X)}$ is in Lemma \ref{lf3}) 
\bal\lb{ct5}
&\|\Rem_{m,\d\f'} F(\d\f'+H)\|_{h,T_{j+1}(\f',X)}
\le
\|\Rem_{m,\x} F(\x+H)\|_{\r h,T_j(\x,X)}
\Big|_{\x=\d\f'}
\cr
&\le
2\lft(1+(\r h)^{-1}\|\x\|_{\CC^2_j(X)}\rgt)^{m+1}
\sup_{t\in[0,1]}\| F(t\x+H)\|_{\r h,T^{\ge m+1}_j(t\x,X)}
\Big|_{\x=\d\f'}
\notag\\
&\le
2\lft(1+(\r h)^{-1}\|\x\|_{\CC^2_j(X)}\rgt)^{m+1} (2\r)^{m+1}
\sup_{t\in[0,1]}\|F(t\x+H)\|_{\frac h2,T_j(t\x,X)}
\Big|_{\x=\d\f'}
\notag\\
&
\le
2\lft(1+(\r h)^{-1}\|\x\|_{\CC^2_j(X)}\rgt)^{m+1} (2\r)^{m+1}
\sup_{t\in[0,1]}\|F(t\x+\z)\|_{\frac h2+\D,T_j(t\x,X)}
\Big|_{\x=\d\f'}
\eal
To obtain the third line we used that
$\|\cdot\|_{s h,T^{\ge m+1}_j(\f,X)}\le s^{m+1}\|\cdot\|_{h,T_j(\f,X)}$. 
As  $X\in \SS_j$, 
\be\lb{lng}
L\|\d\f'\|_{\CC^2_j(X)}
\le C \max_{p=1,2}\|\nabla^p_{j+1}\f'\|_{L^\io(X^*)}.
\ee
Besides, since $G_j$ depends upon the derivatives of the fields, 
$G_j(t\d\f'+\z,X)=G_j(t\f'+\z,X)$.
Therefore
\bal\lb{israel}
&\|\Rem_{m,\d\f'} F(\d\f'+H)\|_{h,T_{j+1}(\f',X)} 
\notag\\
&\le
C^{m+1} L^{-(m+1)}
\|F\|_{h,T_j(X)}
\lft(1+\max_{p=1,2}\|\nabla^p_{j+1}\f'\|_{L^\io(X^*)}\rgt)^{m+1}
\notag\\
&\qquad\times
\sup_{t\in[0,1]} G_j(t\f'+\z,X).
\eal
Finally, Theorem \ref{by} is proven once \pref{israel} is plugged into 
\pref{ct5a} and last part of Lemma \ref{l6.53} 
is used for the integration $\EEE_j$. 
%
%\small
\bibliographystyle{/home/pierluigi/.archivio/biblio/stile3}
\bibliography{/home/pierluigi/.archivio/biblio/bibliografia} 
\end{document}